\documentclass[12pt]{article}
\usepackage{amssymb,amsmath,amsthm,mathrsfs,latexsym,amsxtra,graphicx,appendix,setspace,subfig,fix-cm,color,cite,adjustbox,tabularx,comment,afterpage}
\usepackage{hyperref}
\numberwithin{equation}{section}

\usepackage{titlesec}
\titleformat{\paragraph}
{\normalfont\normalsize\bfseries}{\theparagraph}{1em}{}
\titlespacing*{\paragraph}{0pt}{3.25ex plus 1ex minus .2ex}{1.5ex plus .2ex}

\usepackage{tikz}
\usetikzlibrary{positioning}
\usetikzlibrary{intersections} 
\newlength{\PicScale}
\definecolor{indiagreen}{rgb}{0.07, 0.53, 0.03}

\newcommand{\nn}{\nonumber}

\usepackage{longtable, multirow, array}

\newcolumntype{M}[1]{>{\centering\arraybackslash}m{#1}}

\newcommand{\bea}{\begin{eqnarray}}
\newcommand{\eea}{\end{eqnarray}}
\newcommand{\be}{\begin{equation}}
\newcommand{\ee}{\end{equation}}
\newcommand{\bal}{\begin{align}}
\newcommand{\eal}{\end{align}}

\newcommand{\bz}{{\mathbb Z}}

\newcommand{\zb}{{\bar z}}

\renewcommand{\d}{{\rm d}}

\newcommand{\pa}{\partial}
\newcommand{\pab}{\bar{\partial}}

\newcommand{\ti}[1]{\tilde{#1}}

\begin{document} 

\begin{titlepage}
\thispagestyle{empty}
\begin{center}

\begin{flushright}
MITP-23-082
\end{flushright}

\hfill \\
\hfill \\
\vskip 0.75in
{\Large 
{\bf Conformal Perturbation Theory for $n$-Point Functions:\\[7pt]Structure Constant Deformation}
}\\

\vskip 0.4in

{\large Benjamin A. Burrington${}^{a}$ and Ida G.~Zadeh${}^{b}$
}\\
\vskip 4mm

${}^{a}$
{\it Department of Physics and Astronomy, Hofstra University, Hempstead, NY 11549, USA} \vskip 1mm
${}^{b}$
{\it {\it PRISMA+} Cluster of Excellence and Mainz Institute for Theoretical Physics, Johannes Guttenberg-Universit\"at Mainz, Staudinger Weg 9, 55128 Mainz, Germany} \vskip 1mm

\end{center}

\vskip 0.35in

\begin{center} {\bf Abstract } \end{center}
We consider conformal perturbation theory for $n$-point functions on the sphere in general 2D CFTs to first order in coupling constant.  We regulate perturbation integrals using a canonical hard disk excisions of size $\epsilon$ around the fixed operator insertions, and identify the full set of counter terms which are sufficient to regulate all such integrated $n$-point functions. We further explore the integrated 4-point function which computes changes to the structure constants of the theory.  Using an $sl(2)$ map, the three fixed locations of operators are mapped to $0$, $1$, and $\infty$.  We show that approximating the mapped excised regions to leading order in $\epsilon$ does not lead to the same perturbative shift to the structure constant as the exact in $\epsilon$ region. We explicitly compute the correction back to the exact in $\epsilon$ region of integration in terms of the CFT data. We consider the compact boson, and show that one must use the exact in $\epsilon$ region to obtain agreement with the exact results for structure constants in this theory.

\vfill


\end{titlepage}

\setcounter{page}{1}
\setcounter{tocdepth}{2}

\tableofcontents
\thispagestyle{empty}

\vspace{2cm}
\newpage

\section{Introduction and summary}\label{section_intro}
\setcounter{page}{1}
\hypersetup{pageanchor=false}

Conformal field theories (CFTs) with exactly marginal operators (moduli) form continuous families of CFTs. The geometrical description of a family of CFTs is given by the moduli space, whose coordinates are the exactly marginal couplings of the member CFTs. CFTs in a moduli space\footnote{The moduli space is sometimes referred to as the conformal manifold.} have the same central charge and have an enhanced symmetry group, which may be further enhanced at particular points or loci in the moduli space. In two dimensions, non-linear sigma-models with target spaces being tori or Calabi-Yau manifolds are among the well-known examples of continuous families of CFTs \cite{Wendland:2000ye,Nahm:1999ps}. Various aspects of the moduli spaces of 2-dimensional CFTs have been developed in the literature, often in the context of string theory, but also with applications in condensed matter \cite{Narain:1985jj,Narain:1986am,Dijkgraaf:1987vp,Ginsparg:1987eb,Seiberg:1988pf,Dixon:1988ac,Cecotti:1990kz,Friedan:2012hi,Gomis:2016sab,Baggio:2017aww,Balthazar:2022hzb,Ji:2019ugf}. 

In many cases, for instance Calabi-Yau sigma-models, generic points in the moduli space are not well characterized, i.e. the specific CFTs at those points are unknown.  Regions on these moduli spaces with further enhanced symmetries such as the orbifold loci or Gepner points \cite{Gepner:1987vz,Gepner:1987qi} offer special points where the CFTs are known. One may hope that the known CFTs along with knowledge of geometric aspects of the moduli space may be used to uncover features of the CFTs at generic points, especially for the nearby CFTs. Exploring the changes to a CFT by deforming the theory by an exactly marginal operator is known as conformal perturbation theory. The goal of this work is to formulate the conformal perturbation theory of $n$-point correlation functions in general 2-dimensional CFTs. Our main motivation is to develop a systematic study of the deformation of the structure constants of a CFT.

Together with the conformal dimensions, the structure constants form the complete set of the parameters of the CFT. While the deformation of the conformal dimensions has been extensively studied, the deformation of the structure constants is somewhat less explored. The latter requires the conformal perturbation theory of the 3-point functions, which we consider at first order, as a special case of our $n$-point function perturbation framework.

Conformal perturbation theory has wide applications in physics, and in particular in string theory. The conformal field theory framework and the geometric setup of the moduli space were developed in \cite{Kadanoff:1978pv,KADANOFF197939,Zamolodchikov:1987ti,Cardy:1987vr,Kutasov:1988xb,Dijkgraaf:1987jt,Chaudhuri:1988qb,Ranganathan:1992nb,Ranganathan:1993vj,Gaberdiel:2008fn}. Some applications to non-linear sigma-models and WZW models have been studied in \cite{Yang:1988bi,Hassan:1992gi,Eberle:2001jq,Forste:2003km,eberlephd,Fredenhagen:2007rx,Poghossian:2013fda,Behan:2017mwi,Keller:2019suk,Keller:2019yrr,Benjamin:2020flm,Keller:2023ssv}. In the case of sigma-models on K3, understanding generic CFTs plays a crucial role in resolving two long-standing puzzles. One is the emergence of the Mathieu group $\mathbb M_{24}$ in K3 sigma-models and the conformal field theoretic origin of the Mathieu moonshine \cite{Eguchi:2010ej}, and the other is the distribution of rational points in the moduli space of K3 CFTs \cite{Gukov:2002nw}.  

The case of 2-dimensional CFTs with moduli spaces is also of great importance in the context of the holographic principle, namely in the string theory realisations as well as in other formulations of the AdS$_3$/CFT$_2$ correspondence \cite{Maldacena:1997re,Giveon:1998ns,Seiberg:1999xz,Aharony:1999ti,Larsen:1999uk,Jevicki:1998bm,David:2002wn,Gukov:2004ym,Gaberdiel:2010pz,Gaberdiel:2012uj,Gaberdiel:2013vva,Gaberdiel:2014cha,Belin:2014fna,Eberhardt:2017pty,Gaberdiel:2018rqv,Giribet:2018ada,Eberhardt:2018ouy,Belin:2019rba,Eberhardt:2019ywk,Li:2019qzx,Belin:2020nmp,Eberhardt:2020akk,Maloney:2020nni,Afkhami-Jeddi:2020ezh,Eberhardt:2020bgq,Balthazar:2021xeh,Martinec:2022okx}. The dual CFT in many of these holographic dualities is the symmetric product orbifold of a seed CFT, where the seed theory can be e.g. a non-linear sigma-model with $T^4$, K3 or other target spaces, or a supersymmetric minimal model. 

In conformal perturbation theory one must start with a well characterized CFT, and deform away from it on the moduli space. Various methods for computing correlation functions on both sides of the holographic dualities have been developed \cite{Lunin:2000yv,Lunin:2001pw,Pakman:2009zz,Pakman:2009ab,Dei:2019iym,Li:2020zwo,Gaberdiel:2020ycd,AlvesLima:2022elo,Dei:2022pkr,Guo:2022sos,Guo:2022zpn,Guo:2023czj}, and applied to perturbative analysis. In some cases the perturbed quantities may be protected by supersymmetry or other mechanisms, and do not vary when deforming the theory \cite{Lerche:1989uy,deBoer:2008ss,Baggio:2012rr}.  Also of interest are quantities which acquire perturbative corrections. Conformal perturbation theory on the moduli space of the holographic CFTs has been studied from different aspects\cite{Gava:2002xb,Pakman:2009mi,Avery:2010er, Avery:2010hs,Burrington:2012yq,Burrington:2012yn,Gaberdiel:2013jpa,Carson:2014yxa,Berenstein:2014cia,Carson:2014ena,Burrington:2014yia,Gaberdiel:2015uca, Carson:2015ohj,Carson:2016uwf,Burrington:2017jhh,Bashmakov:2017rko,Hampton:2018ygz,Guo:2019ady,Lima:2020boh,Lima:2020nnx,Guo:2020gxm,Lima:2020kek,Eberhardt:2021jvj,Lima:2021wrz,Benjamin:2021zkn,Eberhardt:2021vsx,Apolo:2022fya,Guo:2022and,Guo:2022ifr,Benjamin:2022jin,Burrington:2022dii,Burrington:2022rtr,Apolo:2022pbq,Fiset:2022erp,Hughes:2023apl,Lerche:2023wkj,Hughes:2023fot,Hikida:2023jyc}. Many of these investigations characterize CFTs at nearby points in the moduli space by focusing on changes to the conformal dimensions of operators, i.e. examining the spectrum of operators in the perturbed CFTs. Changes to higher $n$-point function data have been discussed more rarely, however, see \cite{Behan:2017mwi} for general considerations for deforming structure constants, and for specific cases \cite{Lima:2020boh,Lima:2020nnx,Lima:2020kek,Eberhardt:2021vsx,Apolo:2022pbq}.  Our aim here is to consider the problem of the shifts to the structure constants generally, track counter terms precisely, and address how to tackle the integrated 4-point function piecewise.

\subsection{Summary of the results}

To summarize the results, we briefly recall the basic structure of conformal perturbation theory.  As usual in quantum field theory, one may consider the space of theories by changing the action of the theory under consideration, $S_0$, by adding to it a perturbative term: $S=S_0+ \delta S$. 
In the path integral formulation, the partition function and correlation functions of the new theory $S$ are computed by expanding ${\rm exp}(S_0+\delta S)$ order by order in perturbation theory.
To restrict conformal perturbation theory to be along the moduli space, we impose $\delta S=\lambda^\ell\int dz^2\,\mathcal O_{\ell}$. The operators $\mathcal O_{\ell}$ are the exactly marginal operators with conformal dimensions (1,1), which parameterize the tangent space of the moduli space at the point $\lambda^\ell=0$.  The coefficients $\lambda^\ell$ are the coupling constants, $\ell=\{1,\cdots,d\}$, and $d$ is the total number of the moduli. Varying the values of $\lambda^{\ell}$ generates a neighborhood of the unperturbed theory in the moduli space.

Generically, an all-order perturbative expansion is extremely difficult.  To make progress, we pick a specific deformation operator $\mathcal{O}_D$, for which the coupling constant is simply given as $\lambda$, and expand in $\lambda$.  At leading order, $\lambda$ may show up in the correlation functions in two ways: in $\lambda \int dz^2\,\mathcal O_{D}$ from expanding the measure of the path integral; and in $\phi_{i,\lambda}\approx \phi_i + \lambda \delta \phi_i$ from explicit lambda dependence of the fields.  The integrals must be regulated, and we do so by excising holes of radius $\epsilon$ from the region of integration.  This introduces $\epsilon$ dependence that is generically divergent, and must be regulated with $\epsilon$ divergent counter terms appearing in $\delta \phi_i$.  Once these divergences are removed, the $\epsilon\rightarrow 0$ limit may be taken to extract the first order in $\lambda$ correction to the given correlator.

The form of the counter terms $\delta \phi_i$ is essentially the question of the connection on the space of the states of the theory.  This has been studied in detail for the coordinates on the moduli space \cite{Kutasov:1988xb} (see \cite{Balthazar:2022hzb} for a more recent work). In some cases, one might hope that the regulators and connections on the moduli space may be computed exactly, as was done in \cite{deBoer:2008ss,Baggio:2012rr} for theories with extended supersymmetry, relying on the enlarged chiral ring to constrain the connections. 
Here we keep our concerns limited to conformal symmetry, and in fact we most heavily rely on the $sl(2)$ structure of the theory, which we review in appendix \ref{app_sl2}. We shall computate corrections to general 3-point functions, that may not be protected by any non-renormalization theorem. Furthermore, we will concern ourselves beyond the exactly marginal operators which determine the local coordinates on the moduli space and connections on these coordinates. 

We begin in section \ref{sec_2pf} with the perturbation of the 2-point function, which arises from an integrated 3-point function.  In the literature, one often takes the two locations of the inserted operators $\phi_1(z_1)$ and $\phi_2(z_2)$ and maps them under an $sl(2)$ transformation to 0 and $\infty$.  The mapped region of integration, and its first order in $\epsilon$ approximation (indicated by $\xrightarrow{\epsilon}$ below), are
\begin{align} \label{3ptappx}
\Big|\hat{z}+\frac{\epsilon^2}{|z_{12}|^2-\epsilon^2}\Big| >\frac{\epsilon}{|z_{12}|}\,\frac{1}{1-\frac{\epsilon^2}{|z_{12}|^2}}\ , 
\qquad\left|\hat{z}-1\right|< \frac{|z_{12}|}{\epsilon}\ \qquad \xrightarrow{\;\;\;\epsilon\;\;\;} \qquad   \frac{\epsilon}{|z_{12}|}<&|\hat{z}|< \frac{|z_{12}|}{\epsilon}\ .
\end{align}
For the leading order in $\epsilon$ region, which we call the {\it simplified domain}, we obtain the $\epsilon$ divergence, and define an appropriate set of counter terms in section \ref{cpt_h}. These are given by \cite{Dijkgraaf:1987jt}
\begin{equation}\label{int_2pf_ct}
\phi_{i,\lambda}=\phi_i-2\pi\lambda\,\ln(\epsilon^2)\, C_{D,i,i}\,\phi_i+4\pi\lambda\!\!\!\sum_{{p,\,d_p<d_i}}\!\!\frac{\delta_{s_i,s_p}}{(d_i-d_p)\,\epsilon^{(d_i-d_p)}}\, 
C_{D,i,p}\,\phi_p+O(\lambda^2)\ ,
\end{equation}
where the indices $\{i,p\}$ above run over the set of quasi primaries in the theory, $d_i=h_i+\ti{h}_i, s_i=h_i-\ti{h}_i$ denote the total dimension and spin of operators, and the structure constant for equal dimension quasi primary operators has been diagnoalized to $C_{D,i,i}$.  The approximation on the \textsc{rhs} of eq. \eqref{3ptappx} is non-local. One may hope to find the full set of counter terms which cancel the divergences for the original domain on the \textsc{lhs} of eq. \eqref{3ptappx}, before taking the leading order in $\epsilon$ approximation. The effects of changing the shape of the excised region in a deformed theory has been discussed elsewhere, for example \cite{Campbell:1990dz}. In section \ref{cpt_o}, we find the correction to eq. \eqref{int_2pf_ct} coming from the exact in $\epsilon$ domain \eqref{3ptappx}, and obtain the full set of counter terms: 
\begin{align}\label{fullcounterintro}
& \phi_{i,\lambda}=\phi_i-2\pi\lambda\,\ln(\epsilon^2)\, C_{D,i,i}\,\phi_i\\
&+4\pi\lambda\!\!\!\!\!\!\!\!\sum_{\substack{p,\, d_p<d_i \\ |s_i-s_p|<\lfloor d_i-d_p\rfloor}}\!\!\!\sum_{\substack{\ell=|s_i-s_p|\\\;\;\; \ell-|s_i-s_p|\in 2{\mathbb Z}\\}}^{\lfloor d_i-d_p\rfloor}  
\!\!\!\!\frac{C_{D,i,p}}{(d_i-d_p-\ell)\epsilon^{d_i-d_p-\ell}}\;\frac{(1-h_i+h_p)_n\,(1-\ti{h}_i+\ti{h}_p)_{\ti{n}}}{n!\,(2h_p)_n\,\ti{n}!\,(2\ti{h}_p)_{\ti{n}}}\;  \pa^n \pab^{\ti{n}} \phi_p+O(\lambda^2) \nn 
\end{align}
where $n$ and $\ti{n}$ are a pair of integers $n=(\ell+(s_i-s_p))/2, \ti{n}=(\ell-(s_i-s_p))/2$ defined by the necessarily integer spin difference $s_i-s_j$.

In section \ref{cpt_npf}, we show that the above counter terms, along with the hard disk cutoff for the integrals, regulates {\it all} $n$-point functions needed in first order perturbation theory. This is the first of the two main results of this work, and is obtained by showing that the above terms naturally arise in the operator product expansions (OPEs) computed in section \ref{subsec_ope}. This procedure is effectively the regulator of \cite{Ranganathan:1993vj} used to define ``connection c'', one of the integrable regulators in that work (although written in that work somewhat generically without using much of the structure of the OPE).  In eq. \eqref{fullcounterintro} we have written the counter terms explicitly in terms of the $sl(2)$ highest weight states, i.e. the quasi primaries, and the conformal dimensions and structure constants that characterize the theory.  We note that the sum \eqref{fullcounterintro} only includes the divergent terms, however, extending it beyond this leads to a natural set of $\epsilon$ independent (i.e. order $\epsilon^0$) terms in two families. The first family is
\begin{equation}
2\pi \lambda\, C_{D,i,p}\;\frac{(-1)^{h_i-h_p}}{(h_i-h_p)\,(2h_p)_{h_i-h_p}}\;\pa^{{h}_i-{h}_p}\phi_p \label{dcounterintro}
\end{equation}        
if there exists a set of quasi primary operators $\phi_p$ with $\ti{h}_p=\ti{h}_i$, $h_i-h_p=s_i-s_p\geq 1$ a positive integer ($s_i=h_i-\ti{h}_i$ denotes the spin), and that $\phi_p$ have non-vanishing structure constants with $\mathcal{O}_D$ and $\phi_i$ $(C_{D,i,p}\ne0)$. The second family is
\begin{equation}
2\pi \lambda\, C_{D,i,p}\;\frac{(-1)^{\ti{h}_i-\ti{h}_p}}{(\ti{h}_i-\ti{h}_p)\,(2\ti{h}_p)_{\ti{h}_i-\ti{h}_p}}\;\pab^{\ti{h}_i-\ti{h}_p}\phi_p \label{dbarcounterintro}
\end{equation} 
if there exists a set of quasi primary operators $\phi_p$ with $h_p=h_i$, $\ti{h}_i-\ti{h}_p=-(\ti{s}_i-\ti{s}_p)\geq 1$ a positive integer, and $C_{D,i,p}\ne0$.  It is interesting to note that these operators lie on the Regge trajectory of operators with lower conformal dimension than the operator in question $\phi_i$.  

Whether or not to include these ``borderline operators'' as part of the counter terms, and whether we have correctly identified the coefficients, is a question of how to preserve the correct functional form of the $n$-point functions, i.e. is a question of naturalness \cite{Kutasov:1988xb}.  This is most cleanly addressed in the most constrained correlators: 2-point and 3-point functions.  Knowing what types of functional form can be removed with counter terms is therefore essential to extract meaningful answers.  Our ``borderline'' operators \eqref{dcounterintro} and \eqref{dbarcounterintro} help address this in $n$-point functions with $n$ larger than 2.  The 2-point functions are special because of the simplicity of the functional form, essentially allowing derivatives to not affect the general functional form: a power of a displacement.  The additional counter terms found above therefore seem to not play a role until one addresses 3-point functions and higher.

In section \ref{cpt_3pf} we turn to the perturbed 3-point functions, which are calculated by an integrated 4-point function. We define the function of the cross ratio $f(\zeta, \bar{\zeta})$ by
\begin{align}
&\langle \mathcal{O}_D(z,\zb)\; \phi_i(z_1,\zb_1)\; \phi_j(z_2,\zb_2)\; \phi_k(z_3,\zb_3)\rangle=  \\
&\Big(\frac{z_{23}}{z_{12}z_{13}}\Big)^{h_i}\Big(\frac{z_{13}}{z_{12}z_{23}}\Big)^{h_j}
\Big(\frac{z_{12}}{z_{13}z_{23}}\Big)^{h_k} \Big(\frac{z_{12}z_{13}}{(z_1-z)^2z_{23}}\Big)^{h_D}\;\times\; ({\rm a.h.})\;\times\; f(\zeta,\bar{\zeta})\ , \nn
\end{align}
where $\mathcal{O}_D$ is the deformation operator. The change to the structure constant arises from the constant in $\epsilon$ part of the perturbation integral $-\lambda\int d^2\hat{z} f(\hat{z},\bar{\hat{z}})$ over a regularized domain, see the generalized sum rule in \cite{Behan:2017mwi}.  We explicitly find the regularized domain, which is
\begin{align}
&\Big|\hat{z}-\frac{z_{12}}{z_{23}}\;\frac{\epsilon^2}{|z_{13}|^2-\epsilon^2}\Big|  > \frac{|z_{12}|}{|z_{23}||z_{13}|}\;\frac{\epsilon}{1-\frac{\epsilon^2}{|z_{13}|^2}}\ , \label{fulldomainintro} \\[2pt]
&\Big|\hat{z}-\Big(1+\frac{z_{13}}{z_{23}}\;\frac{\epsilon^2}{|z_{12}|^2-\epsilon^2}\Big)\Big| > \frac{|z_{13}|}{|z_{23}||z_{12}|}\;\frac{\epsilon}{1-\frac{\epsilon^2}{|z_{12}|^2}}\ ,\qquad\qquad
\Big|\hat{z}+\frac{z_{12}}{z_{23}}\Big|< \frac{|z_{12}||z_{13}|}{|z_{23}|\;\epsilon}\ .\nn 
\end{align}
We find the above domain difficult to analyze, and so consider its first order in $\epsilon$ approximation:
\begin{equation}
|\hat{z}|>\frac{|z_{12}|\,\epsilon}{|z_{23}||z_{13}|}\ , \qquad |\hat{z}-1|>\frac{|z_{13}|\,\epsilon}{|z_{23}||z_{12}|}\ ,
\qquad |\hat{z}|< \frac{|z_{12}||z_{13}|}{|z_{23}|\,\epsilon}   \label{appxdomainintro}
\end{equation}
which we call the simplified domain.  The perturbation integral over the simplified domain is also difficult to compute, even given the conformal block structure of $f(\hat{z},\bar{\hat{z}})$.  We find that the corrections needed to go from the simplified domain \eqref{appxdomainintro} to the exact domain \eqref{fulldomainintro} can be calculated in terms of the CFT data: this is the second of the two main results of this work.

We denote the $\epsilon^0$ part of $-\lambda\int d^2\hat{z} f(\hat{z},\bar{\hat{z}})$ as ${\mathcal I}_{{\rm simp},0}$, simply dropping terms divergent in $\epsilon$: a ``minimal subtraction'' scheme.  As shown in section \ref{cpt_npf}, the divergent terms in $-\lambda\int d^2\hat{z} f(\hat{z},\bar{\hat{z}})$ must combine with divergent parts of the integral over the corrections to the simplified domain, and these must cancel against the counter terms \eqref{fullcounterintro}.  We are able to show that the constant in $\epsilon$ part of $-\lambda\int d^2\hat{z} f(\hat{z},\bar{\hat{z}})$ over the full domain \eqref{fulldomainintro}, which we call $\mathcal{I}_0$, is given by
\begin{align}
\mathcal{I}_0= \mathcal{I}_{{\rm simp},0}+\mathcal I_{i,0}^{\rm hol}+\mathcal I_{i,0}^{\rm ahol}+\mathcal I_{j,0}^{\rm hol}+\mathcal I_{j,0}^{\rm ahol}+\mathcal I_{k,0}^{\rm hol}+\mathcal I_{k,0}^{\rm ahol}
\end{align} 
where each of the $\mathcal I_{\{i,j,k\},0}^{\rm (a)hol}$ may be calculated explicitly, for example:
\begin{align}\label{Ik0intro}
&\mathcal I_{k,0}^{\rm hol}=-2\pi \lambda \kern -2em\sum_{\substack{ p \\ \ti{h}_p=\ti{h}_k,\,h_{k}-h_{p} \in {\mathbb Z}^+}} \kern -2em
\frac{(h^k_p-1)!}{(2h_p)_{h^k_p}}\;C_{p,D,k}\,C_{p,j,i}  
\Big(P^{(-h^{j,k}_i,-h^{i,k}_j)}_{h^k_p} 
\Big(\frac{z_{13}}{z_{12}}+\frac{z_{23}}{z_{12}}\Big)-\frac{ (-h^{j,k}_i+1)_{h^k_p}}{(h^k_p)!}\Big)\ .
\end{align}
Above $P$ is the Jacobi polynomial, and we have used $h^{1,2}_3:=h_1+h_2-h_3$ and $h^1_{2}:=h_1-h_2$ to condense notation. The Jabcobi polynomial above is exactly the functional form that may be removed using eqs. \eqref{dcounterintro} and \eqref{dbarcounterintro} at each point, even agreeing with the coefficient inferred by taking non-divergent terms in the sum \eqref{fullcounterintro}.  While the Jacobi polynomial is canceled, the constant parts represent contributions to the change in the structure constant.  Thus, the change to the structure constant to first order in $\lambda$ is of the form:
\begin{equation}\label{deltaCintro} 
\delta C_{i,j,k} = \mathcal{I}_{{\rm simp},0}^{J}+\mathcal I_{i,0}^{\rm hol,J}+\mathcal I_{i,0}^{\rm ahol,J}+\mathcal I_{j,0}^{\rm hol,J}+\mathcal I_{j,0}^{\rm ahol,J}+\mathcal I_{k,0}^{\rm hol,J}+\mathcal I_{k,0}^{\rm ahol,J}\ ,
\end{equation}
where $\mathcal I_{i,0}^{\rm hol,J}$ means to remove the Jacobi polynomials which may be canceled by counter terms \eqref{dcounterintro} and \eqref{dbarcounterintro}  (other counter terms may be necessary in $\mathcal{I}_{{\rm simp},0}^{J}$ as well).  For instance,
\begin{equation}
\mathcal I_{k,0}^{\rm hol,J}=2\pi \lambda \kern -2em\sum_{\substack{ p \\ \ti{h}_p=\ti{h}_k,\,h_{k}-h_{p} \in {\mathbb Z}^+}} \kern -2em
\frac{(h^k_p-1)!}{(2h_p)_{h^k_p}}\;C_{p,D,k}\,C_{p,j,i}\frac{ (-h^{j,k}_i+1)_{h^k_p}}{(h^k_p)!}
\end{equation}
can be read from eq. \eqref{Ik0intro}. The contribution from $\mathcal{I}_{{\rm simp},0}^{J}$ must be calculated theory by theory over the simplified domain,  and then one must add to it the contributions \eqref{deltaCintro}. In section \ref{sec_bos} we use the compact boson CFT to give an example where contributions from the simplified domain and the corrections above are necessary.

The rest of the paper is organized as follows. In section \ref{sec_2pf} we consider the integrated 3-point function which gives the corrections to the 2-point function.   We find the full set of counter terms \eqref{fullcounterintro} needed to regulate the integral, and compute the anomalous dimension of the quasi primaries. We find that this calculation is unchanged whether one uses the simplified domain or the exact domain of integration \eqref{3ptappx}. In section \ref{cpt_npf}, we show that the same counter terms arise by considering the OPE of the deformation operator with the fixed operators in an arbitrary $n$-point function.  This shows that the counter terms \eqref{fullcounterintro} are sufficient to regulate the integral appearing in the first order perturbative correction to any $n$-point function.  We apply our general perturbation framework to the case of 3-point functions in section \ref{cpt_3pf} and consider the shift to the structure constants of the theory. Section \ref{sec_bos} contains an explicit example of the perturbation of the structure constant for the CFT of a compact boson. We conclude in section \ref{conc}. Several appendices contain proofs of the formulae presented in the main text, as well as other technical details.

\section{Shift of conformal dimensions}\label{sec_2pf}
In this section we will compute the change in conformal dimensions, also known as anomalous dimensions, of quasi primary operators at first order in perturbation theory. To do so, we start in section \ref{cpt_h} by considering the corrections to the 2-point function, which are calculated using an integrated 3-point function.  Two of the operators are placed at $z_1$ and $z_2$ respectively, while the third operator is the exactly marginal deformation operator, with dimensions  $(h,\ti{h})=(1,1)$, added to the Lagrangian of the theory; the position of the deformation operator is integrated over.  This 3-point function is singular when the deformation operator approaches either of the insertions, and so the integral must be regulated.

We adopt a canonical regularization procedure where small disks of radius $\epsilon$ are cut out around $z_1$ and $z_2$.  When the size of the holes is relaxed $\epsilon \rightarrow 0$, the integral diverges, and we must find appropriate counter terms to cancel such divergences.  To do so, we proceed in stages.  First, in section \ref{cpt_h}, we map the two insertions at $z_1$ and $z_2$ to the points $1$ and $\infty$ using an $sl(2)$ transformation.  This transformation also maps the excised disks to excised regions at 0 and $\infty$ as well, and we first consider the location and shape of these regions to only to leading order in $\epsilon$, leading to a simplified region of integration.  This allows us to find a set of counter terms necessary to regulate the divergences in this simplified region, and further allows us to find the anomalous dimension to first order in the coupling constant $\lambda$.  In section \ref{cpt_o} we go beyond leading order in $\epsilon$ by correcting the excised regions at $0$ and $\infty$ to all orders in $\epsilon$, leading to a full set of counter terms necessary to regulate the integrated 3-point function.  These include operators belonging to new quasi primary families, but also include $sl(2)$ descendants (derivatives) of the original counter terms necessary to regulate the integral in the simplified region.  While these new counter terms are necessary, we find that they {\it do not} alter the expression for the computation of the anomalous dimension of operators.  The reader familiar with the standard calculation in the simplified region may safely skip to section \ref{cpt_o}.

\subsection{Simplified domain}\label{cpt_h}

Consider a moduli space of 2-dimensional CFTs with real coordinates $\lambda_\ell$, $\ell=\{1,\cdots,d\}$. $\lambda_\ell$ are the coupling constants of the CFT at each point in the moduli space and we refer to them collectively as $\lambda$. Consider a set of quasi primary operators $\phi_i$, which have been put in an orthonormal basis for the CFT at position $\lambda$ in the moduli space. The two point function of two such operators is given by
\begin{equation}
\langle {\phi_{i,\lambda}}(z_1,\zb_1)\;{\phi_{j,\lambda}}(z_2,\zb_2)\rangle_{\lambda}= \frac{\delta_{ij}}{z_{12}^{2h_i(\lambda)}\zb_{12}^{2\ti{h}_i(\lambda)}}\ . \label{orthoOps}
\end{equation}
Above we have explicitly written the coupling constant dependence of the action by giving the expectation value a subscript lambda.  We have also explicitly written the coupling constant dependence of the operators themselves with the $\lambda$ subscripts. The normalization of operators are fixed to be the same at all points $\lambda$, leaving the only coupling constant dependence in the conformal weights.  We may therefore perturb for small $\lambda$, and the \textsc{rhs} of eq. \eqref{orthoOps} reads:
\be
\frac{\delta_{ij}}{z_{12}^{2h_i(\lambda)}\zb_{12}^{2\ti{h}_i(\lambda)}}=\Big(1-2\lambda\Big(\frac{\pa h_i}{\pa \lambda}\Big|_{\lambda=0} \ln(z_{12})+\frac{\pa \tilde{h}_i}{\pa \lambda}\Big|_{\lambda=0} \ln(\zb_{12})\Big)+\cdots\Big)\langle {\phi_i}(z_1,\zb_1)\; {\phi_j}(z_2,\zb_2)\rangle_{\lambda=0}\ . \label{expand2ptRHS}
\ee

The \textsc{lhs} of eq. (\ref{orthoOps}) may be expressed using path integral formulation and deforming away from the $\lambda=0$ theory by an exactly marginal operator $\mathcal{O}_D$: 
\begin{align}
\langle \phi_{i,\lambda}(z_1,\zb_1)\; \phi_{j,\lambda}(z_2,\zb_2)\rangle_{\lambda}= \frac{\int d[O]\; e^{-S_{\rm free}-\lambda\int d^2z\, \mathcal{O}_D(z,\zb)}\; \phi_{i,\lambda}(z_1,\zb_1)\; \phi_{j,\lambda}(z_2,\zb_2)}{\int d[\phi]\; e^{-S_{\rm free}-\lambda\int d^2z\, \mathcal{O}_D(z,\zb)}} \label{PIDefDef}
\end{align}
where $d[O]$ refers to the fundamental fields appearing in the action $S_{\rm free}$.  Expanding to leading order in lambda, we find: 
\bea\label{PIexpand}
&&\!\!\!\!\!\!\!\!\!\!\!\!\!\!\!\!\langle \phi_{i,\lambda}(z_1,\zb_1)\; \phi_{j,\lambda}(z_2,\zb_2)\rangle_{\lambda} = \frac{\int d[O]\; e^{-S_{\rm free}}\Big(1-\lambda\int d^2z \,\mathcal{O}_D(z,\zb)+O(\lambda^2)\Big)\;\phi_{i,\lambda}(z_1,\zb_1)\; \phi_{j,\lambda}(z_2,\zb_2)}{\int d[O]\; e^{-S_{\rm free}}\Big(1-\lambda\int d^2z\, \mathcal{O}_D(z,\zb)+O(\lambda^2)\Big)} \nn\\
&&\!\!\!\!\!\!\!\!\!\!\!\!\!\!\!= \langle\phi_{i,\lambda}(z_1,\zb_1)\; \phi_{j,\lambda}(z_2,\zb_2)\rangle_{{\lambda=0}}-\lambda \int d^2z\, \langle\mathcal{O}_D(z,\zb)\;\phi_{i}(z_1,\zb_1)\; \phi_{j}(z_2,\zb_2)\rangle_{\lambda=0}+O(\lambda^2)\ .\quad
\eea
A couple of remarks are in order. Firstly, since $\langle {\mathcal{O}_D}\rangle_{\lambda=0}=0$, the second term in the denominator vanishes. Secondly, note that the operators $\phi_{i,\lambda}$ in general depend on $\lambda$, which we must eventually expand: the $\lambda^0$ term matches the leading order term in \eqref{expand2ptRHS}, and the $\lambda^1$ terms are counter-terms. In the second line, the correlation functions are calculated in the undeformed theory ($\lambda=0$) and in the second term, operators under the integration should not have any $\lambda$ dependence (such changes to the operators would give rise to $O(\lambda^2)$ terms, and so their subscripts have been removed). The operators that appear in the integrated 3-point function are a set of operators in the $\lambda=0$ theory which satisfy eq. (\ref{orthoOps}) at $\lambda=0$. Similar considerations exist for $\mathcal{O}_D$, which should naturally carry a $\lambda$ subscript as well. For simplicity in subsequent formulas we will drop the subscript $\lambda$ on operators and expectation values when $\lambda=0$.

Let us now focus on the integrated 3-point function in eq. \eqref{PIexpand}:
\begin{align}\label{3pfe1}
\mathcal A&\equiv -\lambda \int d^2z\, \langle \mathcal{O}_D(z,\zb)\; \phi_{i}(z_1,\zb_1)\; \phi_{j}(z_2,\zb_2)\rangle   \\
&\,=-\lambda \int d^2z\, \frac{C_{D,i,j}}{(z-z_1)^{h_i+1-h_j}(z-z_2)^{h_j+1-h_i}(z_1-z_2)^{h_i+h_j-1}\times ({\rm a.h.})}\nn
\end{align}
where a.h. refers to the anti-holomorphic counterparts (note, we make no distinction between up and down indices on operators because of the choice of 2-point function). In the above we have introduced the structure constant $C_{D,i,j}$.  The above integral diverges when the deformation operator $\mathcal O_D$ approaches operator insertions, i.e. at $z\rightarrow z_1$ or $z\rightarrow z_2$, depending on the values of $h_i,\ti{h_i},h_j,\ti{h}_j$.  The integral must then be regulated.  We do so by excising small holes around the operator insertions, allowing only the region $|z-z_1|> \epsilon$ and $|z-z_2| > \epsilon$, for $0<\epsilon\ll1$.

We define new coordinate 
\begin{equation}
\hat{z}\equiv \frac{z-z_1}{z-z_2}.
\end{equation}
Under this map, the excised regions are near $\hat{z}=0$ and $\hat{z}=\infty$, and the region of integration is mapped to
\begin{align}
\Big|\hat{z}+\frac{\epsilon^2}{|z_{12}|^2-\epsilon^2}\Big| >\frac{\epsilon}{|z_{12}|}\,\frac{1}{1-\frac{\epsilon^2}{|z_{12}|^2}} \ ,\qquad\quad\left|\hat{z}-1\right|< \frac{|z_{12}|}{\epsilon}\ .\label{twopointupFirst}
\end{align}
We find this region cumbersome to deal with because the circular boundaries are not concentric, and their radii are not simple functions of the regulation parameter $\epsilon$.  To make progress, we first approximate the region with the leading order behavior in $\epsilon$ for the above region, postponing the higher order corrections to section \ref{cpt_o}.  To leading order, we find 
\begin{align}\label{circle}
\frac{\epsilon}{|z_{12}|}<&|\hat{z}|< \frac{|z_{12}|}{\epsilon},
\end{align}
which we refer to as the {\it simplified domain}.  When restricting \eqref{3pfe1} to the simplified domain, we refer to this integral as $\mathcal A_{\rm s}$.  To make the left-moving/right-moving symmetry obvious, we define $s_i\equiv h_i-\ti{h}_i$ and $d_i\equiv h_i+\ti{h}_i$ which denote respectively the spin and scaling dimension of operator $\phi_i$.  We change to coordinates $\hat{z}=re^{i\phi}$, and recall that the area integral has a Jacobian $d^2 \hat{z}=2r\,dr \d\phi$ (or defining this as our convention\footnote{This gives some differences from what appears in some other references, given that many references define $d^2z_{\rm elsewhere}= rdrd\phi$.  However, all integrals performed here are over positions of deformation operators which always come with factors of $\lambda$, and so this amounts to a scaling of $\lambda$.  To match this other common convention, one may simply replace $\lambda_{\rm us}=\frac{1}{2} \lambda_{\rm elsewhere}$ for easy comparison.\label{footfac2}}), eq. \eqref{3pfe1} over the simplified domain reads 
\begin{align}\label{3pfe1r}
&\mathcal A_{\rm s}=-2\lambda\, C_{D,i,j}\int_{\phi=0}^{\phi=2\pi}d\phi \int_{\frac{\epsilon}{|z_{12}|}}^{\frac{|z_{12}|}{\epsilon}}\frac{dr}{r}\; \frac{(z_{12})^{-(h_i+h_j)}(\zb_{12})^{-(\ti{h}_i+\ti{h}_j)}}{r^{d_i-d_j}}\;e^{-i\phi\left(s_i-s_j\right)}
\end{align}
where $z_{ij}\equiv z_i-z_j$. The integral over  $\phi$ is unobstructed, and so requires that $s_i=s_j$.  This is perhaps not surprising because a weight $(1,1)$ operator should not spoil spin conservation, which we will also see later on as well when considering all orders in $\epsilon$.
Next, consider the two distinct cases: $d_i=d_j$ or $d_i\neq d_j$. In the first case, $d_i=d_j$ and $s_i=s_j$ impose $h_i=h_j$ and $\ti{h}_i=\ti{h}_j$.  We find
\begin{align}
&\mathcal A_{\rm s}=-4\pi \lambda\, C_{D,i,j}\,z_{12}^{-2h_i}\zb_{12}^{-2\ti{h}_i}\big(\ln(z_{12})+\ln(\zb_{12})-2\ln(\epsilon)\big)\ . \label{logz12}
\end{align}
The logarithmic terms appear to match the expansion (\ref{expand2ptRHS}), however, the matrix $C_{D,i,j}$ makes the above have possible $i\neq j$ off diagonal terms.  This is simply because there is an ambiguity in the choice of orthogonal operators as in (\ref{orthoOps}) if there are multiple quasi primary operators with the same conformal dimension.  We choose a basis which diagonalizes $C_{D,i,j}$ as well, given that this must be a symmetric matrix. With this, the regularization dependent term may be absorbed into a counter term, utilizing the first term in the right hand side of eq. (\ref{PIexpand}) to eliminate this divergence.

For the case $d_i\neq d_j$, eq. \eqref{3pfe1r} becomes
\begin{align}\label{2pf_eps1}
&\mathcal A_{\rm s}=-4\pi \lambda\, C_{D,i,j}\,\delta_{s_i,s_j}
\bigg(
\frac{1}{(d_i-d_j)\,\epsilon^{d_i-d_j}}\bigg(\frac{1}{z_{12}^{2h_j}\zb_{12}^{2\ti{h_j}}}\bigg)
+\frac{1}{(d_j-d_i)\,\epsilon^{d_j-d_i}}\bigg(\frac{1}{z_{12}^{2h_i}\zb_{12}^{2\ti{h}_i}}\bigg)
\bigg).
\end{align}
Only one of the terms above is divergent: the first term when  $d_i>d_j$, and the second when $d_j>d_i$.  For now we focus on the first term. In this case, it is natural to associate the divergences with parts of the integral where ${\mathcal O}_D$ approaches $\phi_i$: in such a case, an occurrence of $\phi_j$ in the OPE has a singular coefficient for the case at hand $d_i>d_j$.  We think of these terms as the ones needing to be regularized by counterterms in $\phi_{i,\lambda}$.  For the same reasons, we associate second term in the above equation with counter terms in $\phi_{j,\lambda}$.  Furthermore, divergences of this type will happen for all such quasi primary fields and thus, must be summed over (the $\phi_j$ in the correlator simply projects the sum over terms onto this specific quasi primary).  This operator mixing leads to a sum of ``off diagonal'' terms which may be absorbed by a set of counter terms. All in all, we obtain
\begin{equation}\label{opren}
\phi_{i,\lambda}=\phi_i-2\pi\lambda\,\ln(\epsilon^2)\, C_{D,i,i}\,\phi_i+4\pi\lambda\!\!\!\sum_{{p,\,d_p<d_i}}\!\!\frac{\delta_{s_i,s_p}}{(d_i-d_p)\,\epsilon^{(d_i-d_p)}}\, C_{D,i,p}\,\phi_p+O(\lambda^2)\ . 
\end{equation}

Inserting eqs. \eqref{logz12}-\eqref{opren} in eq. \eqref{PIexpand}, we evaluate the two-point function at the first order perturbation theory. In particular, the counter terms cancel the $\ln(\epsilon)$ terms in (\ref{logz12}), leaving only the $\ln(z_{12})$ and $\ln(\zb_{12})$ terms.  These can be then matched to the log terms in (\ref{expand2ptRHS}), giving the well-known expressions for the anomalous dimension \cite{Dijkgraaf:1987jt,Cardy:1987vr,Kutasov:1988xb} \footnote{Keeping in mind our convention for the area explained in footnote \ref{footfac2} on page \pageref{footfac2}, the right hand side of eq. \eqref{hshifts} must be divided by 2 to match these references.}:
\begin{equation}
\frac{\pa{h_i(\lambda)}}{\pa \lambda}\Big|_{\lambda=0}=\frac{\pa{\ti{h}_i(\lambda)}}{\pa \lambda}\Big|_{\lambda=0}=2\pi C_{D,i,i}\ . \label{hshifts}
\end{equation}

\subsection{Exact domain}\label{cpt_o}

The regulators in the last subsection are non-local.  Going from the allowed region \eqref{twopointupFirst} to the simplified region \eqref{circle} necessarily introduces correction terms in powers of $\epsilon/z_{12}$, and so requires knowledge of the location of both operator insertions.  It is interesting, and perhaps a bit mysterious, that the integral over this simplified region still admits local counter terms which cancel divergences.  

\begin{figure}[ht]
\begin{center}
\includegraphics[width=0.4\textwidth]{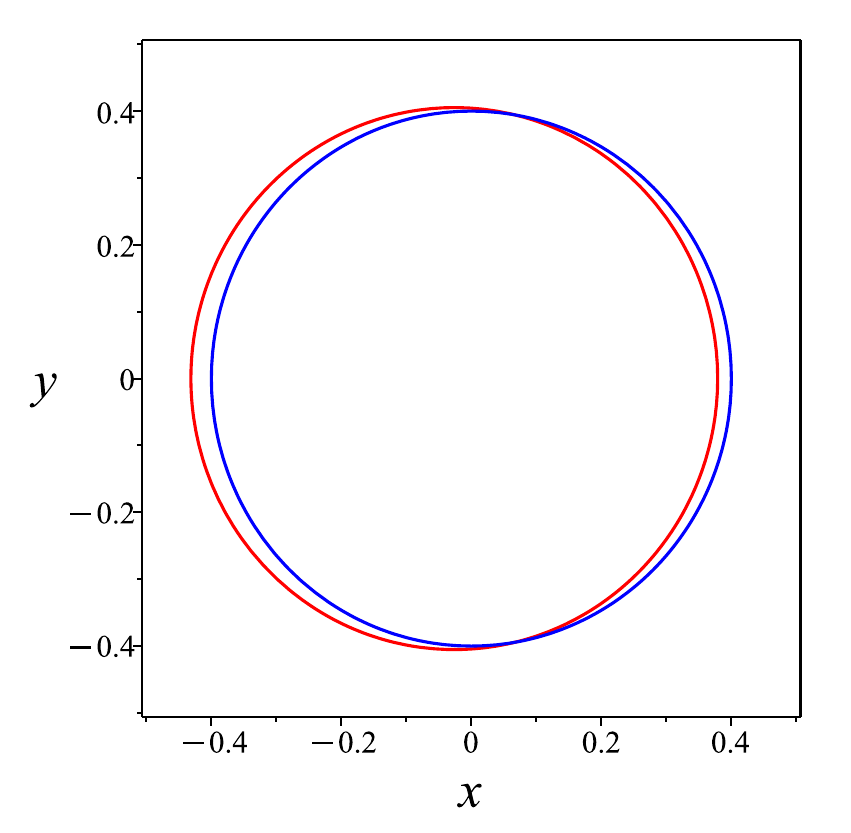}
\end{center}
\caption{\label{crescentfig} The modification of the integrated area near $\hat{z}=0$.  Above we have plotted for $\epsilon=0.8$ and $|z_{12}|=5$, where the blue circle (slightly to the right) is $|\hat{z}|=\epsilon/|z_{12}|$ and the red circle (slightly to the left) is $\Big|\hat{z}+\frac{\epsilon^2}{|z_{12}|^2-\epsilon^2}\Big|= \frac{\epsilon}{|z_{12}|}\,(1-\frac{\epsilon^2}{|z_{12}|^2})^{-1}$ --- see eqs. \eqref{twopointupFirst} and \eqref{circle}.}
\end{figure}

However, we wish to have a fully local regulator, which the original region \eqref{twopointupFirst} supplies. The exact domain \eqref{twopointupFirst} is nearly identical to the simplified domain \eqref{circle} along with corrections to the domain near $\hat{z}=0$ and $\hat{z}=\infty$.  This suggests breaking the integral into three parts: the integral over the simplified domain; the corrected domain near $\hat{z}=0$ associated with operator $i$; the corrected domain near $\hat{z}=\infty$ associated with operator $j$, i.e. 
\begin{align}
{\mathcal A}= {\mathcal A}_{\rm s}+{\mathcal A}_i+{\mathcal A}_j\ . \label{CalAdef}
\end{align}
We find that $\mathcal{A}_i$ and $\mathcal{A}_j$ are integrals defined over crescent regions near $\hat{z}=0$ and $\hat{z}=\infty$, respectively --- see figure \ref{crescentfig}.

Implementing this split of the domain, we find 
\begin{align}
{\mathcal A}_i:=-2\lambda\, C_{D,i,j}\int_0^{2\pi}\!\!\! d\phi\! \int_{\frac{\epsilon}{|z_{12}|}\;\mathcal{R}_0}^{\frac{\epsilon}{|z_{12}|}}\!\frac{dr}{r}\; {Q}(r,\phi), \quad A_j:= -2\lambda\, C_{D,i,j}\int_0^{2\pi} \!\!\!d\phi\! \int_{\frac{|z_{12}|}{\epsilon}}^{\frac{|z_{12}|}{\epsilon}{\mathcal{R}_\infty}}\! \frac{dr}{r}\;{Q}(r,\phi) \label{AiAj}
\end{align}
where the integrand in each term is the same as in eq. \eqref{3pfe1r}:
\begin{align}
{Q}(r,\phi) :=\frac{(z_{12})^{-(h_i+h_j)}(\zb_{12})^{-(\ti{h}_i+\ti{h}_j)}}{r^{d_i-d_j}}\;e^{-i\phi\left(s_i-s_j\right)} \, .
\end{align}
The upper and lower bounds of integration in eq. \eqref{AiAj} are determined by eq. \eqref{twopointupFirst}:
\begin{align}\label{R0inf}
\mathcal{R}_\infty = \sqrt{1 -\frac{\epsilon^2}{|z_{12}|^2}\sin^2(\phi)} +\frac{\epsilon}{|z_{12}|}\cos(\phi)\ ,\quad
\mathcal{R}_0=\frac{1}{{\mathcal{R}}_\infty}=\frac{\sqrt{1 -\frac{\epsilon^2}{|z_{12}|^2}\sin^2(\phi)} -\frac{\epsilon}{|z_{12}|}\cos(\phi)}{1-\frac{\epsilon^2}{|z_{12}|^2}}\ .
\end{align}
Note, one may change variables in ${\mathcal A}_j$ to $r=1/r', \phi=-\phi'$, and along with the symmetry $C_{D,i,j}=C_{D,j,i}$ of the structure constants, we see that $(i\leftrightarrow j) {\mathcal{A}_i}={\mathcal{A}}_j$.  This makes $\mathcal A$ manifestly $i\leftrightarrow j$ symmetric, a property we should expect from a local regulator.  Thus, we may concentrate on $\mathcal{A}_i$ and extract $\mathcal{A}_j$ by symmetry.  

Concentrating on $\mathcal{A}_i$ for the case $d_i\neq d_j$, we find
\begin{align}\label{2pfII}
&\frac{\mathcal{A}_i}{-2\lambda\, C_{D,i,j}}=(z_{12})^{-(h_i+h_j)}(\zb_{12})^{-(\ti{h}_i+\ti{h}_j)}\int_0^{2\pi}\!\!\! d\phi\;
\frac{-|z_{12}|^{d_i-d_j}}{\epsilon^{d_i-d_j}\,(d_i-d_j)}(1-\mathcal{R}_\infty^{d_i-d_j})\cos((s_i-s_j)\phi)
\end{align}
where we have noted that the $\mathcal R_{\infty}$ is symmetric under $\phi\rightarrow-\phi$ and so only the $\cos((s_i-s_j)\phi)$ in $\mathcal{A}_i$ survives. The above expression furnishes an expansion
\begin{equation}
\int_0^{2\pi}\!\!\! d\phi\left(1-\mathcal{R}_\infty^{d_i-d_j}\right)\cos((s_i-s_j)\phi)=\sum_{\ell=\ell_{\rm min}}^{\lfloor{d_i-d_j}\rfloor} a_\ell(d_i-d_j,s_i-s_j)\, \frac{\epsilon^\ell}{|z_{12}|^\ell}+\cdots \label{adef}\, .
\end{equation}  
The constants $a_\ell(d_i-d_j,s_i-s_j)$ are geometrically calculable given a pair of operators with $d_i>d_j$, and are explicitly computed in appendix \ref{app_als}.  

Some general properties of the $a_\ell$ coefficients are easily obtained.  First we consider the range of $\ell$.   Expanding $R_{\infty}^{d_i-d_j}$ in $\epsilon$ above order $\lfloor d_i-d_j\rfloor$ gives terms which go to zero as $\epsilon\rightarrow 0$, and so we truncate the sum there (although these terms are still calculable).  The lower bound $\ell_{\rm min}$ can be found by considering the trig functions appearing in the expansion.  A term of order $\epsilon^n$ may be built from $\epsilon^2 \sin^2(\phi)$ and $\epsilon \cos(\phi)$, or more succinctly as $\epsilon^n(\rm sinusoid)^n$.  Given a spin mismatch $s_i-s_j$, any terms with $\ell<|s_i-s_j|$ will not contribute, as these Fourier modes have not yet been accessed by the expansion of $R_{\infty}$.  This gives $\ell\geq |s_i-s_j|$.  Finally, we note that $\phi\rightarrow \phi+\pi, \epsilon\rightarrow -\epsilon$ is a symmetry of $R_{\infty}$, and so only $\ell-|s_i-s_j|\in 2{\mathbb Z}$ give non-zero $a_\ell$ coefficients (only even/odd powers of $\epsilon$ appear, depending on whether $s_i-s_j$ is even/odd).  It is further evident that $a_0=0$ because $(1-R_\infty^{d_i-d_j})$ is $0$ at order $\epsilon^0$.  

Inserting the expansion \eqref{adef} in eq. \eqref{2pfII}, we find
\begin{align}\label{2pfIals}
\mathcal A_i\sim
&+2\lambda C_{D,i,j}\sum_{\substack{\ell\neq 0,\ell\geq |s_i-s_j| \\
\ell-|s_i-s_j|\in 2{\mathbb{Z}} }}^{\lfloor{d_i-d_j}\rfloor} \frac{1}{d_i-d_j}\,(z_{12})^{-2h_j -\frac{s_i-s_j+\ell}{2}}(\zb_{12})^{-2\ti{h}_j+\frac{s_i-s_j-\ell}{2}}\;
 \frac{a_\ell(d_i-d_j,s_i-s_j)}{\epsilon^{d_i-d_j-\ell}}\ ,
\end{align}
where $\sim$ means up to terms which vanish as $\epsilon\rightarrow 0$.  The integral $\mathcal{A}_j$ contributes to singular terms when the hierarchy of dimensions is reversed, i.e. $d_j>d_i$, as a result of the $i\leftrightarrow j$ symmetry.   

Given the bounds of the sum, we see that $\frac{\ell-(s_i-s_j)}{2}=n\geq 0$ and $\frac{\ell+(s_i-s_j)}{2}=\ti{n}\geq 0$ define two positive integers $n,\ti{n}$, which allows us to write the summand of \eqref{2pfIals} as
\begin{equation}\label{2pfI_lth}
+2\lambda C_{D,i,j}\;\frac{(-1)^{\ell}}{d_i-d_j}\;\frac{1}{(2h_j)_{n}}\;\pa_1^n(z_{12})^{-2h_j}\;\frac{1}{(2\ti{h}_j)_{\ti{n}}}\;\pab_1^{\ti{n}}(\zb_{12})^{-2\ti{h}_j}\; \frac{a_\ell(d_i-d_j,s_i-s_j)}{\epsilon^{d_i-d_j-\ell}}\, .
\end{equation}
The divergences above are associated only with operators obeying $d_i>d_j$.  It is natural to associate these with divergent terms in the OPE as $\mathcal{O}_D$ approaches $\phi_i$, producing operators in the family $\phi_p$ with $d_i>d_p$: the operator $\phi_j$ simply projects onto the operator family $j$ because of the normalization of 2-point functions for quasi primaries.  We therefore group the divergences above with counterterms associated with the operator $\phi_i$, and so the derivatives $\pa_1$ are preferred in this case.  The counter terms associated with $\phi_j$ correspond to the crescents near $\hat{z}=\infty$, and can easily be extracted using the $i\leftrightarrow j$ symmetry of $\mathcal{A}$.  We will approach finding the counter terms using the OPE analysis directly in the section \ref{cpt_npf}. 

Finally, in appendix \ref{app_als} we prove that
\begin{align}\label{als}
a_\ell(d_i-d_j,s_i-s_j)=-2\pi\,\frac{(d_i-d_j)\,(-1)^{\ell}}{(d_i-d_j-\ell)}\; \frac{(1-h_i+h_j)_n\,(1-\ti{h}_i+\ti{h}_j)_{\ti{n}}}{n!\,\ti{n}!}
\end{align}
for all allowed values of $\ell$ (recalling that $\ell\neq 0$, or equivalently $a_0=0$ by eq. \eqref{adef}). Plugging in these results, \eqref{2pfIals} then reads
\begin{align}
\mathcal A_i= -4\pi\lambda C_{D,i,j}\!\!\!\!\!\!\!\! \sum_{\substack{\ell\neq 0,\ell=|s_i-s_j|\\ \ell-|s_i-s_j|\in 2{\mathbb Z}}}^{\lfloor d_i-d_j\rfloor}
\!\frac{(1-h_i+h_j)_n\,(1-\ti{h}_i+\ti{h}_j)_{\ti{n}}}{(d_i-d_j-\ell)\epsilon^{d_i-d_j-\ell}n!\,(2h_j)_n\;\ti{n}!\,(2\ti{h}_j)_{\ti{n}}}\,
\pa_1^n(z_{12})^{-2h_j}\pab_1^{\ti{n}}(\zb_{12})^{-2\ti{h}_j}\ .\nn
\end{align}
Such divergences occur for any quasi primary field $\phi_j$ satisfying $d_i>d_j$. We must sum over all such terms to account for the crescent regions.  Counter terms may then be easily obtained:
\begin{equation}
+4\pi\lambda\!\!\!\!\!\sum_{\substack{p, d_p<d_i \\ |s_i-s_p|\leq \lfloor d_i-d_p\rfloor}}
\sum_{\substack{\ell\neq 0, \ell=|s_i-s_j|\\\;\;\ell-|s_i-s_j|\in 2{\mathbb Z}}}^{\lfloor d_i-d_j\rfloor}
\!\!\!\frac{C_{D,i,p}}{(d_i-d_p-\ell)\epsilon^{d_i-d_p-\ell}}\;\frac{(1-h_i+h_p)_n\,(1-\ti{h}_i+\ti{h}_p)_{\ti{n}}}{n!\,(2h_p)_n\,\ti{n}!\,(2\ti{h}_p)_{\ti{n}}}
\;\pa^n\pab^{\ti{n}}\phi_p\ ,
\end{equation}
which we must add to the previous counter terms \eqref{opren} which cancel the divergences associated with the simplified region.  Interestingly, the previous counter terms of \eqref{opren} are those that exactly complete the above sum by including $\ell=0$ terms when $s_i=s_p$.  Thus, the full counter terms are the above sum dropping the $\ell\neq 0$ restriction.  We therefore see that the above sum includes new quasi primaries in the sum over $p$ that are spin mismatched, but made to match total spin using derivatives.  In addition, the sum also includes the $sl(2)$ descendants of the old counter terms \eqref{opren}.

We have yet to address the case $d_i=d_j$ in $\mathcal A_i$ in equation \eqref{AiAj}. We obtain
\begin{align}
&\int_0^{2\pi}\!\!\!d\phi\! \int_{\frac{\epsilon}{|z_{12}|}\mathcal{R}_0}^{\frac{\epsilon}{|z_{12}|}}\!\frac{dr}{r}\,Q(r,\phi) =(z_{12})^{-(h_i+h_j)}(\zb_{12})^{-(\ti{h}_i+\ti{h}_j)}\int_0^{2\pi}\!\!\! d\phi\; \ln({\mathcal{R}}_\infty)\cos((s_i-s_j)\phi) 
\end{align}
which is finite in the limit $\epsilon\rightarrow 0$, and thus no additional counter terms are required.

The final result for the counter terms, to all orders in $\epsilon$, is of the form
\begin{align}\label{counttotal2pt}
& \phi_{i,\lambda}=\phi_i-2\pi\lambda\,\ln(\epsilon^2)\, C_{D,i,i}\,\phi_i  \\
&+4\pi\lambda\!\!\!\!\!\!\!\!\sum_{\substack{p,\, d_p<d_i \\ |s_i-s_p|<\lfloor d_i-d_p\rfloor}}\!\!\!\sum_{\substack{\ell=|s_i-s_p|\\\;\;\; \ell-|s_i-s_p|\in 2{\mathbb Z}\\}}^{\lfloor d_i-d_p\rfloor}  
\!\!\!\!\frac{C_{D,i,p}}{(d_i-d_p-\ell)\epsilon^{d_i-d_p-\ell}}\;\frac{(1-h_i+h_p)_n\,(1-\ti{h}_i+\ti{h}_p)_{\ti{n}}}{n!\,(2h_p)_n\,\ti{n}!\,(2\ti{h}_p)_{\ti{n}}}\;  \pa^n \pab^{\ti{n}} \phi_p+O(\lambda^2)\ ,\nn 
\end{align}
with the integers $n$ and $\ti{n}$ determined by
\begin{align}
n\equiv \frac{\ell+s_i-s_p}{2}\geq 0\ , &\qquad \ti{n}\equiv \frac{\ell-(s_i-s_p)}{2}\geq 0\ , 
\end{align}
and recalling the requirement that $s_i-s_p\in {\mathbb Z}$, so that the 3-point function is single valued, and so may be integrated.  

We see the above counter terms result from contributions from the integration over the simplified region \eqref{circle} in subsection \ref{cpt_h} along with the new contributions from the crescents in figure \ref{crescentfig} which correct the simplified domain \eqref{circle} to the exact in $\epsilon$ domain \eqref{twopointupFirst}. The $i\leftrightarrow j$ symmetry of \eqref{CalAdef}, shown below eq. \eqref{R0inf}, generates the counter terms for $\phi_j$.  Finally, note that matching the log terms in (\ref{expand2ptRHS}) is unaltered because no additional $\ln(\epsilon/|z_{12}|)$ appear in the calculations of the terms coming from the crescents.  Therefore, the anomalous dimension is {\it not} affected by the higher order corrections in $\epsilon$ to the region of integration, and is given by eq. \eqref{hshifts}.  We will discuss these counter terms in more detail in section \ref{cpt_npf}, where we show that they are sufficient to regulate all integrated $n$-point functions at leading order in perturbation theory, and so have a bearing on the changes to all CFT data.   

For now, we see that we have a powerful calculational tool at hand for calculating anomalous dimensions. First, we map the location of the two fixed operators to zero and infinity.  We implement regulation by restricting the domain of integration to ${\epsilon}/{|z_{12}|}<|\hat{z}|< {|z_{12}|}/{\epsilon}$, which we know is non-local. However, going to the local regulator \eqref{twopointupFirst} only introduces new power-law terms in $\epsilon/|z_{12}|$, which may be canceled by counter terms which are exactly calculable --- i.e. eq. \eqref{counttotal2pt}.  However, it does {\it not} introduce new logarithmic terms.  Thus, the simplified domain of integration (\ref{circle}) is sufficient to read off the anomalous dimension of quasi primary operators at order $\lambda^1$ in perturbation theory.  This is because of the difficulty in generating log terms \cite{Kutasov:1988xb}.  Furthermore, this technique already commonplace in the literature \cite{Dijkgraaf:1987jt,Cardy:1987vr,Kutasov:1988xb}, however the calculation above makes the reasoning more explicit.   We will see in section \ref{cpt_3pf} that corrections to the structure constant are not as robust when attempting a similar simplification for the domain of integration.  Nevertheless, corrections arising from changing the domain of integration will be calculable.

\section{\texorpdfstring{$n$}{TEXT}-point functions and counter terms}\label{cpt_npf}

We will now consider corrections to $n$-point functions of quasi primary operators to the first order in conformal perturbation theory. We regulate the singularities using the OPEs of the deformation operator with the quasi primary operators. This approach was developed in \cite{Ranganathan:1992nb,Ranganathan:1993vj}. Our motivation to use this approach comes from our results for the counter terms in the case of 2-point functions in the previous section. The counter terms \eqref{counttotal2pt} are written entirely in terms of quasi primary fields and their $sl(2)$ descendants, i.e. derivatives. This immediately suggests tackling the OPE coefficients, grouping the operators in terms of the quasi primary ancestors and their $sl(2)$ descendants. We will find that, interestingly, this procedure produces the same counter terms \eqref{counttotal2pt}. Moreover, it proves that these counter terms are sufficient to regulate {\it all} integrated $n$-point functions at leading order $\lambda$ in perturbation theory. This one of the main results of our work: expressing the counter terms explicitly in terms of quasi primaries and derivatives of quasi primaries, with coefficients written in terms of CFT data (namely, conformal dimensions and structure constants). 

\subsection{OPEs of quasi primary fields}\label{subsec_ope}

Let us consider a pair of quasi primary fields $\phi_1$ and $\phi_2$. We place $\phi_2$ at the origin for simplicity, making the state $\mid \phi_2\rangle$, and apply the operator $\phi_1(z,\zb)$ to this state.  This results in the expression for the OPE (written in terms of states):
\begin{equation}
\phi_1(z,\zb) \mid \phi_2\rangle=\sum_p C_{p,1,2}\sum_{n=0}^\infty \sum_{\ti{n}=0}^\infty z^{-h_1-h_2+h_p+n}\,\zb^{-\ti{h}_1-\ti{h}_2+\ti{h}_p+\ti{n}}\;\beta_n^p\, \ti{\beta}_{\ti{n}}^p\; L_{-1}^n\,{\ti{L}}_{-1}^{\ti{n}} \mid \phi_p\rangle\ , \label{OPEoperator}
\end{equation}
where $p$ sums over all quasi primaries $\phi_p$, the sums over $n$ and $\ti{n}$ account for all $sl(2)$ descendants, and $\beta_n^p$ and $\tilde\beta_{\tilde n}^p$ are descent coefficients to be determined. The above formula is just as general as the usual sum over conformal families defined by primaries and thier full Virasoro descendants, see e.g. \cite[eq. (6.165)]{DiFrancesco:1997nk};  it is simply a repackaging of the same information.  The set of all quasi primaries and their $sl(2)$ descendants is a full set of operators in the theory, just as all primaries and their Virasoro descendants is a full set of operators.  The simplicity from considering each quasi primary family separately is that the descendants are constructed using simple powers of $L_{-1}$ and $\ti{L}_{-1}$, where in the full Virasoro case one uses all possible combinations of $L_{-n}$ and $\ti{L}_{-\tilde n}$.  Considering each quasi primary family separately allows us to compute the descent coefficients $\beta_n^p$ and $\tilde\beta_{\tilde n}^p$ in closed form, for all $n$ and $p$, as we shall soon see. This in turn will determine the full set of counter terms in terms of quasi primaries and their descendants. The tradeoff is that the set of structure constants $C_{p,1,2}$ is much larger in eq.\eqref{OPEoperator} as compared to \cite[eq. (6.165)]{DiFrancesco:1997nk}, since there are many more quasi primaries than primaries in a conformal field theory. 

We introduce the notation
\be
\mid \phi_i^{(n,\ti{n})}\rangle\equiv L_{-1}^n \ti{L}_{-1}^{\ti{n}}\mid\phi_i\rangle\ .\label{QPdescendants}
\ee
The normalization of these states is given by 
\be
\langle \phi_i^{(n,\ti{n})} \mid \phi_j^{(n',\ti{n}')}\rangle =\delta_{i,j}\, \delta_{n,n'}\,\delta_{\ti{n},\ti{n}'}\;  n!\,(2h_i)_{n}\;\ti{n}!\,(2\ti{h}_i)_{\ti{n}} \label{descentinner}
\ee
which follows from the $sl(2)$ algebra, and may be proved with a simple induction on $n$ and $\ti{n}$.\footnote{The $\delta$ functions arise from orthogonality of the quasi primary families, and from the matching of conformal dimensions within a given family.} The quasi primary fields have normalization 1.

To compute the coefficient $\beta^{p}_n$, we simply project onto the state $(L_{-1})^n|\,\phi_p\rangle$. Eq. \eqref{OPEoperator} reads:
\begin{equation}
\beta^p_{n}=\frac{1}{n!\,(2h_p)_{n}}\;\frac{\langle \phi_p\mid(L_{1})^n\, \phi_1(z,\zb)\mid \phi_2\rangle}{z^{-h_1-h_2+h_p+n}\,
\langle \phi_p \mid \phi_1(1,1)\mid \phi_2\rangle}\;\zb^{\ti{h}_1+\ti{h}_2-\ti{h}_p}\ . \label{betathreept}
\end{equation}
Similar expression is obtained for $\tilde\beta_{\tilde n}^p$. We evaluate this expression in appendix \ref{app_beta} and show that
\begin{equation}\label{betas}
\beta^p_n=\frac{(h_1-h_2+h_p)_{n}}{n!\, (2h_p)_{n}}\ ,\qquad\qquad \ti{\beta}^p_{\ti{n}}=\frac{(\ti{h}_1-\ti{h}_2+\ti{h}_p)_{\ti{n}}}{\ti{n}!\, (2\ti{h}_p)_{\ti{n}}}\ .
\end{equation}
We may then write the operator product \eqref{OPEoperator} as
\begin{align}\label{opesl2}
\phi_1(z,\zb)\, \phi_i(z_i,\zb_i)&=\sum_p C_{1,i,p}\sum_{n,\tilde n=0}^\infty \frac{(h_1-h_i+h_p)_{n}}{n!\, (2h_p)_{n}}\;
\frac{(\ti{h}_1-\ti{h}_i+\ti{h}_p)_{\ti{n}}}{\ti{n}!\, (2\ti{h}_p)_{\ti{n}}}  \\
& \qquad \qquad \times (z-z_i)^{-h_1-h_i+h_p+n}\,(\zb-\zb_i)^{-\ti{h}_1-\ti{h}_i+\ti{h}_p+\ti{n}}\; \pa^n \pab ^{\ti{n}} \phi_p(z_i,\zb_i)\ .\nn
\end{align}
When considering the case $\phi_1=\mathcal{O}_D$ for perturbation theory, one simply replaces $h_1=\ti{h}_1=1$.  We recall that the location $z$ will be integrated over in perturbation theory, and in such an integration we again see that the above expansion contains divergences as $z\rightarrow z_i$ for the case $d_i-d_p\geq 0$.  This suggests the operators $\pa^n \pab ^{\ti{n}} \phi_p$ as those to include as counter terms.  

\subsection{Perturbation of \texorpdfstring{$n$}{TEXT}-point functions}\label{subsec_npf}

\begin{figure}[ht]
\begin{center}
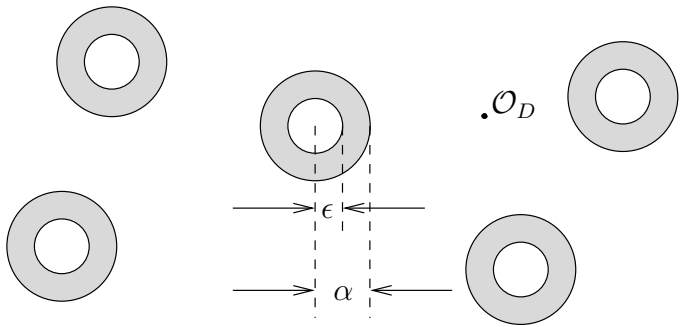
\end{center}
\caption{The full region of integration for the location of the deformation operator is the area outside of the disks of size $\epsilon$.  The integral over this region diverges as $\epsilon\rightarrow 0$.  We break this full region of integration into the region outside of the disks of radius $\alpha$, plus the annuli of outer radius $\alpha$ and inner radius $\epsilon$ centered at each operator.} \label{annulifigure}  
\end{figure}

Let us now consider perturbation of an $n$-point function of quasi primary fields. To the first order in perturbation theory, we have
\begin{align}\label{npointdivint}
\langle \phi_{1,\lambda} (z_1,\zb_1) \cdots \phi_{n,\lambda}(z_n,\zb_n)\rangle_{\lambda} &= \langle \phi_{1,\lambda} (z_1,\zb_1) \cdots \phi_{n,\lambda}(z_n,\zb_n)\rangle_{\lambda=0}\\
&-\lambda\int dz\, d\zb\;\langle {\mathcal O}_D(z,\zb)\;\phi_{1} (z_1,\zb_1)\; \cdots\; \phi_{n}(z_n,\zb_n)\rangle+O(\lambda^2)\ .\nn
\end{align}
The integral in the second line is divergent.  We regulate the integral in stages by considering first a cutoff radius of $\alpha$ around each operator.  The integration of the deformation operator outside these disks of radius $\alpha$ is necessarily finite.  However, we consider a second cutoff inside of these disks of size $\epsilon<\alpha$, and integrate over the entire domain outside of the disks of size $\epsilon$  --- see figure \ref{annulifigure}.  This integral can be broken into two pieces, the integral over the area outside of the disks of size $\alpha$, plus the integral over the annuli around each operator: each annulus has outer radius $\alpha$ and inner radius $\epsilon$, centered at the location of each operator.  The divergence of the integral over the entire domain diverges as $\epsilon\rightarrow 0$, and this divergence is clearly captured by the integral over the annuli: this is the only part of the integral that changes as $\epsilon\rightarrow 0$.  Furthermore, these annuli may be considered in isolation because they are disjoint and well separated.  Therefore, we may use the OPE between the deformation operator and each inserted operator to determine the behavior of the $n$-point function at each annulus. 

We adapt cylindrical coordinates at each annulus $(z-z_i)=r e^{i\phi}$, and show that the divergences are completely captured in the lower bound of integration in the $r$ variable.  The annuli themselves, however, are rotationally symmetric. For simplicity, we center the $i^{\rm th}$ annulus at 0, and consider the divergences at this point.  The perturbation integral in this annulus is given by
\begin{align}
\,\langle \phi_1(z_1,\zb_1)\; \phi_2(z_2,\zb_2)\; \cdots\Big(\lambda \int_{i} d^2z\; {{\mathcal O}_D}(z,\zb)\; \phi_i(0,0)\Big) \cdots\; \phi_n(z_n,\zb_n)\rangle \label{divexpt}
\end{align}
where we group ${\mathcal O}_D$ and $\phi_i$ together since these are the only operators which become close in the integral for the $i^{\rm th}$ annulus. Note that the $\phi_i$ fields can not have any $\lambda$ dependence at the first order in perturbation. Defining $\mathcal B_i\equiv \lambda \int_{i} d^2z\; {{\mathcal O}_D}\; \phi_i$ and using the OPE \eqref{opesl2}, we obtain
\begin{align}\label{opeBi}
 &\mathcal B_i= 2 \lambda \int_\epsilon^\alpha\!\! r dr\! \int_0^{2\pi}\!\!\! d\phi\; \mathcal{O}_D (z,\zb)\; \phi_i(0,0) 
= 2\lambda \sum_p C_{D,i,p}\!\sum_{n,\ti{n}=0}^\infty\frac{(1-h_i+h_p)_{n}}{n!\, (2h_p)_{n}}\; \frac{(1-\ti{h}_i+\ti{h}_p)_{\ti{n}}}{\ti{n}!\, (2\ti{h}_p)_{\ti{n}}}\nn  \\
&\qquad\qquad\qquad\qquad\qquad\times\int_\epsilon^\alpha\!\! dr\! \int_0^{2\pi}\!\! d\phi\; \frac{1}{r^{d_i-d_p+1-n-\ti{n}}}\; e^{(-s_i+s_p+n-\ti{n})i\phi}\; \pa^n \pab ^{\ti{n}} \phi_p(0,0)\ .
\end{align}
The Fourier mode must cancel and so the spin difference is given by the integer $\Delta s\equiv s_i-s_p=n-\ti{n}$ (this must be integer for the deformation to be well-defined).  Define $\ell\equiv n+\ti{n}$. The $r$ integration is then $\int dr\, r^{-(d_i-d_p+1-\ell)}$.  This only contains divergences from the lower bound of integration if $d_i-d_p+1-\ell\geq 1$.  

Consider nesting the sum in eq. \eqref{opeBi} with the sum over $p$ as the last sum performed. Thus, the sum over $n$ and $\ti{n}$ is performed knowing both $\phi_i$ and $\phi_p$, and so $s_i-s_p$ is known. If $s_i \geq s_p$, then $n=s_i-s_p+f$, $\ti{n}=:f$, where $f$ is an integer which steps up both $n$ and $\ti{n}$, keeping their difference fixed.  Therefore, $\ell=n+\ti{n}=s_i-s_p+2f$ has a minimum value of $s_i-s_p$. Moreover, $\ell$ and $s_i-s_p$ are both even or both odd.  Similarly, if $s_i \leq s_p$ then $\ti{n}=s_p-s_i+f$ and $n=:f$. Then $\ell=n+\ti{n}=s_p-s_i+2f$ has a minimum $\ell=s_p-s_i$, with similar restrictions.  All in all, $\ell\geq|\Delta s|$, and $\ell-|\Delta s|\in 2{\mathbb{Z}}$.

Note that if $d_i=d_p$ in eq. \eqref{opeBi}, then the only divergent term possible is $n=\ti{n}=0$, implying $\Delta s=0$.  This is the $h_i=h_p,\ti{h}_i=\ti{h}_p$ block, which is assumed to be diagonalized --- see below eq. \eqref{logz12}.  To keep only the divergent terms in $\epsilon$ in eq. \eqref{opeBi}, we limit the sums and evaluate only on the lower bound of the integral over $r$. We find:
\begin{align}\label{opeBi_div}
 \mathcal B_i &\sim -4 \pi \lambda\, C_{D,i,i}\, \ln(\epsilon)\, \phi_i(0,0) \\
&+4\pi \lambda \kern-2em \sum_{\substack{p,\, d_p<d_i \\ |s_i-s_p|\le\lfloor d_i-d_p\rfloor}}\;
\sum_{\substack{\ell=|s_i-s_p|\\ \ell-|s_i-s_p|\in 2{\mathbb Z}\\}}^{\lfloor d_i-d_p\rfloor}\!\!\!\!
\frac{C_{D,i,p}}{(d_i-d_p-\ell)\,\epsilon^{d_i-d_p-\ell}}\;
\frac{(1-h_i+h_p)_{n}}{n!\, (2h_p)_{n}}\; \frac{(1-\ti{h}_i+\ti{h}_p)_{\ti{n}}}{\ti{n}!\, (2\ti{h}_p)_{\ti{n}}}\; \pa^n \pab ^{\ti{n}}\phi_p(0,0)\ ,\nn
\end{align}
where $\sim$ denotes keeping only the divergent terms. Reinserting eq. \eqref{opeBi_div} into the expectation value (\ref{divexpt}) gives the divergences explicitly in terms of power law and log divergences in $\epsilon$ multiplying functions that only involve the locations $z_i$ of the fixed operators (i.e. the evaluated correlation function). There is a separate sum of such divergent terms for each annulus in figure \ref{annulifigure}. Altogether, they represent the full divergence of the perturbation integral in eq. \eqref{npointdivint}.

We next write the perturbation to the $n$-point function \eqref{npointdivint}, keeping both counter terms and the integrated correlation function. To order $\lambda^1$ we obtain:
\begin{align}\label{npointdivint_ii}
\langle \phi_{1,\lambda}\; \cdots\; \phi_{n,\lambda}\rangle_\lambda=
 \langle \phi_{1}\; \cdots\; \phi_{n}\rangle_{\lambda=0}&+\sum_{i=1}^n \langle \phi_1\; \cdots \delta \phi_i\;\cdots \phi_n\rangle\\
& -\lambda \int dz\, d\zb\;\langle {\mathcal O}_D(z,\zb)\;\phi_{1} (z_1,\zb_1)\;\cdots\; \phi_{n}(z_n,\zb_n)\rangle+O(\lambda^2)\ ,\nn
\end{align}
where $\delta \phi_i$ refer to the $O(\lambda)$ counter terms which must cancel the divergences that appear in the perturbation integral in the second line. Using eq. \eqref{opeBi_div}, the  counter terms are of the form:
\begin{align}\label{counterTerms2}
&\phi_{i,\lambda} = \phi_i -4 \pi \lambda\, C_{D,i,i}\, \ln(\epsilon)\, \phi_i + \\
&4\pi \lambda \kern-2em \sum_{\substack{p, d_p<d_i \\ |s_i-s_p|\le\lfloor d_i-d_p\rfloor}}\;
\sum_{\substack{\ell=|s_i-s_p|\\ \ell-|s_i-s_p|\in 2{\mathbb Z}\\}}^{\lfloor d_i-d_p\rfloor}\!\!\!\!
\frac{C_{D,i,p}}{(d_i-d_p-\ell)\epsilon^{d_i-d_p-\ell}}\;\frac{(1-h_i+h_p)_{n}}{n!\, (2h_p)_{n}}\;
\frac{(1-\ti{h}_i+\ti{h}_p)_{\ti{n}}}{\ti{n}!\, (2\ti{h}_p)_{\ti{n}}}\; \pa^n \pab ^{\ti{n}}\phi_p+O(\lambda^2)\ ,\nn
\end{align}
where $n=[{\ell+(s_i-s_p)}]/{2}\geq 0$ and $\ti{n}=[{\ell-(s_i-s_p)}]/{2}\geq 0$. This {\it exactly} matches the expression \eqref{counttotal2pt}, which was found using only the integrated 3-point function.  {\it Thus, the counter terms \eqref{counterTerms2} are sufficient not only for the regulation of the integrated 3-point function, but they are sufficient to regulate {\it all} integrated $n$-point functions that appear at leading order in $\lambda$.}

Since this is one of the main results of this work, it is worth considering the counter terms \eqref{counterTerms2} in detail. Given the form of the counter terms \eqref{counterTerms2}, one may be concerned in the case the $d_i-d_p-\ell=0$ in the sum.  Such a term may only be present if $d_i-d_p\in\mathbb Z$, along with the condition ${{d_i-d_p-|s_i-s_j|\in 2{\mathbb Z}}}$ so that such an $\ell$ exists. In this case, the coefficient
\begin{equation}\label{als_ii}
b_\ell \equiv-2\pi\,\frac{(d_i-d_p)\,(-1)^{\ell}}{(d_i-d_p-\ell)\,n!\,\ti{n}!}\,\Big(1-\frac{d_i-d_p+(s_i-s_p)}{2}\Big)_{\frac{\ell+s_i-s_p}{2}}\Big(1-\frac{d_i-d_p-(s_i-s_p)}{2}\Big)_{\frac{\ell-(s_i-s_p)}{2}}\nn
\end{equation}
seems ill-defined, because the denominator goes to 0.  However, one may consider how the above expression arises in the integrated 3-point of the last section.  There, integrals defining the $a_{\ell}$ \eqref{als} are manifestly well defined, as is the integral over the simplified region, which combine to give the above coefficient $b_{\ell}$.  We expect, therefore, that there is no issue.  One can see this directly by rewriting the Pochhammer symbols appearing in $b_{\ell}$ in terms of $|s_i-s_p|$ and using the identity $(\alpha)_{m}=(-1)^m(1-\alpha-m)_{m}$.  Doing so, we find:
\begin{equation}
b_\ell=-2\pi\,\frac{(d_i-d_p)}{(d_i-d_p-\ell)\,n!\,\ti{n}!}\,\Big(\frac{d_i-d_p-\ell}{2}\Big)_{\frac{\ell+|s_i-s_p|}{2}}\Big(\frac{d_i-d_p-\ell}{2}\Big)_{\frac{\ell-|s_i-s_p|}{2}}\ .
\end{equation}
We first consider the case $(\ell+|s_i-s_p|)/2=0$. This would require both $s_i-s_p=0$ and $\ell=0$.  Terms with $\ell=0$ are excluded from the corrections coming from the crescents, and so this term must come from the integration over the simplified domain.  In this case, can the denominator be $0$?  This would require that $d_i-d_p-\ell=d_i-d_p=0$.  However, if $s_i=s_j$ and $d_i=d_j$, then we have that $h_i=h_j$ and $\ti{h}_i=\ti{h}_j$ which is the case treated separately that leads to the log term.  Thus, we conclude that $(\ell+|s_i-s_j|)/2\geq 1$ for there to be a problematic $d_i-d_p-\ell=0$ denominator.  However, in such a case we may expand the first pochhammer symbol (because the subscript is $(\ell+|s_i-s_j|)/2\geq 1$), leaving 
\begin{equation}\label{alpFinite}
b_\ell=-2\pi\,\frac{(d_i-d_p)}{2\,n!\,\ti{n}!}\,\Big(\frac{d_i-d_p-\ell+2}{2}\Big)_{\frac{\ell+|s_i-s_p|}{2}-1}\Big(\frac{d_i-d_p-\ell}{2}\Big)_{\frac{\ell-|s_i-s_p|}{2}}
\end{equation}
which is manifestly well defined.  In fact, in finding equation $a_\ell$ \eqref{als} in appendix \ref{app_als}, the final lines in the proof for $a_\ell'$ are written in the form above.

There is one other interesting fact about eq. \eqref{counterTerms2}. We have truncated the sum in $\ell$ such that terms with positive powers of $\epsilon$ are simply dropped, which is not strictly necessary given that these terms will vanish as $\epsilon\rightarrow 0$. The negative powers of $\epsilon$ which diverge are tuned to cancel the divergences in the integrated $n$-point function, including the log term.  However, there are borderline cases with coefficient $\epsilon^0$, i.e. when $d_i-d_p-\ell=0$, which are exactly the problematic terms from the last paragraph.  We have already shown that these arise only when $\ell\geq 1$, allowing us to write the expression \eqref{alpFinite}.  Using the expression \eqref{alpFinite}, we see that this expression is finite if the second Pochhammer symbol has subscript $0$, and vanishes otherwise.  Thus, such terms only show up when $\ell=|s_i-s_p|=\pm\Delta s \geq 1$.  

Consider the case where $\ell=s_i-s_p\geq 1$ (i.e. $\tilde{n}=0$), along with the condition $d_i-d_p-\ell=0$, so that we are considering an $\epsilon^0$ term.  Combining the two conditions yields $\ti{h}_i=\ti{h}_p$.  Thus, $d_i-d_p=s_i-s_p=h_i-h_p=\ell\in\mathbb Z^{+}$. We find that there exists an $\epsilon$-independent term in \eqref{counterTerms2}:
\begin{equation}
2\pi \lambda\, C_{D,i,p}\;\frac{(-1)^{h_i-h_p}}{(h_i-h_p)\,(2h_p)_{h_i-h_p}}\;\pa^{{h}_i-{h}_p}\phi_p \label{epsilon0dterm}
\end{equation}        
only if there exist quasi primary operators $\phi_p$ with $\ti{h}_p=\ti{h}_i$, $h_i-h_p=s_i-s_p\geq 1$ a positive integer, and that $\phi_p$ have non-vanishing structure constants with $\mathcal{O}_D$ and $\phi_i$. Similar considerations for the case $\ell=-(s_i-s_p)\geq 1$ (i.e. $n=0$) give that there exists an $\epsilon^0$ counter term: 
\begin{equation}
2\pi \lambda\, C_{D,i,p}\;\frac{(-1)^{\ti{h}_i-\ti{h}_p}}{(\ti{h}_i-\ti{h}_p)\,(2\ti{h}_p)_{\ti{h}_i-\ti{h}_p}}\;\pab^{\ti{h}_i-\ti{h}_p}\phi_p
 \label{epsilon0dbarterm}
\end{equation} 
only if there exist quasi primaries $\phi_p$ satisfying $h_p=h_i$, $\ti{h}_i-\ti{h}_p=-(s_i-s_p)\geq 1$ a positive integer, and $C_{D,i,p}\ne0$.  We see these as ``Regge trajectories'' of quasi primaries with $d_p<d_i$ which may have their spins and dimensions ``fixed'' to agree with $\phi_i$ using only $\pa^{d_i-d_j}$ or $\pab^{d_i-d_j}$.

The above argument may be viewed with some suspicion, given that there are some fine cancellations between the Pochhammer symbols in the numerator and the $(d_i-d_p-\ell)$ factor in the denominator in eq. \eqref{counterTerms2}.  This might seem to indicate an order of limits issue.  However, one should view this as arising from a natural regulation scheme, allowing the dimensions of the fields to vary continuously, as they do on the moduli space of conformal field theories.  Thus, it prescribes a way of relaxing an $\epsilon^{\delta}$ term with $\delta$ small to $\delta=0$, resulting in the sought-after term. This result provides guidance about the types of counter terms that are available and natural at order $\epsilon^0$.  This becomes particularly important when computing the shift to the structure constants.  In the next section, we will adopt a similar approach to section \ref{sec_2pf}, where the operators are mapped to fixed points, and the domain of integration approximated.  The domain is then corrected to give the result for the local regulator.  We will find that the corrections introduced in this process have functions which are exactly those that get canceled with the above type of operators, in fact, with the same coefficients.  Thus, operators \eqref{epsilon0dterm} and \eqref{epsilon0dbarterm} do appear to play a role in the correct calculation of the shift in structure constants.  We will follow this up in section \ref{sec_bos} with an explicit example using the compact boson, where such terms are shown to be necessary to reproduce the correct result for the shift in a specific structure constant.

\section{Structure constant deformation}\label{cpt_3pf}

We start with the 3-point function of quasi primary fields on the moduli space
\begin{equation}
\langle \phi_{i,\lambda}(z_1,\zb_1)\; \phi_{j,\lambda}(z_2,\zb_2)\; \phi_{k,\lambda}(z_3,\zb_3)\rangle_\lambda = C_{i,j,k} (\lambda)
\Big(\frac{z_{23}}{z_{12}z_{13}}\Big)^{h_i(\lambda)}\Big(\frac{z_{13}}{z_{12}z_{23}}\Big)^{h_j(\lambda)}\Big(\frac{z_{12}}{z_{13}z_{23}}\Big)^{h_k(\lambda)} \times\, ({\rm a.h.})\ .,\label{3ptLambda}
\end{equation}
where $C_{i,j,k}(\lambda)$ is the structure constant. Starting at the base point $\lambda=0$, perturbations of 3-point functions come from both changes to the structure constant $\delta C_{i,j,k}$ and changes to the conformal dimensions $\delta h_i$, $\delta \ti{h}_i$.  The aim of this section is to consider $\delta C_{i,j,k}$ at first order in perturbation theory.

We start by expanding the right hand side of \eqref{3ptLambda} for small $\lambda$:
\begin{align}\label{Expand3pt}
&C_{i,j,k,\lambda}\Big(\frac{z_{23}}{z_{12}z_{13}}\Big)^{h_i(\lambda)}\Big(\frac{z_{13}}{z_{12}z_{23}}\Big)^{h_j(\lambda)}\Big(\frac{z_{12}}{z_{13}z_{23}}\Big)^{h_k(\lambda)} \times\; ({\rm a.h.}) = \\
&\Bigg\{C_{i,j,k}\bigg(1+\lambda \bigg[\frac{\pa h_i}{\pa \lambda}\Big|_{\lambda=0} \ln\Big(\frac{z_{23}}{z_{12}z_{13}}\Big)+
\frac{\pa h_j}{\pa \lambda}\Big|_{\lambda=0} \ln\Big(\frac{z_{13}}{z_{12}z_{23}}\Big)+\frac{\pa h_k}{\pa \lambda}\Big|_{\lambda=0} \ln\Big(\frac{z_{12}}{z_{13}z_{23}}\Big)\bigg]
 + \bigg[{\rm a.h.}\bigg] \bigg)  \nn \\
&\qquad\qquad\;+ \lambda\, \frac{d C_{i,j,k,\lambda}}{d \lambda}\Big|_{\lambda=0}\Bigg\}\Big(\frac{z_{23}}{z_{12}z_{13}}\Big)^{h_i(0)}\Big(\frac{z_{13}}{z_{12}z_{23}}\Big)^{h_j(0)}\Big(\frac{z_{12}}{z_{13}z_{23}}\Big)^{h_k(0)} \times\; ({\rm a.h.}) +O(\lambda^2) \ , \nn
\end{align}
where $C_{i,j,k}\equiv C_{i,j,k}(\lambda)|_{\lambda=0}$. In what follows, we will simply call $\lambda (\pa C_{i,j,k}/\pa \lambda)=\delta C_{i,j,k}$.  Using path integral formulation, the left hand side of eq. (\ref{3ptLambda}) reads:
\begin{align} \label{expand3ptPI}
\langle \phi_{i,\lambda}(z_1,\zb_1)\; \phi_{j,\lambda}(z_2,\zb_2)\; \phi_{k,\lambda}(z_3,\zb_3)\rangle_{\lambda} &= 
  \langle \phi_{i,\lambda}(z_1,\zb_1)\; \phi_{j,\lambda}(z_2,\zb_2)\; \phi_{k,\lambda}(z_3,\zb_3)\rangle\\
 &-\lambda\! \int\! d^2z\,\langle \mathcal{O}_D(z,\zb)\; \phi_i(z_1,\zb_1)\; \phi_j(z_2,\zb_2)\; \phi_k(z_3,\zb_3)\rangle+O(\lambda^2)\ .\nn
\end{align}
We again regularize the integral by cutting out small disks of radius $\epsilon$ around the insertion points $z_i$, $i=1,2,3$.  The $\epsilon$ dependence in the integral in eq. \eqref{expand3ptPI} must be canceled by the counter terms appearing in the first line, leading to the $\epsilon$-independent answer \eqref{Expand3pt}.  

The rest of this section analyzes eq. \eqref{expand3ptPI}, focusing on extracting the $\epsilon^0$ terms which contribute to $\delta C_{i,j,k}$. In section \ref{eps0ct} we calculate the $\epsilon^0$ contributions arising from counter terms; these counter terms have already been found in eqs. \eqref{epsilon0dterm} and \eqref{epsilon0dbarterm}.  This helps us identify what functions of $z_{ij}$  appearing in the integral in eq. \eqref{expand3ptPI} at order $\epsilon^0$ may be eliminated with counter terms.  In section \ref{eps0pi} we map the integral in eq. \eqref{expand3ptPI} using an $sl(2)$ transformation which sends the points $z_i$ to $0$, $1$ and $\infty$.  This allows us to identify a single integral which leads to the terms of order $\lambda^1$ appearing in eq. \eqref{Expand3pt}: the integration parameter is the cross ratio appearing in the 4-point function in eq. \eqref{expand3ptPI}, and integrand is the function of cross ratios which naturally appears in the 4-point function.  Furthermore, this allows us to identify the exact in $\epsilon$ domain over which this integral is evaluated, along with a simplified domain of integration which approximates it.  The integral over the simplified domain is difficult to analyze, even when written in terms of $sl(2)$ conformal blocks.  However, in section \ref{sub_3pf_cres} we are able to show that the integral over the simplified domain is {\it insufficient} for finding $\delta C_{i,j,k}$.  We do so by explicitly finding the corrections to $\delta C_{i,j,k}$ arising from correcting the simplified domain of integration to the exact in $\epsilon$ domain, which are expressed in terms of the CFT data.

\subsection{Contributions from counter terms of order \texorpdfstring{$\epsilon^0$}{TEXT}}\label{eps0ct}
Let us consider the counter terms associated with $\phi_{k,\lambda}$ in the first term on the \textsc{rhs} of eq. \eqref{expand3ptPI}.  Replacing $\phi_{k,\lambda}$ by a sum of $\epsilon^0$ counter terms \eqref{epsilon0dterm} gives the $\epsilon^0$ contribution to the first term:
\begin{align}
2\pi \lambda\, \sum_{\substack{p, \ti{h}_p=\ti{h}_k \\
h_k-h_p\in {\mathbb Z}^+}} C_{D,k,p}\;\frac{(-1)^{h_k-h_p}}{(h_k-h_p)\,(2h_p)_{h_k-h_p}}\;
\pa_3^{{h}_k-{h}_p}\langle \phi_i(z_1,\zb_1)\; \phi_j(z_2,\zb_2)\; \phi_p(z_3,\zb_3)\rangle\ ,
\end{align}  
where the sum is over all $\phi_p$ allowed. We divide this expression by the pre-factor on the \textsc{rhs} of eq. (\ref{Expand3pt}), for later convenience.  We find:
\begin{align}
\mathcal C_{k,0}^{\rm hol}&\equiv \sum_{\substack{p, \ti{h}_p=\ti{h}_k  \\
h_k-h_p\in {\mathbb Z}^+}}  \frac{2\pi \lambda\,C_{D,k,p}\;\frac{(-1)^{h_k-h_p}}{(h_k-h_p)(2h_p)_{h_k-h_p}}\,\pa_3^{{h}_k-{h}_p}
\langle \phi_i(z_1,\zb_1)\; \phi_j(z_2,\zb_2)\; \phi_p(z_3,\zb_3)\rangle}{{z_{12}^{-(h_i+h_j-h_k)}z_{13}^{-(h_i+h_k-h_j)}z_{23}^{-(h_j+h_k-h_i)}}\;\times\; ({\rm a.h.})}\nn\\
&=2\pi \lambda\, \sum_{\substack{p, \ti{h}_p=\ti{h}_k \label{Ck} \\
h_k-h_p\in {\mathbb Z}^+}} C_{D,k,p}\,C_{i,j,p}(-1)^{h_k-h_p}\,\frac{(h_k-h_p-1)!}{(2h_p)_{h_k-h_p}}\Big(\frac{z_{13}}{z_{12}}\Big)^{(h_k-h_p)} \\
&\qquad \qquad\qquad\qquad\qquad\qquad \times \sum_{q=0}^{h_k-h_p} \frac{(h_i+h_p-h_j)_q}{q!} \frac{(h_j+h_p-h_i)_{h_k-h_p-q}}{(h_k-h_p-q)!} \Big(\frac{z_{23}}{z_{13}}\Big)^{q}\ .\nn
\end{align}
We label this term as $\mathcal C_{k,0}^{\rm hol}$ because it arises from the holomorphic derivative family of {$\mathcal C$}ounter terms for $\phi_k$ at order $\epsilon^0$, i.e. eq. \eqref{epsilon0dterm}.  Define the complex parameter $w\equiv {z_{23}}/{z_{13}}$, which represents the only independent parameter written in terms of $z_1,z_2,z_3$, and is both dimensionless and translation invariant. This allows us to write
\begin{align}
\mathcal C_{k,0}^{\rm hol}=2\pi \lambda\!\!\! \sum_{\substack{p, \ti{h}_p=\ti{h}_k \nn \\
h_k-h_p\in {\mathbb Z}^+}}\!\!\! C_{D,k,p}\,C_{i,j,p}\,\frac{(h_k-h_p-1)!}{(2h_p)_{h_k-h_p}}\,\Big(\frac{1}{1-w}\Big)^{(h_k-h_p)}  P^{(-h_j-h_k+h_i,2h_p-1)}_{h_k-h_p}\big(2(1-w)-1\big)\ ,\nn
\end{align}
where $P^{(\alpha,\beta)}_{\gamma}(x)$ is the Jacobi polynomial and is defined in appendix \ref{app_3pf_jacobi}. Using the identity
\begin{equation}
P_{\gamma}^{(\alpha,\beta)}(2/u-1)=\frac{1}{u^\gamma} P_\gamma^{(\alpha,-\beta-\alpha-2\gamma-1)}(2u-1)\nn
\end{equation}
with $u\equiv (1-w)$, we obtain
\begin{align}\label{Ck_i}
\mathcal C_{k,0}^{\rm hol}= 2\pi \lambda \sum_{\substack{p, \ti{h}_p=\ti{h}_k  \\
h_k-h_p\in {\mathbb Z}^+}} C_{D,k,p}\,C_{i,j,p}\,\frac{(h_k-h_p-1)!}{(2h_p)_{h_k-h_p}}\; P^{(-h_j-h_k+h_i,-h_i-h_k+h_j)}_{h_k-h_p} \Big(\frac{z_{13}}{z_{12}}+\frac{z_{23}}{z_{12}}\Big)\ .
\end{align}
There are the anti-holomorphic terms as well, which occur from eq. (\ref{epsilon0dbarterm}). Thus, at each insertion point, there are two sums of such operators: the terms arising from holomorphic derivatives of quasi primaries, and the anti-holomorphic derivatives of quasi primaries.  Similar arguments hold for the counter term insertions at $z_1$ and $z_2$, and are summarised in eq. \eqref{3pt_ct_123}.

\subsection{Domains of integration and integrand}\label{eps0pi}
We now consider the perturbation integral in eq. (\ref{expand3ptPI}). The 4-point function may be written as usual in terms of a fixed function times a function of the cross ratio $f(\zeta,\bar{\zeta})$:
\begin{align}\label{4ptconvention}
\hat{\mathcal I}&\equiv -\lambda \int d^2z\;
\langle \mathcal{O}_D(z,\zb)\; \phi_i(z_1,\zb_1)\; \phi_j(z_2,\zb_2)\; \phi_k(z_3,\zb_3)\rangle  \\
&=-\lambda \int d^2z\; \Big(\frac{z_{23}}{z_{12}z_{13}}\Big)^{h_i}\Big(\frac{z_{13}}{z_{12}z_{23}}\Big)^{h_j}
\Big(\frac{z_{12}}{z_{13}z_{23}}\Big)^{h_k} \Big(\frac{z_{12}z_{13}}{(z_1-z)^2z_{23}}\Big)^{h_D}\;\times\; ({\rm a.h.})\;\times\; f(\zeta,\bar{\zeta}) \nn
\end{align}
where $h_D=\ti h_D=1$, and the cross ratio is defined as $\zeta\equiv ({z_{12}(z-z_3)})/(z_{23}(z_1-z))$.  Our particular choice for the definition of the function of the cross ratio above will become clear momentarily.  To regulate the integral, we excise disks around $z_i$, and limit the integration parameter $z$ to the domain $|z-z_1|>\epsilon, |z-z_2|>\epsilon, |z-z_3|>\epsilon$.  We may take this integral to a more standard form by performing an $sl(2)$ transformation
\begin{equation}\label{zhat}
\hat{z}=-\frac{z_{12}\,(z-z_3)}{z_{23}\,(z-z_1)}
\end{equation}
which maps $\{z_1,z_2,z_3\}\to\{0,1,\infty\}$, and further $\zeta=\hat{z}$. The integral \eqref{4ptconvention} reads
\begin{align}\label{4ptconvention_hat}
\hat{\mathcal I}=-\lambda\; z_{12}^{-(h_i+h_j-h_k)}z_{13}^{-(h_i+h_k-h_j)}z_{23}^{-(h_j+h_k-h_i)}
\times({\rm a.h.}) \int d^2\hat{z}\; f(\hat{z},\bar{\hat{z}}) \, .
\end{align}

The allowed region in the $\hat{z}$ plane is given by three simultaneous constraints:
\begin{align}
&\Big|\hat{z}-\frac{z_{12}}{z_{23}}\;\frac{\epsilon^2}{|z_{13}|^2-\epsilon^2}\Big|  > \frac{|z_{12}|}{|z_{23}||z_{13}|}\;\frac{\epsilon}{1-\frac{\epsilon^2}{|z_{13}|^2}}\ ,\label{4ptallowedfull_first}\\[2pt]
&\Big|\hat{z}-\Big(1+\frac{z_{13}}{z_{23}}\;\frac{\epsilon^2}{|z_{12}|^2-\epsilon^2}\Big)\Big| > \frac{|z_{13}|}{|z_{23}||z_{12}|}\;\frac{\epsilon}{1-\frac{\epsilon^2}{|z_{12}|^2}}\ ,\qquad\qquad
\Big|\hat{z}+\frac{z_{12}}{z_{23}}\Big|< \frac{|z_{12}||z_{13}|}{|z_{23}|\;\epsilon}\ .\nn 
\end{align}
As in section \ref{sec_2pf}, we find the above domain difficult to deal with directly.  Therefore, we turn to the leading order terms in $\epsilon$, finding:
\begin{equation}
|\hat{z}|>\frac{|z_{12}|\,\epsilon}{|z_{23}||z_{13}|}\ , \qquad |\hat{z}-1|>\frac{|z_{13}|\,\epsilon}{|z_{23}||z_{12}|}\ ,
\qquad |\hat{z}|< \frac{|z_{12}||z_{13}|}{|z_{23}|\,\epsilon} \label{4ptallowed0}
\end{equation}
which we refer to as the {\it simplified domain} of integration.  We will consider the higher order terms in $\epsilon$ later in section \ref{sub_3pf_cres}. The powers of $z_{ij}$ multiplying the integration on the \textsc{rhs} of \eqref{4ptconvention_hat} are just the powers expected from the unperturbed 3-point function, which also appear in eq. \eqref{Expand3pt}.  We strip these powers off for easy comparison to the terms of order $\lambda$ inside of the curly brackets on the \textsc{rhs} of eq. \eqref{Expand3pt}, and so define:
\begin{align}\label{perturbstruct}
\mathcal I \equiv -\lambda \int d^2\hat{z}\; f(\hat{z},\bar{\hat{z}})= \mathcal I_{\rm s}+ \mathcal I_{\rm i}+\mathcal I_{\rm j}+ \mathcal I_{\rm k}
\end{align}
where the subscript `s' refers to the simplified domain of integration, i.e. eq. \eqref{4ptallowed0}.  This is the generalized sum rule from \cite{Behan:2017mwi}, writing the domain of integration piecewise.  The additional integrals ${\mathcal I}_i, {\mathcal I}_j, {\mathcal I}_k$, which we explore in the next subsection, give the corrections from the simplified domain \eqref{4ptallowed0} to the exact in $\epsilon$ domain \eqref{4ptallowedfull_first}.  The corrections to the domain of integration are again given by double crescent regions (see figure \ref{crescentfig}), and are given subscripts $i,j,k$ for the operator insertion near which they are located.  All four integrals have the same integrand, and the four separate terms simply represent dividing up the domain of integration.

We next turn to the integrand.  The function $f(\hat{z},\bar{\hat{z}})$ is naturally written in terms of conformal blocks.  Given that the counter terms (\ref{counterTerms2}) are written in terms of the $sl(2)$ descendants of quasi primary fields, it will be necessary to write $f(\hat{z},\bar{\hat{z}})$ in terms of $sl(2)$ conformal blocks for direct comparison (as opposed to using Virasoro conformal blocks).  Recalling eqs. \eqref{OPEoperator}-\eqref{descentinner} and \eqref{betas}, and the discussion in appendix \ref{app_sl2} that the set of all quasi primaries and their $sl(2)$ descendants form a basis, we consider the 4-point function: 
\begin{equation}\label{sl2_4pf}
G_{3,4}^{2,1}(z,\bar{z})\equiv \langle \phi_1\mid\phi_2(1,1)\; \phi_3(z,\bar{z}) \mid \phi_4\rangle\ .
\end{equation}
We insert a complete set of orthonormal states
\begin{align}\label{blockpower}
G_{3,4}^{2,1}(z,\zb)&=\sum_p\!\sum_{n,\ti{n}=0}^\infty \frac{1}{n!\,(2h_p)_{n}}\;\frac{1}{\ti{n}!\,(2\ti{h}_p)_{\ti{n}}}\;
\langle \phi_1\mid\phi_2(1,1)\mid \phi_p^{(n,\ti{n})}\rangle\;\langle\phi_p^{(n,\ti{n})} \mid  \phi_3(z,\zb) \mid \phi_4\rangle\\
&=\sum_p z^{h_p-h_3-h_4}\;\zb^{\ti{h}_p-\ti{h}_3-\ti{h}_4}\;C_{p,3,4}\;C_{p,2,1} \sum_{n=0}^\infty \frac{(h_2-h_1+h_p)_{n}\;(h_3-h_4+h_p)_{(n)}}{(2h_p)_{n}}\; \frac{z^n}{n!}\nn \\
&\qquad \qquad \qquad \qquad \qquad \quad\qquad\qquad\;  \times 
\sum_{\ti{n}=0}^\infty \frac{(\ti{h}_2-\ti{h}_1+\ti{h}_p)_{\ti{n}}\;(\ti{h}_3-\ti{h}_4+\ti{h}_p)_{(\ti{n})}}{(2\ti{h}_p)_{\ti{n}}}\; \frac{\zb^{\ti{n}}}{\ti{n}!}\ ,\nn
\end{align}
and observe that the $sl(2)$ conformal blocks may be written in terms of ordinary hypergeometric functions (see appendix \ref{app_3pf_hyper} for definitions) \cite{Dolan:2003hv,Osborn:2012vt,Dolan:2000ut} \footnote{We thank Scott Collier for useful discussions about this point, and pointing us to these references.}:
\begin{align}\label{3pf_hyper}
G_{3,4}^{2,1}(z,\zb)&= \sum_p \;C_{p,3,4}\; C_{p,2,1}\;z^{-h^{3,4}_p}\;\zb^{-\ti{h}^{3,4}_p} 
\,_2F_1\bigg(\genfrac{}{}{0pt}{}{h^{p,2}_1,h^{p,3}_4}{2h_p};z\bigg)
\,_2F_1\bigg(\genfrac{}{}{0pt}{}{\ti{h}^{p,2}_1,\ti{h}^{p,3}_4}{2\ti{h}_p};\zb\bigg)
\end{align}
where we define the symbols 
\begin{equation}\label{hshbars}
h^{i,j}_k= h_i+h_j-h_k\ , \qquad h^{i}_j=h_i-h_j\ , \qquad \ti{h}^{i,j}_k=\ti{h}_i+\ti{h}_j-\ti{h}_k\ , \qquad \ti{h}^i_j=\ti{h}_i-\ti{h}_j\ .
\end{equation}
Crossing symmetry imposes
\begin{equation}\label{cross_sym}
G_{3,4}^{2,1}(z,\zb)=G_{3,2}^{4,1}(1-z,1-\zb)=\frac{1}{z^{2h_3}\zb^{2\ti{h}_3}}G_{3,1}^{2,4}(1/z,1/\zb)\ .
\end{equation}
The corresponding expressions in terms of hypergeometric functions are outlined in eq. \eqref{app_3pf_cross}.  To make contact with perturbation theory, we set $1\rightarrow i$, $2\rightarrow j$, $3\rightarrow D$, and $4\rightarrow k$, and note that $h_D=\ti{h}_D=1$. This gives
\begin{align}\label{3pf_hyper_d}
& G_{D,k}^{j,i}(z,\zb) = \sum_p \;C_{p,D,k}\; C_{p,j,i}\;z^{h^p_k-1}\;\zb^{\ti{h}^p_k-1}  \,_2F_1\bigg(\genfrac{}{}{0pt}{}{h^{p,j}_i,h^p_k+1}{2h_p};z\bigg) \;
_2F_1\bigg(\genfrac{}{}{0pt}{}{\ti{h}^{p,j}_i,\ti{h}^p_k+1}{2\ti{h}_p};\zb\bigg)\ .
\end{align}
The crossing symmetry expressions are summarised in eq. \eqref{app_3pf_cross_d}. 

Restoring $z\rightarrow \hat{z}$, we have ${ G_{D,k}^{j,i}(\hat{z},\bar{\hat{z}})=f(\hat{z},\bar{\hat{z}})}$, which must now be integrated over the allowed region in the perturbation integral \eqref{perturbstruct}.  The utility of eqs. \eqref{3pf_hyper_d} and the crossing symmetric expressions \eqref{app_3pf_cross_d} is that they make the divergences in the integral quite clear.
Note that although the series of descendants which contribute to the 4-point function is infinite, {\it only a finite number of terms give divergences}, which effectively truncates the hypergeometric series to a finite number of terms when considering the singularity structure\footnote{Although more care may be needed in cases where there is a conserved current in the spectrum, or if the theory develops a continuum \cite{Balthazar:2022hzb}.   However, in such cases there may be additional ``charge conservation'' constraints for correlators.  This would limit counter terms to be within the same charge sector, possibly still limiting to a finite number of operators.}.
We may use the crossing symmetric expressions to extract the singularity structure at other points as well.

The integral $\mathcal I_{\rm s}$ \eqref{perturbstruct} is difficult to compute in general. Consider the integrand $G_{D,k}^{j,i}$ \eqref{3pf_hyper_d}. While we have analytic expressions for the divergent terms at $\hat z=0$ in the local coordinates, the full expression becomes cumbersome when attempting to evaluate it near other points, especially $\hat z=1$ and $\infty$ where the hypergeometric functions in \eqref{3pf_hyper_d} have branch points.  We know that the branch cut discontinuities must cancel, at least in the sum over $p$, leading to the crossing symmetric expressions which are manifestly single valued around these points.  Thus, unlike the case of the integrated 3-point function, see eqs. \eqref{logz12} and \eqref{2pf_eps1}, we are not able to obtain a universal expression for $\mathcal I_{\rm s}$, i.e. an expression which holds for all conformal field theories. The method for integrating the 4-point function may be addressed theory by theory, and in particular cases may be solved exactly (some examples are discussed in section \ref{sec_bos}). 

Speaking generally, the integral $\mathcal I_{\rm s}$ has an $\epsilon^0$ piece which contributes to the shift $\delta C_{i,j,k}$. We may ask whether this contribution from the simplified domain (\ref{4ptallowed0}) and the calculation from the exact in $\epsilon$ domain \eqref{4ptallowedfull_first} give the same $\epsilon^0$ piece.  If so, then $\delta C_{i,j,k}$ is unaffected, and one can safely use the simpler cutoffs (\ref{4ptallowed0}).  We will now turn to this point and find that, interestingly, the constant pieces from the simplified domain (\ref{4ptallowed0}) {\it need to be corrected} when using the full expressions \eqref{4ptallowedfull_first}.  This leads to corrections to the shift in structure constant $\delta C_{i,j,k}$. We find that these corrections are {\it universal}: they can be exactly calculated and expressed in terms of the original CFT data, namely the unperturbed structure constants and conformal dimensions.

\subsection{Corrections to the simplified domain to all orders in \texorpdfstring{$\epsilon$}{TEXT}}\label{sub_3pf_cres}

We now turn our attention to the integrals $\mathcal I_i, \mathcal I_j, \mathcal I_k$ in eq. \eqref{perturbstruct}.  As discussed above, the evaluation of $\mathcal I_{\rm s}$ is difficult, given that the different crossing symmetric expressions for the 4-point function are useful only in a fixed neighborhood around the point for which they were adapted.  The three integrals over crescents, however, are localized near the respective operators.  The double crescents in each case lie order $\epsilon$ away from the operator in question. 
Thus, using crossing symmetry, one inserts the form of the function $f(\hat{z},\bar{\hat{z}})$ adapted to the point in question \eqref{app_3pf_cross_d} to compute the integral over the double crescent near this point.  This will be sufficient to find the order $\epsilon^0$ contributions, which is our primary purpose in this subsection.

We tackle the hole at $z_3$ first, corresponding to the hole at $\hat{z}=0$. We identify the bounds of integration using eq. (\ref{4ptallowedfull_first}). Define the real coordinates $\hat{z}\equiv r e^{i(\phi +\phi_{12}-\phi_{23}+\pi)}$ where the phases $\phi_{12}$ and $\phi_{23}$ are defined by: $z_{12}=|z_{12}|e^{i\phi_{12}}$, $z_{23}=|z_{23}|e^{i\phi_{23}}$. Plugging these into (\ref{4ptallowedfull_first}), we find that the new contour in the $\hat z$ plane is defined by:
\begin{equation}
r=\frac{|z_{12}|}{|z_{23}||z_{13}|}\; \frac{1}{R_{\infty,13}}\; \epsilon\ , \qquad 
R_{\infty,ij}\equiv \sqrt{\Big(1-\frac{\epsilon^2}{|z_{ij}|^2}\Big)+\frac{\epsilon^2}{|z_{ij}|^2} \cos^2(\phi)} +\frac{\epsilon}{|z_{ij}|}\cos(\phi)\ 
\end{equation}
where we have defined $R_{\infty,ij}$ for later use.  The correction $\mathcal I_{\rm k}$ at $\hat z=0$ is given by
\begin{align}
\mathcal I_{k}\equiv -\lambda \int_{{\rm cres}_0} \!\!\!\!\!d^2\hat{z}f(\hat{z},\bar{\hat{z}})=-2\lambda
\int_{0}^{2\pi}\!\!\!d\phi\! \int_{\frac{\epsilon\, |z_{12}|}{|z_{23}||z_{13}|}\frac{1}{R_{\infty,13}}}^{\frac{\epsilon\, |z_{12}|}{|z_{23}||z_{13}|} } 
\!\!r dr\; f(\hat{z},\bar{\hat{z}})\ .
\end{align}
The form of $f(\hat{z},\bar{\hat{z}})$ is the one adapted to the singularity at $\hat{z}=0$, given by eq. \eqref{3pf_hyper_d}.  We find:
\begin{align}\label{crescentcorrectfull}
&\mathcal I_{k}=-\lambda \sum_p\!\sum_{n,\ti{n}=0}^\infty C_{p,D,k}\;C_{p,j,i}\; e^{i\pi(s_k-s_p-(n-\ti{n}))}\;
\frac{(h^{p,j}_i)_n\,(h^p_k+1)_n}{(2 h_p)_n\, n!}\;
\frac{(\ti{h}^{j,p}_i)_{\ti{n}}\,(\ti{h}^p_k+1)_{\ti{n}}}{(2\ti{h}_p)_{\ti{n}}\, \ti{n}!} \\
&\quad\times\; e^{-i(\phi_{12}-\phi_{23})(s_k-s_p-(n-\ti{n}))}\;
2\int_{0}^{2\pi}\!\!\!d\phi \!\int_{\frac{\epsilon|z_{12}|}{|z_{23}||z_{13}|} \frac{1}{R_{\infty,13}}}^{\frac{\epsilon|z_{12}|}{|z_{23}||z_{13}|} }\;
\frac{dr}{r^{d_k-d_p+1-(n+\ti{n})}}\;e^{-i\phi(s_k-s_p-(n-\ti{n}))}\ .\nn
\end{align}
The integral in the second line is the same as the integrals in eq. \eqref{AiAj} upon substituting $d_i-d_j\rightarrow d_k-d_p-(n+\ti{n})$ and $s_i-s_j\rightarrow s_k-s_p- (n-\ti{n})$.  The case $d_k-d_p-(n+\ti{n})=0$ produces a term $\log(R_{\infty,13})$, and represents a purely perturbative expansion in $\epsilon$.  The singular terms correspond to $d_k-d_p-(n+\ti{n})>0$.  With this restriction, and using our analysis in appendix \ref{app_als}, we evaluate the integral:
\begin{align}\label{4pointcorrectioncrescent}
& \int_{0}^{2\pi}\!\!\!d\phi\! \int_{r=\frac{\epsilon\, |z_{12}|}{|z_{23}||z_{13}|} \frac{1}{R_{\infty,13}}}^{\frac{\epsilon\, |z_{12}|}{|z_{23}||z_{13}|} }
\!\!\!\!\!dr\,\frac{e^{-i\phi((s_k-s_p)-(n-\ti{n}))}}{r^{d_k-d_p+1-(n+\ti{n})}} =
2\pi\!\!\!\!\!\!\!\!\!\!\!\!\!\!\!\sum_{\substack{\ell \geq |(s_k-s_p)-(n-\ti{n})| \\ \ell\neq 0, \\ \ell-|(s_k-s_p)-(n-\ti{n})|\in 2{\mathbb{Z}}}}
\!\!\!\!\!\!\!\left(\frac{|z_{23}||z_{13}|}{|z_{12}|\epsilon}\right)^{d_k-d_p-(n+\ti{n})}\!\!\left(\frac{\epsilon}{|z_{13}|}\right)^\ell  \\[10pt]
&\!\!\!\!\! \times(-1)^\ell \frac{\left(1-\frac{d_k-d_p-(n+\ti{n})+s_k-s_p-(n-\ti{n})}{2}\right)_{\frac{\ell+(s_k-s_p-(n-\ti{n}))}{2}} 
\Big(1-\frac{d_k-d_p-(n+\ti{n})-(s_k-s_p)+(n-\ti{n})}{2}\Big)_{\frac{\ell-(s_k-s_p-(n-\ti{n}))}{2}}}{ \Big(\frac{\ell+(s_k-s_p-(n-\ti{n}))}{2}\Big)!\;
\Big(\frac{\ell-(s_k-s_p-(n-\ti{n}))}{2}\Big)!\;(d_k-d_p-(n+\ti{n})-\ell)}\ . \nn
\end{align}
As discussed in section \ref{cpt_o}, the above is a pure power law series with no logarithmic terms.

The singular terms in eq. \eqref{4pointcorrectioncrescent} are grouped in with the singularities in $\mathcal I_{\rm s}$ (\ref{4ptallowed0}), and cancel against counter terms \eqref{counterTerms2}. The finite terms that vanish as $\epsilon\rightarrow 0$ are negligible.  However, there may exist terms that go like $\epsilon^0$ and do not vanish.  Such terms require $\ell=d_k-d_p-(n+\ti{n})$, and $d_k-d_p\in\mathbb Z^+$. Note that at least one of the Pochhammer symbols in the sum \eqref{4pointcorrectioncrescent} has a non-zero subscript (otherwise $\ell=0$, which is an excluded term in the sum), and thus contains the factor $(d_k-d_p-(n+\ti{n})-\ell)$, which cancels the one in the denominator. If the subscripts of both Pochhammer symbols are non-zero, then the constant term is simply 0. Thus, we have either $\ell=d_k-d_p-(n+\ti{n})=s_k-s_p-(n-\ti{n})$ or $\ell=d_k-d_p-(n+\ti{n})=-(s_k-s_p-(n-\ti{n}))$.  

We first consider the case $\ell=d_k-d_p-(n+\ti{n})=s_k-s_p-(n-\ti{n})$, i.e. the bottom term in the sum \eqref{4pointcorrectioncrescent}. This gives $\ti{h}_k-\ti{h}_p=\ti{n}$ and $\ell={h}_k-{h}_p-{n}\in\mathbb Z^+$, and imposes $0\leq n\leq h_k-h_p-1$.  Plugging into (\ref{crescentcorrectfull}), we have $(1-\ti{h}_k+\ti{h}_p)_{\ti{n}}=(1-\ti{n})_{\ti{n}}=0$ unless $\ti{n}=0$.  Thus, to have a non-zero entry requires $\ti{h}_k=\ti{h}_p$, $\ti{n}=0$, and $d_k-d_p=s_k-s_p=h_k-h_p$.  This leads to one of the two $\epsilon$-independent pieces of eq. (\ref{crescentcorrectfull}), which we refer to as $\mathcal I_{k,0}^{\rm hol}$, where `hol' denotes  that the conditions on $h_k$, $\tilde h_k$, $h_p$, $\tilde h_p$ exactly correspond to those of holomorphic counter terms \eqref{epsilon0dterm}. We find:
\begin{align} 
&\mathcal I_{k,0}^{\rm hol}=-2\pi \lambda\!\!\!\!\!\!\!\sum_{\substack{ p \\ \ti{h}_p=\ti{h}_k\\ h_{k}-h_{p} \in {\mathbb Z}^+}}
\!\!\!\!\!\!\!\!\sum_{n=0}^{h_k-h_p-1}\!\!\!C_{p,D,k}\,C_{p,j,i}\; (-1)^{h^k_p-n}\,
\frac{(h^{j,p}_i)_n\,(h^p_k+1)_n}{(h^k_p-n)\,(2 h_p)_n\, n!}\,
\Big(\frac{z_{23}}{z_{12}}\Big)^{h^k_p-n}\ ,\nn
\end{align}
where we have reunited the phases $\phi_{23}$ and $\phi_{12}$ with the appropriate magnitudes, e.g. $z_{12}=|z_{12}|e^{i\phi_{12}}$. Furthermore, we read the above $\mathcal I_{k,0}^{\rm hol}$ as meaning the contribution to eq. \eqref{expand3ptPI} arising from the $\mathcal I$ntegral, at order $\epsilon^0$, which does not vanish because $\ti{h}_p=\ti{h}_k$ in the sum \eqref{4pointcorrectioncrescent}, i.e. that only the holomorphic weights $h_k, h_p$ may differ.  Defining $q\equiv h_k-h_p-n$ and using $q$ as the new summation index, we have:
\begin{align} \label{Ccres0hol}
\mathcal I_{k,0}^{\rm hol}=-2\pi \lambda\!\!\!\!\!\!\! \sum_{\substack{ p \\ \ti{h}_p=\ti{h}_k\\ h_{k}-h_{p} \in {\mathbb Z}^+}}
\!\!\!\!\!\!\!\sum_{q=1}^{h_k-h_p}\!C_{p,D,k}\,C_{p,j,i}\, (-1)^{q-1}\, \frac{(h^{p+j}_i)_{h^k_p-q}\,(h^p_k+1)_{h^k_p-q-1}}{(2 h_p)_{h^k_p-q} \,(h^k_p-q)!}\,
\Big(\frac{z_{23}}{z_{12}}\Big)^{q}\ .
\end{align}
Using the definition for Jacobi polynomials \eqref{app_3pf_jacobi}, we finally obtain: 
\begin{align}\label{3ptinegralcorrectS}
&\mathcal I_{k,0}^{\rm hol}=-2\pi \lambda \kern -2em\sum_{\substack{ p \\ \ti{h}_p=\ti{h}_k,\,h_{k}-h_{p} \in {\mathbb Z}^+}} \kern -2em
\frac{(h^k_p-1)!}{(2h_p)_{h^k_p}}\;C_{p,D,k}\,C_{p,j,i}  
\Big(P^{(-h^{j,k}_i,-h^{i,k}_j)}_{h^k_p} 
\Big(\frac{z_{13}}{z_{12}}+\frac{z_{23}}{z_{12}}\Big)-\frac{ (-h^{j,k}_i+1)_{h^k_p}}{(h^k_p)!}\Big)\ .
\end{align}
The resulting sum is {\it almost} the same form as the one we found for the counter terms \eqref{Ck_i}! Perhaps this is not too surprising, given that the counter terms and the expression for the conformal blocks both arise from the OPE (although integrated over regions that look dissimilar).  Furthermore, in both cases the calculation is at first done for general conformal weights, giving manifestly well defined expressions, and then relaxed to obtain the $\epsilon^0$ term.

The conformal block calculation \eqref{3ptinegralcorrectS} has yielded exactly the same Jacobi polynomial as in \eqref{Ck_i}, with only the $q=0$ term (the second term on the \textsc{rhs}) subtracted off --- c.f. eq. (\ref{Ccres0hol}).  This explicitly shows that in the exact region (\ref{4ptallowedfull_first}), the counter terms (\ref{epsilon0dterm}) must be considered, at least when comparing the results between the simplified and full domain of integration.  In the simplified region (\ref{4ptallowed0}), however, these counter terms may not be required (in the compact boson example in section \ref{sec_bos}, we indeed perform calculations where they are not required). This is {\it not} to say that we may simply drop Jacobi polynomials when they show up in $\mathcal I_{\rm s}$, rather, any Jacobi polynomials that appear in $\mathcal I_{\rm s}$ must be combined with those from the crescent calculations, and then the full set of  Jacobi polynomials are removed with counter terms. As a result, the order $\epsilon^0$ term has been shifted by a constant because the leading term in the Jacobi polynomial from the crescents is not included in the sum \eqref{3ptinegralcorrectS}. Thus, the $q=0$ term represents a further shift to the structure constant.

Recall that at the same location $z=z_3$ $(\hat z=0)$, one should also consider the case $\ell=-(s_k-s_p-(n-\ti{n}))$ as well, see below eq. \eqref{4pointcorrectioncrescent}. This will lead to a similar expression to \eqref{3ptinegralcorrectS} with holomorphic weights and variables  flipped to antiholomorphic counterparts.  This amounts to requiring the counter terms (\ref{epsilon0dbarterm}). Furthermore, there are  contributions from the crescent regions at $z=z_2$ $(\hat z=1)$ and $z=z_1$ $(\hat z=\infty)$, which are obtained using crossing symmetry \eqref{cross_sym}. For the double crescent region at $\hat z=1$ we define $(1-\hat z)\equiv re^{i(\phi+\phi_{13}-\phi_{23})}$, and find:
\begin{align}\label{3ptinegralcorrect_1}
&\mathcal I_{j,0}^{\rm hol}= -2\pi\lambda\kern-2em 
\sum_{\substack{ p \\ \ti{h}_p=\ti{h}_j,\, h_{j}-h_{p} \in {\mathbb Z}^+}}
\kern-2em \frac{(h_j-h_p-1)!}{(2h_p)_{h_j-h_p}}\;C_{p,D,j}\,C_{p,k,i} \Big((-1)^{h^j_p}P^{(-h^{j,i}_k,-h^{j,k}_i)}_{h^j_p}
\Big(\frac{z_{21}}{z_{13}}+\frac{z_{23}}{z_{13}}\Big)-\frac{(-h^{j,k}_i+1)_{h^j_p}}{(h^j_p)!}\Big)\ .
\end{align}
Comparing to the counter term insertions in the second equality in eq. \eqref{3pt_ct_123}, we observe that the shifted hole at $z=z_2$ act somewhat differently.  For the double crescent region at $\hat z=\infty$, we use $\hat z\equiv r e^{{i(\phi+\phi_{12}-\phi_{23}+\pi)}}$ and obtain:
\begin{align}\label{3ptinegralcorrect_inf}
&\mathcal I_{i,0}^{\rm hol}= -2\pi\lambda\kern-2em
\sum_{\substack{ p \\ \ti{h}_p=\ti{h}_i,\, h_{i}-h_{p} \in {\mathbb Z}^+}}\kern-2em
\frac{(h^i_p-1)!}{(2h_p)_{h^i_p}}\;C_{p,D,i}\,C_{p,j,k}  \Big(P^{(-h^{i,j}_k,-h^{i,k}_j)}_{h^i_p}
\Big(\frac{z_{12}}{z_{23}}+\frac{z_{13}}{z_{23}}\Big)-\frac{(-h^{i,j}_k+1)_{h^i_p}}{(h^i_p)!}\Big)\ ,
\end{align}
which agrees with the counter terms in the last equality in eq. \eqref{3pt_ct_123}. Thus, at all three insertion points, shifting from the simplified domain (\ref{4ptallowed0}) to the $\epsilon$-exact domain (\ref{4ptallowedfull_first}) introduces shifts to the functional form, which can be compensated for by counter terms, but also shifts by calculable constants, given the CFT data.  Together with anti-holomorphic derivative insertions, there are in total six constant terms that modify $C_{i,j,k}$, see eq. \eqref{deltaCintro} in the introduction.

In the above calculation, the shifts arising from the crescents near $\hat{z}=0$ and $\hat{z}=\infty$ have the opposite signs of those arising from the counter terms.  However, for the crescents at $\hat{z}=1$, half the terms from the crescents match the signs from the counter term calculation, and the other half appear with negative signs.  We expect all insertions to be treated on the same footing for a local regulator.  Thus, we interpret this to mean that the calculation of $\mathcal I_{\rm s}$ contains no occurrences of the Jacobi polynomials germane to $\hat{z}=0, \hat{z}=\infty$, and are introduced by the crescents, making them natural, and included the same way at both points.  To have the same effect on the operator at $\hat{z}=1$ we should have that some Jacobi polynomials appear in $\mathcal I_{\rm s}$ with coefficient twice what appears above, and the others do not.  The shift from the crescents at $\hat{z}=1$ will then shift them to have all the same coefficient, making the counter terms identical at all points.  We expect this on the grounds that the local regulator should have all counter terms show up symmetrically under interchange of operators, or equivalently, one may choose which operators to map to $\hat{z}=0,1,\infty$ respectively. We will consider this in the next section for the CFT of a compact boson and show that the inclusion of these counter terms is needed with exactly the coefficients \eqref{epsilon0dterm}-\eqref{epsilon0dbarterm}.

\section{Compact boson}\label{sec_bos}

We now consider the well-known theory of a compact free boson. The shifts to the conformal dimensions and structure constant can be exactly computed in this theory. Hence, it provides a concrete setup to compare our conformal perturbation theory analysis with exact results, appropriately expanded for small $\lambda$. In particular, in subsection \ref{freebos_3pf} we present an example where the integral $\mathcal I_{\rm s}$ \eqref{perturbstruct} leads to an order $\epsilon^0$ piece which is insufficient to give the correct change to a certain structure constant. To obtain a matching between the perturbative and exact results, we modify the order $\epsilon^0$ piece from $\mathcal I_{\rm s}$ by including the contributions from the crescent regions, i.e. the constants appearing in eqs. \eqref{3ptinegralcorrectS}-\eqref{3ptinegralcorrect_inf}, and by cancelling the Jacobi polynomials with a counter term.  This results in an exact matching at leading order in $\lambda$.

We start with the action of the unperturbed theory, $S_{\rm free}=\frac{1}{4\pi} \int d^2z\; \pa x\, \pab x$, where $x$ is the compact bosonic field, $x=x+2\pi R$, and $R$ is the radius of the circle on which $x$ takes its values. We choose conventional string theory normalizations \cite[ch. 8]{Polchinski:1998rq} with $\alpha'=2$. Let us define $x(z,\zb)=x_L(z)+x_R(\zb)$, with
\begin{align}
x_L(z)=x_{0,L}-i\alpha_0\ln(z)+i\sum_{m\neq 0}\frac1m\,\frac{\alpha_m}{z^{m}} \ ,\qquad\quad x_R(\zb)=x_{0,R}-i\ti{\alpha}_0\ln(\zb)+i\sum_{m\neq 0}\frac1m\,\frac{\ti{\alpha}_m}{\zb^{m}}\ ,\nn
\end{align}
and standard commutation relations $[\alpha_m,\alpha_n]=m\delta_{m,-n}$, $[ x_{0,L},\alpha_0]=i$, together with similar expressions for the right movers. The OPEs take the standard form $ x_L(z) x_L(0)=-\ln(z)+\cdots$, and $ x_R(\zb) x_R(0)=-\ln(\zb)+\cdots$, where $\cdots$ denotes the non-singular terms. The momentum and winding sectors are given by a pair of integers $m,w$ giving the eigenvalues $(k_L,k_R)$ of the operators $(\alpha_0,\ti{\alpha}_0)$ respectively. We define the momentum vector $\vec{k}\equiv (k_L,k_R)$, where
\begin{equation}
k_L=\frac{m}{R}+\frac{wR}{2}\ , \qquad\qquad k_R=\frac{m}{R}-\frac{wR}{2}\ ,
\end{equation}
and $\vec k$ is a vector in the even self-dual lattice of signature $(1,1)$.

The operators $1$, $\pa x$ and $\pab x$ are Virasoro primaries with dimensions $(h,\tilde h)=(0,0)$, $(1,0)$ and $(0,1)$, respectively. Moreover, there are the vertex operators
\begin{equation}\label{vo}
V_{\vec{k}}(z,\zb)\equiv \exp\Big(\frac{i\pi}{2}(k_L-k_R)(\alpha_0+\ti{\alpha}_0)\Big):\exp(ik_L x_L(z)+ik_R x_R(\zb)):\ ,
\end{equation}
which are Virasoro primaries with dimensions $(k_L^2/2,k_R^2/2)$. A few normalized low-lying quasi primaries are of the form:
\begin{align} 
&\!\!\!\!\!\!\!\!\!\!\!\!\!\sqrt{2}\,T  = -\frac{1}{\sqrt{2}}\; \pa  x \pa x\ , \qquad \sqrt{2}\,\tilde T=-\frac{1}{\sqrt{2}}\;\pab  x\pab x \label{quasidefine}  \\
V_{\vec{k},2,0}&=\frac{(\pa x \pa x + i k_L \pa^2 x)}{\sqrt{2(k_L^2+1)}}\;c_{\vec{k}}\;e^{ik_L x_L+ik_R x_R}\ ,\quad
V_{\vec{k},0,2}=\frac{(\pab x \pab x + i k_R \pab^2 x)}{\sqrt{2(k_R^2+1)}}c_{\vec{k}}e^{ik_L x_L+ik_R x_R}\ ,\nn \\
V_{\vec{k},2,2}&=\frac{(\pa x \pa x + i k_L \pa^2 x)}{\sqrt{2(k_L^2+1)}}\;\frac{(\pab x \pab x + i k_R \pab^2 x)}{\sqrt{2(k_R^2+1)}}\;c_{\vec{k}}\;e^{ik_L x_L+ik_R x_R}\ ,\nn
\end{align}
where we introduce $c_{\vec{k}}\equiv \exp(\frac{i\pi}{2}(k_L-k_R)(\alpha_0+\ti{\alpha}_0))$ for the cocycle, and for simplicity drop the normal ordering product symbol $:~:$, keeping it implicit. In this notation, the vertex operators are given by $V_{\vec{k}}=V_{\vec{k},0,0}$. As an example, the OPE of $V_{\vec{k},2,0}$  with the stress tensor is
\begin{equation}
T(z)\; V_{\vec{k},2,0}(0,0)\sim \frac{1}{z^4}\; \frac{2 k_L^2-1}{\sqrt{2(k_L^2+1)}}\; e^{ik_L x_L(0)+ik_R x_R(0)} +
\frac{1}{z^2}\Big(\frac{k_L^2}{2}+2\Big)V_{\vec{k},2,0}(0,0) + \frac{1}{z}\; \pa V_{\vec{k},2,0}(0,0)\ ,\nn
\end{equation}
showing that it is a quasi primary with holomorphic dimension $h={k_L^2}/{2}+2$. Note that $V_{\vec{k},2,0}$ is a Virasoro primary when $k_L^2=1/2$, which is possible, e.g. at the self dual radius $R=\sqrt{2}$ with $(m,w)=(1,0)$ or $(0,1)$. The operators $V_{\vec{k},a,b}$ have been scaled so that their 2-point functions are normalized:
\begin{equation}
\langle V_{\vec{k},a,b}(z,\zb)\; V_{-\vec{k},a',b'}(0,0)\rangle = e^{i\pi mw}\frac{\delta_{a,a'}\;\delta_{b,b'}}{z^{k_L^2+2a}\; \zb^{k_R^2+2b}}\ .\label{2ptMomWind}
\end{equation}
We then define fields which diagonalize the 2-point function with only positive momenta:\footnote{``Positive'' momentum is determined by the pair $(m,w)$: it is positive when $m>0$. If $m=0$, the momentum is positive if $w>0$, splitting the lattice in half. Each non-zero momentum  $\vec{k}=(k_L,k_R)$ is  then represented in only one such pair $W_{\vec{k}}^{\pm}$.  The simultaneous case $m=w=0$ is neither positive nor negative. It represents the zero momentum sector, and is considered separately.}
\begin{equation}\label{diagbasis}
W_{\vec{k},a,b}^+(z,\zb)  = \frac{e^{i\pi\frac{mw}2}}{\sqrt{2}}(V_{\vec{k},a,b}+V_{-\vec{k},{a,b}})\ , \qquad\quad
W_{\vec{k},a,b}^-(z,\zb)  = \frac{e^{i\pi\frac{mw}2}}{\sqrt{2}i}(V_{\vec{k},a,b}-V_{-\vec{k},{a,b}})\ .
\end{equation}

Let us now perturb the theory by the exactly marginal operator:
\begin{equation}\label{fbpert}
S=S_{\rm free}+\frac{\lambda}{4\pi} \int d^2z\; \pa x\,\pab x=(1+\lambda)\,S_{\rm free}\ .
\end{equation}
It is straightforward to compute the spectrum of the theory.  Define $ X\equiv \sqrt{1+\lambda}\,x$. The periodicity then becomes $ X= X+2\pi \sqrt{1+\lambda}\,R$. The perturbation simply gives a shift to the compactification radius, with the new radius being $R'=\sqrt{1+\lambda}\,R$. The exact spectrum of $S$ has operators $1$, $\pa X$ and $\pab X$ with dimensions $(0,0)$, $(1,0)$ and $(0,1)$, and has the vertex operators:
\begin{equation}\label{voprime}
V'_{\vec{k'}}=\exp\Big(\frac{i\pi}{2}(k'_L-k'_R)(\alpha_0+\ti{\alpha}_0)\Big):\exp(ik'_L X_L(z)+ik'_R X_R(\zb)):
\end{equation}
with $k'_L={m}/{R'}+{wR'}/{2}$ and $k'_R={m}/{R'}-{wR'}/{2}$. The corresponding quasi primary operators and the $W^\pm$ operators are defined  as in eqs. \eqref{quasidefine} and \eqref{diagbasis}.

The compact boson thus provides a simple example which is solvable, and where all conformal dimensions and 3-point functions can be computed exactly. This allows us to expand these quantities in $\lambda$ to calculate the corrections they acquire, and to compare with the perturbative analysis developed in the previous sections. As a warm up exercise, we compute the anomalous dimensions of the vertex operators $V_{\vec k}$ to first order in $\lambda$ in appendix \ref{freebos_2pf}. We verify that the results of the exact and perturbation theory approaches match --- see also \cite[section 2]{Benjamin:2020flm}. Below, we compute the shifts to a specific structure constant which obtains order $\epsilon^0$ contributions from two sources: $\mathcal I_{\rm s}$ from the simplified domain; and $\mathcal I_k$ from the crescent region correction.

\subsection{Structure constant deformation}\label{freebos_3pf}

We found in subsection \ref{sub_3pf_cres} that the corrections to the perturbation integral over the simplified domain \eqref{4ptallowed0} are rather complicated at the insertion point $z_2$, but straightforward at $z_1$ and $z_3$ --- see the discussion below eq. \eqref{3ptinegralcorrect_inf}. According to eq. \eqref{3ptinegralcorrect_1}, the complication at $z_2$ depends on there being a quasi primary of dimensions lower than that of $\phi_j(z_2,\zb_2)$, i.e. $(h_j, \tilde h_j)$, which one can take derivatives of to make an operator of dimensions $(h_j,\tilde h_j)$. This suggests that the simplest calculation to do is to consider an operator at $z_2$ for which there is no candidate exchange operator satisfying the constraints. This is easily satisfied in the 3-point function $\langle V_{-\vec{k}}(z_1,\zb_1)\;\pa x\,\pab x(z_2,\zb_2)\;V_{\vec{k}}(z_3,\zb_3) \rangle$, where $V_{\vec{k}}$ are the vertex operators \eqref{vo}, and the operator at $z_2$ happens to be the modulus used to perturb the theory. This example turns out to be too simple because all three operators are the lowest conformal dimension operators in their momentum class and there are no additional terms arising from crescent corrections to the simplified domain. Therefore, a {\it minimal subtraction} scheme, namely evaluating the perturbation integral $\mathcal I_{\rm s}$ \eqref{perturbstruct} over the simplified domain, is sufficient to compute $\delta C_{i,j,k}$. The computations are presented as a warm up 3-point function deformation exercise in appendix \ref{app_freebos_3pf}. An even simpler case where the structure constant is unchanged is computed in \cite{Behan:2017mwi}.

We now consider the more interesting correlator $\langle W^\pm_{\vec{k},0,0}(z_1,\zb_1)\;\pa x\,\pab x(z_2,\zb_2)\,W^\pm_{\vec{k},2,0}(z_3,\zb_3)\rangle$, where $W^\pm$ are the diagonalized vertex operators \eqref{diagbasis}. We observe from eq. \eqref{3ptinegralcorrectS} that  there are constant terms contributing from the $\epsilon$-exact domain (\ref{4ptallowedfull_first}). We proceed by first calculating the 3-point function in the perturbed theory with $R'=\sqrt{1+\lambda}\,R$:
\begin{equation}
\langle W^{'\pm}_{\vec{k}',0,0}(z_1,\zb_1)\;\pa X\,\pab X(z_2,\zb_2)\;W^{'\pm}_{\vec{k}',2,0}(z_3,\zb_3)\rangle=\frac{-2k'_Lk'_R}{\sqrt{2(k^{'2}_L+1)}}\;
\frac{1}{z_{12}^{-1}\,z_{13}^{k^{'2}_L+1}\,z_{23}^3}\;\frac{1}{\zb_{12}\;\zb_{13}^{k^{'2}_R-1}\;\zb_{23}}\ .
\end{equation}
We denote the structure constant $C'_{0,2}$ and expand it in $\lambda$:
\begin{equation}\label{unbalancedStructExpand}
C'_{0,2}=\frac{-2k'_Lk'_R}{\sqrt{2(k^{'2}_L+1)}}= C_{0,2}+
\lambda\bigg(\frac{k_L^2}{\sqrt{2(k_L^2+1)}}+\frac{k_R^2}{(k_L^2+1)\sqrt{2(k_L^2+1)}}\bigg)+O(\lambda^2)\ .
\end{equation}
The second term on the \textsc{rhs} is the shift $\delta C_{0,2}=C'_{0,2}-C_{0,2}$ at order $\lambda^1$.

We next wish to compute this in perturbation theory. The perturbation integral \eqref{4ptconvention} reads:
\begin{align}\label{bos3pf_pertint}
\mathcal I=-\lambda \int d^2z\;\langle W^\pm_{\vec{k},0,0}(z_1,\zb_1)\;\pa x\pab x(z_2,\zb_2)\;
W^\pm_{\vec{k},2,0}(z_3,\zb_3)\;\frac{1}{4\pi}\,\pa x\pab x(z,\zb)\rangle
\end{align}
where the factor $\frac1{4\pi}$ is the normalization of the perturbation term \eqref{fbpert}. Using the $\hat z$ co-ordinates \eqref{zhat}, and starting with the simplified domain of integration (\ref{4ptallowed0}), the integral \eqref{perturbstruct} reads:
\begin{align}
&\mathcal I_{\rm s}=-{\lambda}
\int_{\rm simp}\!\!\!\!\!d\hat{z}\,f(\hat{z},\bar{\hat{z}})\ ,\quad
f(\hat{z},\bar{\hat{z}})=\frac{1}{4\pi\sqrt{2(k_L^2+1)}}\Big(-\frac{2}{\hat{z}^2}+\frac{2k_L^2}{\hat{z}^3}+\frac{2k_L^2}{\hat{z}}\Big)
\bigg(\frac{1}{(\bar{\hat{z}}-1)^2}+\frac{k_R^2}{\bar{\hat{z}}}\bigg)\ .\nn
\end{align}
To extract the order $\epsilon^0$ piece of the integral, we define
\be\label{glgr}
g_L(\hat{z})\equiv \frac{2}{\hat{z}}-\frac{k_L^2}{\hat{z}^2}\ , \qquad g_R(\bar{\hat{z}})\equiv-\frac{1}{(\bar{\hat{z}}-1)}\ ,
\ee
and obtain
\begin{equation}\label{fzzhat}
\mathcal I_{\rm s}=\frac{-\lambda}{4\pi\sqrt{2(k_L^2+1)}}\;
\int_{\rm simp}\!\!\!\!\! d^2z\;\Big(\pa g_L(\hat{z})+\frac{2k_L^2}{\hat{z}}\Big)\Big(\pab g_R(\bar{\hat{z}})+\frac{k_R^2}{\bar{\hat{z}}}\Big)\ .
\end{equation}
We denote each term on the \textsc{rhs} as $\mathcal I^a_{\rm s}$, $a=\{1,\cdots,4\}$. Three of them may be integrated using the divergence theorem. Consider
\begin{equation}\label{C1simpbos}
\mathcal I^1_{\rm s}\equiv \frac{-\lambda}{4\pi\sqrt{2(k_L^2+1)}}
\int_{\rm simp}\!\!\!\!\!d\hat{z}\,d\bar{\hat{z}}\;\pa g_L(\hat{z})\,\pab g_R(\bar{\hat{z}})=
\frac{\lambda\,i}{4\pi\sqrt{2(k_L^2+1)}}\int_{\pa\, {\rm simp}}\!\!\!\!d\hat{z}\;\pa g_L(\hat{z})\,g_R(\bar{\hat{z}})\ .\nn
\end{equation}
At the boundary of the hole at $\hat{z}=0$, $\hat{z}$, $\bar{\hat{z}}$ and $d\hat{z}$ contain factors of $\epsilon$.  The order $\epsilon^0$ term must arise from terms of the form $\frac{1}{\hat{z}} \left(\frac{\hat{z}}{\bar{\hat{z}}}\right)^q$. However, only the $q=0$ term survives due to the factors of $e^{i\phi}$ describing the contour. We must then find the coefficient of the $1/\hat{z}$ term in the expansion of $g_L$ near $\hat{z}=0$, and extract the constant term in the expansion of $g_R$  near $\bar{\hat{z}}=0$.  According to eq. \eqref{glgr}, no such term exists. Thus, there is no constant term arising from the hole at $\hat{z}=0$. Similar arguments apply to holes at $\hat{z}=1$ and $\infty$, and yield the same conclusion.

We next consider one of the other terms in eq. \eqref{fzzhat}:
\begin{equation}\label{C2simpbos}
\mathcal I^2_{\rm s}\equiv \frac{-\lambda}{4\pi\sqrt{2(k_L^2+1)}} \int_{\rm simp}\!\!\!\!\!
d\hat{z}\,d\bar{\hat{z}}\;\frac{2k_L^2}{\hat{z}}\;\pab g_R(\bar{\hat{z}})=
\frac{\lambda\,i\;2k_L^2}{4\pi\sqrt{2(k_L^2+1)}}\int_{\pa\, {\rm sim}}\!\!\frac{d\hat{z}}{\hat z}\,g_R(\bar{\hat{z}})\ .
\end{equation}
For the contour at $\hat{z}=0$, the factor $1/\hat{z}$ makes a contribution. Furthermore, $g_R(\bar{\hat{z}})=1 + O(\bar{\hat z})$ near $\bar{\hat{z}}=0$. The contours for the excisions at finite points in the $\hat{z}$ plane are clockwise. This then gives a contribution of $\lambda k_L^2/\sqrt{2(k_L^2+1)}$. Similar analysis shows that there are no contributions from the excised hole at $\hat{z}=1$ and $\infty$. We denote the $\epsilon^0$ term of the integral \eqref{C2simpbos} as $\mathcal I^{2}_{\rm s,0}= \lambda{k_L^2}/{\sqrt{2(k_L^2+1)}}$.
An analogous computation for
\begin{align}
\mathcal I^{3}_{\rm s}\equiv \frac{-\lambda}{4\pi\sqrt{2(k_L^2+1)}}
\int_{\rm sim}\!\!\!\!d\hat{z}\, d\bar{\hat{z}}\;\pa g_L(\hat{z})\;\frac{k_R^2}{\bar{\hat{z}}}\nn
\end{align}
yields $\mathcal I^{3}_{\rm s,0}=0$. Finally, we compute
\begin{align}
\mathcal I^{4}_{\rm s}&\equiv \frac{-\lambda}{4\pi\sqrt{2(k_L^2+1)}} \int_{\rm sim}\!\!\! d\hat{z}\,d\bar{\hat{z}}\;
\frac{2\, k_L^2\, k_R^2}{\hat{z}\,\bar{\hat z}}=\frac{-4\lambda\,k_L^2\,k_R^2}{4\pi\sqrt{2(k_L^2+1)}}
\int_{0}^{2\pi} d\phi \int_{\frac{\epsilon |z_{12}|}{|z_{23}||z_{13}|}}^{\frac{|z_{12}||z_{23}|}{\epsilon |z_{23}|}} \frac{dr}{r}\nn\\
&\qquad\qquad\qquad\qquad\qquad+\,\frac{4\lambda\,k_L^2k_R^2}{4\pi\sqrt{2(k_L^2+1)}} 
\int_{\phi=0}^{2\pi}d\phi \int_{0}^{\frac{\epsilon |z_{13}|}{|z_{23}||z_{12}|}} r dr\frac{1}{(r e^{i\phi}+1)(re^{-i\phi}+1)}\ ,\nn
\end{align} 
The last term represents the excision of the disc near $\hat{z}=1$, and is an integral of a finite function over a vanishing domain. Therefore, it is purely perturbative in $\epsilon$, and may be safely ignored. The first integral yields:
\begin{align}
\mathcal I^{4}_{\rm s} = \frac{2\lambda\,k_L^2k_R^2}{\sqrt{2(k_L^2+1)}}\left( \ln\left(\frac{\epsilon |z_{23}|}{|z_{12}||z_{13}|}\right) + \ln\left(\frac{\epsilon |z_{12}|}{|z_{13}||z_{23}|}\right)\right)\ .\nn
\end{align}
Recalling eq. \eqref{unbalancedStructExpand}, the coefficient of the log terms is found to be $\lambda(-k_Lk_R/2)\,C_{0,2}$.  Note, this provides another way to read off the anomalous dimension of the vertex operators, i.e. $-k_Lk_R/2$ --- see eq. \eqref{Expand3pt} and appendix \ref{freebos_2pf}. There is, however, no contribution of $O(\epsilon^0)$. Thus, $\mathcal I^{4}_{\rm s,0}=0$.

All in all, the constant term of the perturbation integral \eqref{fzzhat} at first-order in $\lambda$ is:
\begin{align}\label{bos3pf_simp}
\delta C_{0,2}^{\rm simp}\equiv\sum_{a=1}^4\mathcal I^{a}_{\rm s,0}=
\lambda\frac{k_L^2}{\sqrt{2(k_L^2+1)}}\ .
\end{align}
This {\it does not} agree with the shift computed exactly in eq. (\ref{unbalancedStructExpand}) at first order in $\lambda$.  In fact, it only gives the first piece of the expression in eq. (\ref{unbalancedStructExpand}).

So far we have computed the perturbation integral \eqref{bos3pf_pertint} in the simplified domain (\ref{4ptallowed0}). In section \ref{cpt_3pf} we showed that, given our regularization scheme, the correct computation of the shift $\delta C_{i,j,k}$ is obtained by integrating over the exact domain (\ref{4ptallowedfull_first}). As such, we must add the contributions of the crescent regions around the insertion points, if appropriate crossing channels exist. We shall now use the formulae derived in subsection \eqref{sub_3pf_cres} to compute the constant term contributions from the exact domain.

Following the conventions of section \ref{cpt_3pf}, the 3-point function consists of $\phi_i=W_{\vec{k},\pm,0,0}$, $\phi_j=\pa x\pab x$ and $\phi_k=W_{\vec{k},\pm,2,0}$. The only quasi primary that is a valid crossing channel for this correlator is $\phi_p= W_{\vec{k},\pm, 0,0}$, which is part of the OPE between the deformation operator $\frac{1}{4\pi} \pa x \pab x$ and $W_{\vec{k},\pm, 2,0}$ at  $z_3$ ($\hat z=0$). 
Thus, the only constant term contribution comes from the crescent located at $\hat z=0$, i.e. $\mathcal I^{\rm hol}_{k,0}$ \eqref{3ptinegralcorrectS}, with a single term in the sum over $p$ for $\phi_p= W_{\vec{k},\pm, 0,0}$. As argued in subsection \ref{subsec_npf}, the Jacobi polynomial in eq. \eqref{3ptinegralcorrectS} is removed by adding the counter term \eqref{epsilon0dterm} to $\phi_k= W_{\vec{k},\pm,2,0}$. The contribution from the perturbation theory integral $\mathcal I^{\rm hol}_{k,0}$ is combined with the appropriate counter term $\mathcal C^{\rm hol}_{k,0}$, giving the shift to the structure constant as
\begin{equation}
\delta C_{0,2}^{\rm cres}\equiv \mathcal I^{\rm hol}_{k,0}+\mathcal C^{\rm hol}_{k,0}=2\pi\lambda\; 
\frac{(h_k-h_p-1)!}{(2h_p)_{h_k-h_p}}\; C_{p,D,k}\, C_{p,j,i}\; \frac{(-h_j-h_k+h_i+1)_{h_k-h_p}}{(h_k-h_p)!}\ .\nn
\end{equation}
We have $h_k= \frac{k_L^2}{2}+2$, $h_j=1$, $h_i=h_p=\frac{k_L^2}{2}$. The 3-point functions $C_{p,D,k}=\frac{-2k_L k_R}{4\pi\sqrt{2(k_L^2+1)}}$ and $C_{p,j,i}= -k_L k_R$ are computed in appendix \ref{app_freebos_3pf}. We find:
\begin{align}\label{bos3pf_cres}
\delta C_{0,2}^{\rm cres}=\lambda\frac{k_R^2}{(k_L^2+1)\sqrt{2(k_L^2+1)}}\ .
\end{align}
The contribution of constant terms from the simplified domain \eqref{bos3pf_simp}, together with the additional contribution from the crescents \eqref{bos3pf_cres}, leads to the total shift:
\begin{equation}
\delta C_{0,2}^{\rm pert}=\delta C_{0,2}^{\rm simp}+\delta C_{0,2}^{\rm cres}=
\lambda\bigg(\frac{k_L^2}{\sqrt{2(k_L^2+1)}}+\frac{k_R^2}{(k_L^2+1)\sqrt{2(k_L^2+1)}}\bigg)
\end{equation}
obtained from perturbation theory. This {\it does} agree with the order $\lambda^1$ shift (\ref{unbalancedStructExpand}) exactly. The extra contributions from the crescents and counter terms are necessary, even in simple theories such as the compact boson.  

It should be noted that the above analysis is insensitive to whether the theory is at the self-dual radius: one may choose to analyze the primaries $(m,w)=(1,0)$ or $(0,1)$ at this radius and still obtain meaningful results.  In fact, the perturbation here may be thought of as a symmetry breaking ``Higgs'' mechanism, where several short representations of the Virasoro algebra with null vectors combine to make a long representation without null vectors.

\section{Conclusions}\label{conc}
We derived the full set of counter terms \eqref{counterTerms2} sufficient to regulate the integrated $n$-point functions of quasi primary operators which appear in first order conformal perturbation theory for general 2-dimensional CFTs, using a hard disk regularization. We show that these counter terms do not affect the usual computation of the anomalous dimensions.  We then considered the perturbation of 3-point functions, computing the corrections to the structure constant when the domain of integration for \eqref{4ptconvention_hat} is shifted from the simplified domain \eqref{4ptallowed0} to the exact domain \eqref{4ptallowedfull_first}. Explicit expressions for these corrections are derived in terms of the CFT data --- eqs. \eqref{3ptinegralcorrectS}-\eqref{3ptinegralcorrect_inf} and their anti-holomorphic counterparts.

Thus, our procedure to compute the shift of structure constants is as follows.   
\begin{enumerate}
\item For the 3-point function at hand, construct the relevant 4-point function by including the deformation operator in the correlator.  Extract the function of cross ratios from this 4-point function, $f(\zeta,\bar{\zeta})$, where our convention for this function is given in eq. \eqref{4ptconvention}.
\item Evaluate $-\lambda \int d^2\hat{z} f(\hat{z},\bar{\hat{z}})$ over the simplified domain \eqref{4ptallowed0}, keeping only the $\epsilon^0$ term.  The function $f(\hat{z},\bar{\hat{z}})$ contains no $\epsilon$ dependence, and so the simplified domain's simple dependence on $\epsilon$ helps identify the $\epsilon^0$ terms.  This contribution must be evaluated theory by theory.
\item In addition, one must compute the structure constants for the operators in the Regge trajectory of lower conformal dimension to construct the constant parts of \eqref{3ptinegralcorrectS}-\eqref{3ptinegralcorrect_inf} and their holomorphic counter parts.  In these expressions one simply drops the Jacobi polynomials: the relevant counter terms to do so have been identified. \footnote{These shifts may also be computed directly by evaluating integrals, e.g. \eqref{crescentcorrectfull} and similar expressions, and evaluating the $\epsilon^0$ term.  The types of integrals encountered have been evaluated in section \ref{sec_2pf} and appendix \ref{app_als}, although, one must group the results into the Jacobi polynomials to identify the constant term correctly.}
\item Adding the contributions to the $\epsilon^0$ term from step 2 and step 3 furnishes the full first order in $\lambda$ change to the structure constant. 
\end{enumerate}

It would be interesting to consider CFTs for which the full shift to the structure constants \eqref{deltaCintro} can be exactly computed.  The only obstruction here is step 2, while the corrections from the crescent regions have been computed in this work. However, step 2 seems to be quite difficult to obtain in general, given that each crossing symmetric form of the conformal block decomposition is only well adapted to part of the domain of integration \cite{Behan:2017mwi}.  Additional structure is needed.  It might be in particular interesting to explore families of rational conformal field theories where the simplified forms of the correlation functions may allow for performing the perturbation integrals analytically. One of our main motivations for a systematic study of the perturbation of the structure constants is to explore this deformation in the context of holography, in particular in the moduli space of the D1-D5 brane system. We hope to report on this elsewhere \cite{bbfuture}.  Orbifold CFTs provide a natural framework where the correlators on the sphere can have simple forms, and may therefore allow explicit evaluation of $-\lambda \int d^2\hat{z} f(\hat{z},\bar{\hat{z}})$ over the simplified domain \eqref{4ptallowed0}.

\section*{Acknowledgements}
We thank the Mainz Institute for Theoretical Physics for hospitality where part of this work was done. We are grateful to Luis Apolo, Scott Collier, and Christoph Keller for helpful discussions. We wish to thank the participants of the workshop ``Exact Results and Holographic Correspondences" at MITP for many interesting conversations.  BAB is thankful for funding support from Hofstra Univeristy including startup funds and faculty research and development grants, and for support from the Scholars program at KITP, which is supported in part by grants NSF PHY-1748958 and PHY-2309135 to the Kavli Institute for Theoretical Physics (KITP), where some of this work was completed.  IGZ is supported by the Cluster of Excellence Precision Physics, Fundamental Interactions, and Structure of Matter (PRISMA+, EXC 2118/1) within the German Excellence Strategy (Project-ID 390831469).

\appendix

\section{\texorpdfstring{$sl(2)$}{TEXT} algebra}\label{app_sl2}

In the main body we have leaned heavily on the $sl(2)$ structure of the theory, and so this is worth reviewing.  The $sl(2)$ subalgebra of the Virasoro algebra is generated by $L_{-1}, L_0, L_1$; quasi primaries of weight $h_i$ are defined via $L_1\mid \phi_i\rangle =0$ and $L_0\mid \phi_i\rangle =h_i \mid \phi_i\rangle$ (we will use $\phi_i$ for quasi primaries and often use ${\mathcal O}_i$ for primaries).  In the space of quasi primaries, we diagonalize each eigenspace of $L_0$ under the 2-point function.  Each of these operators is regarded as the highest weight (lowest conformal dimension) operator in a representation of the $sl(2)$ algebra.  As usual for any subalgebra, many such representations of the $sl(2)$ subalgebra compose a single representation of the Virasoro algebra.  This by itself is enough to make the following assertion: any operator in the CFT may be written as a sum of quasi primairies and their derivatives.  We next prove this statement directly.  

We start by dividing the state space by how many applications of $L_1$ it takes to annihilate a state.  A state that is annihilated by $L_1^{k+1}$, but not annihilated by $L_1^k$, is in ``class $k$'', and an operator in class 0 is quasi primary. Consider a primary $\Phi$, and a state of dimension $h$ in class $k$, $\mid \Phi_{k} \rangle$, in this conformal family. Then $L_1^k\mid \Phi_{k} \rangle=\mid Q \rangle$ where $\mid Q \rangle $ is a quasi primary of dimension $h_Q=h-k$ ($>0$ for unitary theories, which we assume).  We may write this state as
\be
\mid \Phi_{k} \rangle = \left(\mid \Phi_{k} \rangle - \frac{1}{k!(2h_Q)_{k}} L_{-1}^{k} \mid Q \rangle\right) + \frac{1}{k!(2h_Q)_{k}} L_{-1}^{k} \mid Q \rangle\nn
\ee
where $(~)_n$ denotes the Pochhammer symbol $(\alpha)_m\equiv \prod_{i=0}^{m-1}(\alpha+i)=\Gamma(\alpha+m)/\Gamma(\alpha)$.  Using $L_1^k L_{-1}^k \mid Q \rangle= k!(2h_Q)_k \mid Q \rangle$ for $Q$ quasi primary, the state in parentheses is annihilated by $L_1^k$ by construction, and so is an operator in class $k'$ with $k'<k$.
The second operator is just the derivative of a quasi primary $L_{-1}^{k} \mid Q \rangle=\mid \pa^k Q\rangle$.  Iterating this procedure on the state in parentheses one eventually arrives at class $0$.  Formalizing this gives an inductive proof that any operator in the CFT may be written as a sum of quasi primaries and derivatives of quasi primaries.  We do likewise on the anti-holomophic side, and to simplify terminology, call those operators annihilated by both $L_1$ and $\ti{L}_1$ quasi primary.  The set of distinct quasi primaries and their $sl(2)\times sl(2)$ descendants provides a minimal spannin set of states/operators in the CFT.

Another important note is that the global Ward identities are constructed only using the $sl(2)$ generators. Under $sl(2)$ transformations the quasi primaries transform as tensors.  This is enough to conclude that the usual expression for the global Ward identities apply to the $n$-point functions of quasi primaries.  More concretely, the global ward identities arise from applying $L_n$ with $n=\{-1,0,1\}$ at infinity, giving 0, and then pulling the contour inward.  We recall $L_{-n}=\oint \frac{dz}{2\pi i} z^{1-n}\,T(z)$, limiting the powers of $z$ in the measure to $0,1,2$ for the $sl(2)$ operators. After deforming the contours in and expanding around each operator insertion, this leads to powers of $(z-z_i)^{1-n'}$ with $n'=\{-1,0,1\}$ as well.  If the operator insertions are quasi primary, only $n'={-1,0}$ survive, leaving expressions written in terms of the conformal dimensions and derivatives of the fields.  This yields the familiar Ward identities:
\begin{align} \label{gward} 
&\sum_i \pa_i \langle \phi_1(z_1) \cdots \phi_n(z_n)\rangle =0\ ,\qquad\qquad
\sum_i (z_i \pa_i + h_i) \langle \phi_1(z_1) \cdots \phi_n(z_n)\rangle =0\nn\\
&\sum_i (z_i^2 \pa_i+2h_i z_i) \langle \phi_1(z_1) \cdots \phi_n(z_n)\rangle =0 \nn
\end{align}  
for $\phi_i$ quasi primary, which are the same as those stated for primary operators.  Similar considerations apply to the anti-holomorphic side.  Thus, the form of the 2-point and 3-point functions for quasi primaries is the same as the form of 2-point and 3-point functions of primaries:
\begin{align}
\langle \phi_i(z_1,\zb_1) \phi_j (z_2,\zb_2)\rangle&=\frac{\delta_{ij}}{z_{12}^{2h_i}\zb_{12}^{2\ti{h}_i}}\ ,\nn \\
\langle \phi_i(z_1,\zb_1) \phi_j (z_2,\zb_2) \phi_k(z_3,\zb_3)\rangle&=
\frac{C_{i,j,k}}{z_{12}^{h_1+h_2-h_3}z_{13}^{h_1+h_3-h_1}z_{23}^{h_2+h_3-h_1}\zb_{12}^{\ti{h}_1+
\ti{h}_2-\ti{h}_3}\zb_{13}^{\ti{h}_1+\ti{h}_3-\ti{h}_1}\zb_{23}^{\ti{h}_2+\ti{h}_3-\ti{h}_1}}\ .\nn
\end{align}

We consider the conformal dimensions $h_i, \ti{h}_i$ and the structure constants $C_{i,j,k}$ for the quasi primary fields the data of the CFT.  This is an over complete set of data in the CFT: given the $h_i, \ti{h}_i$ and the structure constants $C_{i,j,k}$ for the primary fields, the $h_i, \ti{h}_i$ and the structure constants $C_{i,j,k}$ for the quasi primaries may be determined.  However, this poses a problem when dealing with special points in the moduli space.  At some special points, certain quasi primary fields may become primary, for example at special values of the moduli in Narain lattices.  When deforming away from  these special points in the moduli space, different (short) representations of the Virasoro algebra may merge to become one (long) representation.  We have seen one such example when considering the compact boson CFT in section \ref{sec_bos}.  From this viewpoint, the quasi primaries and the data associated with them may seem more natural to consider on the moduli space.  Indeed, one can see that the $sl(2)$ descendants are never null states:
\be
\langle Q \mid L_{1}^k L_{-1}^{k} \mid Q \rangle=k!(2h_Q)_k\nn
\ee
which is always positive.  Thus, these representations always remain ``intact'', and descent relations which calculate correlators of $sl(2)$ descendants in terms of the quasi primaries are always available.

\section{\texorpdfstring{$a_{\ell}$ coefficients}{aa2}}\label{app_als}

In this appendix we prove that the coefficients $a_{\ell}$ which show up in the expansion of the perturbation integral \eqref{adef} in terms of $\epsilon$, are given by the expression \eqref{als}.  For this, we will need several ingredients.  First, we will need the generalized binomial expansion, defined through the power series $(1+x)^\alpha=\sum_{m=0}^\infty \binom{\alpha}{m} x^m$, where $m$ is an integer and $\alpha$ is an arbitrary number. The coefficients of $x^m$ in the sum are easily found by applying derivatives on the power series:
\begin{equation}\label{binom}
 \binom{\alpha}{m} = \frac{1}{m!} \left(\frac{d^m}{dx^m} (1+x)^\alpha\right)\bigg|_{x=0}=\frac{(\alpha-m+1)_m}{m!}=\frac{\Gamma(\alpha+1)}{\Gamma(m+1)\Gamma(\alpha-m+1)}
\end{equation}
where we have used the relation between Pochhammer symbols $(q)_m=q (q+1) (q+2) \cdots (q+m-1)$ and gamma functions: $(q)_m= {\Gamma(q+m)}/{\Gamma(q)}$. The latter expression may not be used when $q$ is a non-positive integer. If $q$ is 0 or a negative integer, one must regulate the $\Gamma$ functions, while the Pochhammer symbol is always well defined (and always agrees with the regulation $q\rightarrow q+\epsilon$ in an $\epsilon\rightarrow 0$ limit). The generalized binomial coefficient can be written as
\begin{align}\label{pascal}
\binom{\alpha}{m}& =\frac{1}{m!} \Big(\frac{d^m}{dx^m} (1+x)^\alpha\Big)\Big|_{x=0} 
=\frac{1}{m!} \bigg(\Big(\frac{\pa}{\pa x}+\frac{\pa}{\pa y}\Big)^m (1+x)^\beta(1+y)^{\alpha-\beta}\bigg)\Big|_{x=0,y=0}\\
&=\sum_{n=0}^m \frac{(\beta-n+1)_{n}}{n!}\; \frac{(\alpha-\beta-(m-n)+1)_{m-n}}{(m-n)!}\ ,\nn
\end{align}
where $\beta$ is an arbitrary number and the final result is just the familiar ``Pascal's triangle'' recurrence relation for binomial coefficients when $\alpha$ and $\beta$ are integers.  It continues to be valid for generalized binomial coefficients.\footnote{One may also prove eq. \eqref{pascal} by using the recurrence relations for $\Gamma$ functions and shifting the bounds of the sums; however, the above is more efficient.} Similarly, we can write the generalized binomial series $(y+x)^\alpha=\sum_{m=0}^\infty \binom{\alpha}{m} y^{\alpha-m}x^m$ which is natural in the case $|x|<|y|$.  If $\alpha$ is a positive integer, the series naturally truncates, with all formulae above remaining valid because $(\alpha-m+1)_m=0$ for any integer $m\geq \alpha+1$.

Next, we will use various relationships between factorials and Pochhammer symbols. For any integer $n\geq 1$, we have $n!=n!! (n-1)!!$, where the $!!$ indicates the double factorial, producing a product over every other descending integer greater than zero.  For an even number $N=2n$, we have $N!!=(2n)!!=2^n n!$ and for an odd number $N=2n+1$,  $N!!=(2n+1)!!=2^{n+1} \left(\frac{1}{2}\right)_{n+1}$. Furthermore, we may break Pochhammer symbols in the middle as $(a)_m= a_{m-n}\, (a+(m-n))_n$. One may also break Pochhammer symbols into ``skipping by two''. If the subscript of the Pochhammer symbol is even $N=2n$, we have
\begin{equation}
(a)_{N}=(a)_{2n}=2^{2n} \Big(\frac{a^{\,}}{2}\Big)_{n} \Big(\frac{a+1}{2}\Big)_{n}\ , \label{pochspliteven}
\end{equation}
and if the subscript is odd $N=2n+1$,
\begin{equation}
(a)_{N}=(a)_{(2n+1)}=2^{2n+1} \left(\frac{a^{\,}}{2}\right)_{n+1} \left(\frac{a+1}{2}\right)_n\ .
\end{equation}

With these preparations, we are ready to address the evaluation of the perturbation integral \eqref{adef}. The first term in the integrand vanishes and thus we need to evaluate the integral
\begin{equation}\label{Rinfint}
\int_0^{2\pi} d\phi\; \mathcal{R}_\infty^{d_i-d_j}\cos((s_i-s_j)\phi)\equiv \sum_{\ell} a'_{\ell}\; \bigg(\frac{\epsilon}{|z_{12}|}\bigg)^\ell\ .
\end{equation}
Note that $a'_\ell=-a_\ell$ defined in eq. \eqref{adef}. Define $\delta\equiv \frac{\epsilon}{|z_{12}|}$. Using eq. \eqref{R0inf}, we have $\mathcal{R}_\infty=\sqrt{1-\delta^2\sin^2(\phi)}+\delta \cos(\phi)$. Reintroduce the exponential in the integrand and note that the integrand only depends on $|s_i-s_j|$ --- see the discussion below eq. \eqref{2pfII}. Eq. \eqref{Rinfint} reads
\begin{align}
&\mathcal I\equiv \int_0^{2\pi}\!\!\d\phi \;\Big(\sqrt{(1-\delta^2)+\delta^2\cos^2(\phi)}+\delta \cos(\phi) \Big)^{d_i-d_j}e^{-i|\Delta s| \phi}.
\end{align}
where $\Delta s\equiv s_i-s_j$. We now use the generalized binomial expansion and write
\begin{align}
&\mathcal I= \sum_{a=0}^\infty \sum_{b=0}^\infty  \int_0^{2\pi}\!\! d\phi\;  \binom{d_i-d_j}{a}\binom{\frac{d_i-d_j-a}{2}}{b} \delta^{a+2b}\cos^{a+2b}(\phi)(1-\delta^2)^{\frac{d_i-d_j-a-2b}{2}} e^{-i|\Delta s| \phi}\ .\nn
\end{align}
Furthermore, taking the exponential form $\cos(\phi)=\frac{e^{i\phi}+e^{-i\phi}}{2}$ we have
\begin{align}
\mathcal I = \sum_{a=0}^\infty \sum_{b=0}^\infty \sum_{c=0}^{a+2b}  \int_0^{2\pi}\!\!\! d\phi\; \binom{d_i-d_j}{a}\binom{\frac{d_i-d_j-a}{2}}{b} \binom{a+2b}{c}  \frac{\delta^{a+2b}}{2^{a+2b}}\; e^{i(a+2b-2c)\phi}(1-\delta^2)^{\frac{d_i-d_j-a-2b}{2}} e^{-i|\Delta s| \phi}\ .\nn
\end{align}
The Fourier modes must now match, giving $a+2b-2c=|\Delta s|$.  This is only satisfiable if $a$ and $|\Delta s|$ are both even or both odd.  Thus, the sum over $a$ becomes restricted to $a-|\Delta s| \in 2{\mathbb Z}$.  Furthermore, it must be that $a+2b\geq a+2b-2c=|\Delta s|$.  Thus, the $a$ and $b$ sums are restricted with a lower bound.  As long as $a+2b\geq |\Delta s|$ and $a-|\Delta s| \in 2{\mathbb Z}$, there is a member of the sum over $c$ which matches the mode. The sums are then truncated to
\begin{align}
\mathcal I = \sum_{\substack{a=0,\,b=0 \\ a+2b \geq |\Delta s| \\ a-|\Delta s|\in 2{\mathbb Z}}}^\infty 2\pi  \binom{d_i-d_j}{a}\binom{\frac{d_i-d_j-a}{2}}{b} \binom{a+2b}{\frac{a+2b-|\Delta s|}{2}}  \frac{\delta^{a+2b}}{2^{a+2b}}\,(1-\delta^2)^{\frac{d_i-d_j-a-2b}{2}}\ .\nn
\end{align}
We expand one further time to write
\begin{align}
\mathcal I= \sum_{\substack{a=0,\,b=0 \\ a+2b \geq |\Delta s| \\ a-|\Delta s|\in 2{\mathbb Z}}}^\infty \sum_{d=0}^\infty 2\pi  \binom{d_i-d_j}{a}\binom{\frac{d_i-d_j-a}{2}}{b} \binom{a+2b}{\frac{a+2b-|\Delta s|}{2}} \binom{\frac{d_i-d_j-a-2b}{2}}{d}\frac{(-1)^d}{2^{a+2b}}\;{\delta^{a+2b+2d}}\ .\nn
\end{align}
The power of $\delta$ is $a+2b+2d\geq |\Delta s|$. Hence, all terms with $a'_\ell$, $\ell< |\Delta s|$, in the expansion \eqref{Rinfint} vanish --- see also the discussion below eq. \eqref{adef}.  We wish to sum over all $a,b,d$ which satisfy $a+2b+2d=\ell$ to find the coefficient of $\delta^\ell$, i.e. the coefficient $a'_{\ell}$ we seek in eq. \eqref{Rinfint}. This equation is only solvable if $a$ and $\ell$ are both even or both odd.  Thus, $a, |\Delta s|, \ell$ are all even or all odd.  Under this restriction, we solve for $d=({\ell-(a+2b)})/{2}\geq 0$ where the restriction is necessary for the positive integer to appear in the sum over $d$. This imposes $\ell\geq a+2b$ and we find $\ell\geq a+2b \geq |\Delta s|$. All in all, using eq. \ref{Rinfint}, we find
\begin{align}
a'_{\ell}& = 2\pi\!\!\!\!\!\!\!\! \sum_{\substack{a=0,\,b=0 \\ \ell \geq a+2b \geq |\Delta s| \\ a-|\Delta s|\in 2{\mathbb Z}}}^\infty\!\!\!\!\!
\binom{d_i-d_j}{a}\binom{\frac{d_i-d_j-a}{2}}{b} \binom{a+2b}{\frac{a+2b-|\Delta s|}{2}} \binom{\frac{d_i-d_j-a-2b}{2}}{\frac{\ell-(a+2b)}{2}}\;{2^{-(a+2b)}}{(-1)^{\frac{\ell-(a+2b)}{2}}}\ ,\label{fullbinomexpress}
\end{align}
where $\ell-|\Delta s|\in 2{\mathbb Z}$ is understood.  It is now helpful to consider a geometrical description of the sums in the space spanned by $a$ and $b$.
\begin{figure}[ht]
\begin{center}
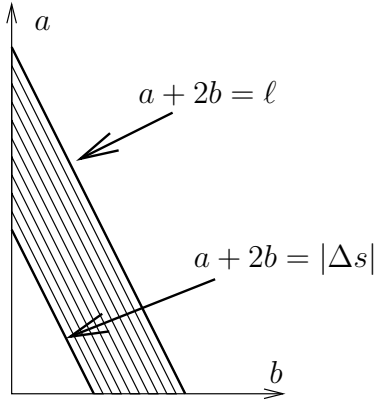
\end{center}
\caption{The shaded trapezoid region represents the terms in the sum over $a$ and $b$ that contribute to $a'_{\ell}$ \eqref{fullbinomexpress}.  The restrictions $a\geq 0$, $b\geq 0$ and  $|\Delta s|\leq a+2b \leq \ell$ define this region. Only those integers with $a-|\Delta s|\in 2{\mathbb Z}$ and $\ell-|\Delta s|\in 2{\mathbb Z}$ in the shaded region are allowed.}
\label{figab}
\end{figure}

Since $\ell$ and $|\Delta s|$ must either be both even or both odd, we may take $\ell\equiv 2f+|\Delta s|$. The integer $f$ describes the thickness of trapezoidal region in the $(a,b)$ space, as depicted in figure \ref{figab}. We wish to induct on $f$.  Start with the base case $f=0$. This yields $\ell=|\Delta s|$ and $a+2b=|\Delta s|$, and restricts the sum over $a$. The trapezoid of figure \ref{figab} thus degenerates to a line given by $a=|\Delta s|-2b\geq 0$.  This truncates the sum in $a$ and adds the restriction $0\leq b \leq \lfloor \frac{|\Delta s|}{2}\rfloor$:
\begin{align}\label{f0}
a'_{\ell=|\Delta s|} = 2\pi\sum_{b=0}^{\lfloor \frac{|\Delta s|}{2}\rfloor}  \binom{d_i-d_j}{|\Delta s|-2b}\binom{\frac{d_i-d_j-|\Delta s|+2b}{2}}{b}\, \frac{1}{2^{|\Delta s|}}\ .
\end{align}
We break this into two separate cases when $|\Delta s|$ is even and odd. For the even case define $|\Delta s|=2S$, $S\in\mathbb Z^{+}$. Note that $\Delta s=0$ is excluded and will be discussed shortly. We obtain:
\begin{align}
a'_{\ell=|\Delta s|} 
= 2\pi  \sum_{b=0}^{S}  \frac{2^{2S-2b}\Big(\frac{d_i-d_j-(2S-2b)+1}{2}\Big)_{S-b}\Big(\frac{d_i-d_j-(2S-2b)+2}{2}\Big)_{S-b}}{(2S-2b)!}\; 
\frac{\left(\frac{d_i-d_j-2S+2}{2}\right)_b}{b!}\;\frac{1}{2^{2S}} \nn 
\end{align}
where have we used eq. (\ref{pochspliteven}). Using the identity
\begin{align}
\frac{1}{(2S-2b)!}= \frac{\left(\frac{2S-2b+1}{2}\right)_b}{2^{S-b}\,(S-b)!\,2^{S-b}\left(\frac{1}{2}\right)_{S}}\nn
\end{align}
and combining the Pochhammer symbols under the sum we find:
\begin{align}
a'_{\ell=2S} &= \frac{2\pi}{2^{2S}\big(\frac{1}{2}\big)_S} \Big(\frac{d_i-d_j-2S+2}{2}\Big)_S \; \sum_{b=0}^S
\frac{\Big(\frac{d_i-d_j-(2S-2b)+1}{2}\Big)_{S-b}}{(S-b)!}\;\frac{\left(\frac{2S-2b+1}{2}\right)_b}{b!}\ .\nn
\end{align}
Plug into eq. (\ref{pascal}) the following: $n=b$, $m=S$, $\beta=\frac{2S-1}{2}$, $\alpha=\frac{d_i-d_j+2S-2}{2}$. Using eq. \eqref{binom}, the sum reads
\begin{align}
a'_{\ell=2S} = \frac{2\pi}{2^{2S}\big(\frac{1}{2}\big)_S} \Big(\frac{d_i-d_j-2S+2}{2}\Big)_S \frac{\big(\frac{d_i-d_j}{2}\big)_S}{S!}\ .\nn
\end{align}
The top term in product for the first Pochhammer symbol, $(d_i-d_j)/2$, is the same as the bottom term in the second symbol.  This term occurs twice, and can be factored. The two Pochhammer symbols may be then combined into one. Using $2^{2S}S!\left(\frac{1}{2}\right)_S=(2S)!=(|\Delta s |!)$, we finally obtain:
\begin{align}\label{f0ev}
a'_{\ell=|\Delta s|=2S} = \frac{2\pi}{|\Delta s |!}\; \frac{(d_i-d_j)}{2}\;\Big(\frac{d_i-d_j-|\Delta s|}{2}+1\Big)_{|\Delta s|-1}\ .
\end{align}
For the odd case, $|\Delta s|=2S+1$, $S\in\mathbb Z_{\ge0}$, in eq. \eqref{f0}, performing similar manipulations shows that $a'_{\ell=|\Delta s|=2S+1}$ has the same expression as in the even case \eqref{f0ev}.  Let us now consider $|\Delta s|=0$. Plugging this into eq. \eqref{fullbinomexpress}, or directly into the defining integral \eqref{Rinfint}, we find that $a'_{\ell=|\Delta s|=0}=2\pi$. The expression \eqref{f0ev} remains valid for $\Delta s=0$ if we take the Pochhammer symbol with negative integer subscript to mean the product on reciprocals\footnote{As is commonly done, and is equivalent to the gamma function representation of Pochhammer symbols.}:
\begin{equation}
a'_{\ell=|\Delta s|=0}= \frac{2\pi}{0!}\;\frac{(d_i-d_j)}{2}\,\Big(\frac{d_i-d_j}{2}+1\Big)_{-1}=2\pi\;\frac{(d_i-d_j)}{2}\;\frac{1}{ \frac{(d_i-d_j)}{2}}=2\pi\ .\nn
\end{equation}

Let us next consider a non-degenerate trapezoidal region in figure \ref{figab} by taking $f\neq0$ (recall that $\ell=|\Delta s|+2f$). The sum (\ref{fullbinomexpress}) can be broken into sums over lines of the form $a+2b=|\Delta s|+2g$, where $g\in\{0,\cdots,f\}$.  The second two binomial coefficients in eq. (\ref{fullbinomexpress}) are constant over the line. We thus may think of the double sum in (\ref{fullbinomexpress}) as a sum over distinct lines making up the trapezoidal region in figure \ref{figab}. Consider the line $a+2b=|\Delta s|+2g$.  Over this line, we are summing terms $\binom{d_i-d_j}{a}\binom{\frac{d_i-d_j-a}{2}}{b} {2^{-(a+2b)}}$ that satisfy $a+2b = |\Delta s|+2g$.  This is just identical to the previous calculation \eqref{f0ev} with the substitution $|\Delta s|\rightarrow |\Delta s|+2g$.  The remaining coefficients in the sum (\ref{fullbinomexpress}) differ only for different lines: for a line specified by $g$, the coefficient is $\binom{|\Delta s|+2g}{g} \binom{\frac{d_i-d_j-(|\Delta s|+2g)}{2}}{f-g}(-1)^{{f-g}}$. Thus we are left with a single sum over $g$, which specifies the sum over distinct lines in the trapezoidal region:
\begin{align}
a'_{\ell}&=\sum_{g=0}^f\frac{ 2\pi }{(|\Delta s|+2g)!}\; \frac{(d_i-d_j)}{2}\,\Big(\frac{d_i-d_j-(|\Delta s|+2g)}{2}+1\Big)_{|\Delta s|+2g-1} \\
& \qquad \qquad\qquad \times  \binom{|\Delta s|+2g}{g} \binom{\frac{d_i-d_j-(|\Delta s|+2g)}{2}}{f-g}(-1)^{{f-g}}\ .\nn
\end{align}
Expanding out the binomial coefficients, we find
\begin{align}
a'_{\ell}=\sum_{g=0}^f 2\pi\; \frac{1}{(|\Delta s|+g)!g!(f-g)!}\; \frac{(d_i-d_j)}{2}\,\Big(\frac{d_i-d_j-(|\Delta s|+2f)}{2}+1\Big)_{|\Delta s|+f+g-1}\!\!(-1)^{f-g}\ . \nn
\end{align}
The Pochhammer symbol with $g=0$ is common to all members of the sum and so may be factored by splitting off this part of the Pochhammer symbol:
\begin{align}
a'_{\ell}&=2\pi\;\frac{(d_i-d_j)}{2}\; \frac{\left(\frac{d_i-d_j-(|\Delta s|+2f)}{2}+1\right)_{|\Delta s|+f-1}}{(|\Delta s|+f)!}\; (-1)^f\nn \\
&  \times \sum_{g=0}^f \frac{\left(-\frac{d_i-d_j+|\Delta s|}{2}-g+1\right)_g}{g!}\;\frac{(\ell-f-(f-g)+1)_{f-g}}{(f-g)!}\ , \nn
\end{align}
where we have used the identity
\begin{align}
\frac{1}{(|\Delta s|+g)!}=\frac{(\ell-2f+g+1)_{f-g}}{(|\Delta s|+f)!}\ .\nn
\end{align}
Define $\alpha=-\frac{d_i-d_j-\ell}{2}, \beta=-\frac{d_i-d_j+|\Delta s|}{2}, m=f, n=g$, and use eq. \eqref{pascal} to obtain
\begin{align}
a'_{\ell}&=2\pi\;\frac{(d_i-d_j)}{2}\; \frac{\Big(\frac{d_i-d_j-(|\Delta s|+2f)}{2}+1\Big)_{|\Delta s|+f-1}}{(|\Delta s|+f)!}\; (-1)^f\;\frac{\Big(-\frac{d_i-d_j-\ell}{2}-f+1\Big)_f}{f!}\nn \\
&=2\pi\;\frac{(d_i-d_j)}{(d_i-d_j-\ell)}\; \frac{\Big(\frac{d_i-d_j-\ell}{2}\Big)_{\frac{\ell+|\Delta s|}{2}}}{(\frac{\ell+|\Delta s|}{2})!}\;\frac{\Big(\frac{d_i-d_j-\ell}{2}\Big)_{\frac{\ell-|\Delta s|}{2}}}{(\frac{\ell-|\Delta s|}{2})!} \nn \\
&=(-1)^\ell\; 2\pi\;\frac{(d_i-d_j)}{(d_i-d_j-\ell)} \;\frac{\Big(1-\frac{d_i-d_j+\Delta s}{2}\Big)_{\frac{\ell+\Delta s}{2}}}{(\frac{\ell+\Delta s}{2})!}\;\frac{\Big(1-\frac{d_i-d_j-\Delta s}{2}\Big)_{\frac{\ell-\Delta s}{2}}}{(\frac{\ell-\Delta s}{2})!}\ .
\end{align}
This concludes the proof of eq. \eqref{als}.

\section{OPE coefficients \texorpdfstring{$\beta^p_{n}$}{TEXT} and \texorpdfstring{$\tilde\beta_{\tilde n}^p$}{TEXT}}\label{app_beta}

In this appendix we compute the coefficients $\beta^p_{n}$ and $\tilde\beta_{\tilde n}^p$ in the OPE of quasi primary fields $\phi_1$ and $\phi_2$, defined in eq. \eqref{OPEoperator}. Expressing the OPE in terms of a sum over quasi primaries $\phi_p$ allows us to derive exact expressions for $\beta^p_{n}$ and $\tilde\beta_{\tilde n}^p$, which solely depend on conformal dimensions of $\phi_1$, $\phi_2$ and $\phi_p$.

Let us consider first $\beta^p_{n}$, derived in eq. \eqref{betathreept}, which we rewrite here for convenience:
\begin{equation}
\beta^p_{n}=\frac{1}{n!\,(2h_p)_{(n)}}\;\frac{\langle \phi_p\mid(L_{1})^n\, \phi_1(z,\zb)\mid \phi_2\rangle}{z^{-h_1-h_2+h_p+n}\,\langle \phi_p \mid \phi_1(1,1)\mid \phi_2\rangle}\;\zb^{\ti{h}_1+\ti{h}_2-\ti{h}_p}\ . \label{app_betathreept}
\end{equation}
The expectation value in the denominator is just the structure constant $C_{p,1,2}$, which is assumed to be non-zero.\footnote{If the structure constant is zero for a given quasi primary $\phi_p$, then this family does not appear in the OPE.}  Consider the expectation value in the numerator. The contour defining $L_1$ is centered at 0, but with radius larger than $z$. One may pull this contour inward, which results in two contours centred at $0$ and $z$.  The contour around 0 produces an $L_1$ acting on $\mid \phi_2 \rangle$, which annihilates this state. The contour around the operator $\phi_1$ at $z$ yields:
\begin{equation}
\oint_{z} \frac{dz'}{2\pi i}\, z'^{2}\, T(z')= \oint_z \frac{dz'}{2\pi i}\, \Big((z'-z)^2+2 z (z'-z)+z^2\Big)\, T(z')=(L_{z,1}+2z L_{z,0}+z^2 L_{z,-1})\nn
\end{equation}
where the $z$ subscript on the $L$ operators denotes that they are centred at $z$.  These generators satisfy the $sl(2)$ algebra. We then find:
\begin{align}\label{ind_form}
(L_{z,1}+2zL_{z,0}+z^2 L_{z,-1})^n\phi_1(z,\zb) &= z^n \sum_{k=0}^n \frac{n!}{k!\,(n-k!)} \,(2h_1+k)_{(n-k)}\,z^k\, (L_{z,-1})^k\, \phi_1(z,\zb) \\
&=z^n \Big(z^{1-2h_1} \pa^n\big( z^{2h_1+n-1} \phi_1(z,\zb)\big)\Big)\ .\nn
\end{align}
This equation is proved by induction and the proof is provided in subsection \ref{app_ind}.

We thus obtain:
\begin{equation}
\langle \phi_p|L_{1}^n \phi_1(z,\zb)\mid \phi_2\rangle=z^{n+1-2h_1}\pa^n\left(z^{2h_1+n-1} \langle \phi_p\mid\phi_1(z,\zb)\mid \phi_2\rangle\right)\ , \label{L1eprexpand}
\end{equation}
which is a differential operator acting on the limiting form of a 3-point function. Plugging in
\begin{equation}
\langle \phi_p\mid\phi_1(z,\zb)\mid \phi_2\rangle= \frac{C_{p,1,2}}{z^{h_1+h_2-h_p}\,\zb^{\ti{h}_1+\ti{h}_2-\ti{h}_p}}\nn
\end{equation}
into eq. (\ref{L1eprexpand}), we find
\begin{equation}
\langle \phi_p|L_{1}^n \phi_1(z,\zb)\mid \phi_2\rangle={C_{p,1,2}}\, (h_1-h_2+h_p)_{n}\;z^{h_p-h_1-h_2+n}\,\zb^{\ti{h}_p-\ti{h}_1-\ti{h}_2}\ .\nn
\end{equation}
We next Include the antiholomorphic sector and obtain
\begin{align}
\langle \phi^{(n,\ti{n})}_p|\phi_1(z,\zb)\mid \phi_2\rangle&={C_{p,1,2}}\langle \phi_p\mid(L_{1})^n\,(\ti{L}_{1})^{\ti{n}}\; 
\phi_1(z,\zb)\mid \phi_2\rangle\nn \\
&={C_{p,1,2}} (h_1-h_2+h_p)_{n}\,(\ti{h}_1-\ti{h}_2+\ti{h}_p)_{\ti{n}}\;z^{h_p-h_1-h_2+n}\,\zb^{\ti{h}_p-\ti{h}_1-\ti{h}_2+\ti{n}}.\nn
\end{align}
Inserting this in eq. (\ref{app_betathreept}), we find
\begin{equation}
\beta^p_n=\frac{(h_1-h_2+h_p)_{n}}{n!\, (2h_p)_{n}}\ .
\end{equation}
A similar projection onto a state with $n=0$ yields
\begin{equation}
\ti{\beta}^p_{\ti{n}}=\frac{(\ti{h}_1-\ti{h}_2+\ti{h}_p)_{\ti{n}}}{\ti{n}!\, (2\ti{h}_p)_{\ti{n}}}\ .
\end{equation}

\subsection{Proof of eq. \eqref{ind_form}}\label{app_ind}
The proof of eq. \eqref{ind_form} by induction is fairly straightforward.  First, the $n=0$ case is trivially satisfied. We assume that the equation holds up to $n\geq1$.  For the case $n+1$ we have
\begin{align}
&(L_{z,1}+2zL_{z,0}+z^2 L_{z,-1})^{n+1}\phi_1(z,\zb)  \\
& =(L_{z,1}+2zL_{z,0}+z^2 L_{z,-1})z^n \sum_{k=0}^n \frac{n!}{k!\,(n-k)!} \,(2h_1+k)_{(n-k)}\,z^k\, (L_{z,-1})^k\, \phi_1(z,\zb)\ .\nn
\end{align}
Recalling that $L_1 L_{-1}^k \phi_1=[L_1,L_{-1}^k]\phi_1 = k(2h_1+(k-1))L_{-1}^{k-1} \phi_1$ for quasi primaries and substituting in the eigenvalue for $L_0$, we find
\begin{align}\label{3sums}
&(L_{z,1}+2zL_{z,0}+z^2 L_{z,-1})^{n+1}\phi_1(z,\zb)\\
=&\,z^n \sum_{k=0}^n \frac{n!}{k!\,(n-k)!} \,(2h_1+k)_{(n-k)}\,z^k\, k(2h_1+k-1)(L_{z,-1})^{k-1}\, \phi_1(z,\zb) \nn \\
+&\,z^{n+1} \sum_{k=0}^n \frac{n!}{k!\,(n-k)!} \,(2h_1+k)_{(n-k)}\,z^k\,(2h_1+2k) (L_{z,-1})^k\, \phi_1(z,\zb) \nn \\
+&\,z^{n+2} \sum_{k=0}^n \frac{n!}{k!\,(n-k)!} \,(2h_1+k)_{(n-k)}\,z^k\, (L_{z,-1})^{k+1}\, \phi_1(z,\zb)\ .\nn
\end{align}
These three sums may be rewritten by shifting the sum indices such that the summand terms all appear as $z^kL_{z,-1}^k$, being sure that no factorials of negative numbers nor negative subscripts on Pochhammer symbols appear in the sum.  In the first sum, we have
\begin{align}
&z^n \sum_{k=0}^n \frac{n!}{k!\,(n-k)!} \,(2h_1+k)_{(n-k)}\,z^k\, k(2h_1+k-1)(L_{z,-1})^{k-1}\, \phi_1(z,\zb) \\ 
&=z^n \sum_{k=1}^{n+2} \frac{n!(n+1-k)(n+2-k)}{(k-1)!\,(n+2-k)!} \,\frac{(2h_1+k-1)_{(n+2-k)}}{(2h_1+n)}\,z^k\, (L_{z,-1})^{k-1}\, \phi_1(z,\zb)\nn
\end{align}
where we have added two new terms in the sum, $k=n+1,n+2$, both of which are 0.  We have also dropped the term in the sum $k=0$ which is 0. This allows us to write $k/k!=1/(k-1)!$.  Moreover, we have recognized $(2h_1+k)_{(n-k)}(2h_1+k-1)=(2h_1+k-1)_{(n+1-k)}=(2h_1+k-1)_{(n+2-k)}/(2h_1+n)$. Shifting the indices of the sum on the second line, the above equation reads
\begin{align}\label{sum1}
&z^n \sum_{k=0}^n \frac{n!}{k!\,(n-k)!} \,(2h_1+k)_{(n-k)}\,z^k\, k(2h_1+k-1)(L_{z,-1})^{k-1}\, \phi_1(z,\zb)\\ 
&=z^{n+1} \sum_{k=0}^{n+1} \frac{n!(n-k)(n+1-k)}{k!\,(n+1-k)!} \,\frac{(2h_1+k)_{(n+1-k)}}{(2h_1+n)}\,z^k\, (L_{z,-1})^{k}\, \phi_1(z,\zb)\ .\nn 
\end{align}

The second sum in eq. \eqref{3sums} may be written as:
\begin{align}\label{sum2}
&z^{n+1} \sum_{k=0}^n \frac{n!}{k!\,(n-k)!} \,(2h_1+k)_{(n-k)}\,z^k\,(2h_1+2k) (L_{z,-1})^k\, \phi_1(z,\zb)\\
&=z^{n+1} \sum_{k=0}^{n+1} \frac{n!(n+1-k)}{k!\,(n+1-k)!} \,\frac{(2h_1+k)_{(n+1-k)}}{(2h_1+n)}\,z^k\,(2h_1+2k) (L_{z,-1})^k\, \phi_1(z,\zb)\ .\nn 
\end{align}
The third sum in eq. \eqref{3sums} sum reads:
\begin{align}\label{sum3}
& z^{n+2} \sum_{k=0}^n \frac{n!}{k!\,(n-k)!} \,(2h_1+k)_{(n-k)}\,z^k\, (L_{z,-1})^{k+1}\, \phi_1(z,\zb)\\
&=z^{n+1} \sum_{k=-1}^n \frac{n!(k+1)}{(k+1)!\,(n-k)!} \,(2h_1+k)_{(n-k)}\,z^{k+1}\, (L_{z,-1})^{k+1}\, \phi_1(z,\zb)\nn \\
&=z^{n+1} \sum_{k=0}^{n+1} \frac{n!k}{k!\,(n+1-k)!} \,\frac{(2h_1+k-1)}{(2h_1+n)}(2h_1+k)_{(n+1-k)}\,z^{k}\, (L_{z,-1})^{k}\, \phi_1(z,\zb)\ .\nn
\end{align}

The three sums \eqref{sum1}-\eqref{sum3} now all have the same overall power $z^{n+1}$, and the same factor of $z^k L_{z,-1}^k$ in the summand.  The coefficients that remain are
\begin{align}
& \frac{n!}{k!(n+1-k)!}(2h_1+k)_{(n+1-k)} \frac{\big((n-k)(n+1-k)+(n+1-k)(2h_1+2k)+k(2h_1+k-1)\big)}{2h_1+n} \nn \\
&= \frac{(n+1)!}{k!(n+1-k)!}(2h_1+k)_{(n+1-k)} 
\end{align}
which completes the proof, i.e. eq. \eqref{ind_form} with $n\rightarrow n+1$ has been proved.  The second line of eq. \eqref{ind_form} follows by expressing $L_{z,-1}^k \phi_1(z,\zb)=\pa^k \phi_1$ and 
\begin{equation}
z^n \Big(z^{1-2h_1} \pa^n\big( z^{2h_1+n-1} \phi_1(z,\zb)\big)\Big)=z_1^n \Big(z_1^{1-2h_1} (\pa_1+\pa_2)^n\big( z_1^{2h_1+n-1} \phi_1(z_2,\zb_2)\big)\Big) \Big|_{z_1=z_2=z}\ ,
\end{equation}
and expanding the operator $(\pa_1+\pa_2)^n$ in a binomial expansion.

\section{Jacobi polynomials and Hypergeometric functions}\label{app_3pf}

\subsection{Constant counter terms}\label{app_3pf_jacobi}
The Jacobi polynomials are defined as
\be\label{jacobi}
P^{(\alpha,\beta)}_{n}(x)\equiv \sum_{m=0}^{n}\;\frac{(\alpha+\beta+n+1)_{m}\,(\alpha+m+1)_{n-m}}{m!\,(n-m)!}\;\Big(\frac{x-1}2\Big)^m\ .
\ee 
In section \ref{eps0ct} we compute the contribution from constant (i.e. order $\epsilon^0$) counter terms for 3-point functions. Below we summarise such counter terms for quasi primaries $\phi_k$, $\phi_j$, $\phi_i$, inserted respectively at $z_3$, $z_2$, $z_1$. Their contribution to the first term on the \textsc{rhs} of eq. \eqref{expand3ptPI} is denoted as $\mathcal C_k$, $\mathcal C_j$ and $\mathcal C_i$ --- see eqs. \eqref{Ck} and \eqref{Ck_i}. To order $\epsilon$ we have:
\begin{align}\label{3pt_ct_123}
\mathcal C_{k,0}^{\rm hol}&=  \sum_{\substack{p, \ti{h}_p=\ti{h}_k \\
h_k-h_p\in {\mathbb Z}^+}}  \frac{2\pi \lambda\,C_{D,k,p}\;\frac{(-1)^{h_k-h_p}}{(h_k-h_p)(2h_p)_{h_k-h_p}}\,\pa_3^{{h}_k-{h}_p}
\langle \phi_i(z_1,\zb_1)\; \phi_j(z_2,\zb_2)\; \phi_p(z_3,\zb_3)\rangle}{{(z_{12})^{-(h_i+h_j-h_k)}(z_{13})^{-(h_i+h_k-h_j)}(z_{23})^{-(h_j+h_k-h_i)}}\;\times\; ({\rm a.h.})}\\
&=\sum_{\substack{p, \ti{h}_p=\ti{h}_k \\
h_k-h_p\in {\mathbb Z}^+}} 2\pi \lambda\, C_{D,k,p}\,C_{i,j,p}\,\frac{(h_k-h_p-1)!}{(2h_p)_{h_k-h_p}}\;P^{(-h_j-h_k+h_i,-h_i-h_k+h_j)}_{h_k-h_p}\Big(\frac{z_{13}}{z_{12}}+\frac{z_{23}}{z_{12}}\Big)\ ,\nn
\end{align}
\begin{align}
\mathcal C_{j,0}^{\rm hol}&=\sum_{\substack{p, \ti{h}_p=\ti{h}_j \\
h_j-h_p\in {\mathbb Z}^+}}  \frac{ 2\pi \lambda\,C_{D,j,p}\;\frac{(-1)^{h_j-h_p}}{(h_j-h_p)\,(2h_p)_{h_j-h_p}}\,\pa_2^{h_j-h_p}\langle \phi_i(z_1,\zb_1)\; \phi_p(z_2,\zb_2)\; \phi_k(z_3,\zb_3)\rangle}
{{(z_{12})^{-(h_i+h_j-h_k)}(z_{13})^{-(h_i+h_k-h_j)}(z_{23})^{-(h_j+h_k-h_i)}}\;\times\;({\rm a.h.})}\nn\\
&= \sum_{\substack{p, \ti{h}_p=\ti{h}_j \\
h_j-h_p\in {\mathbb Z}^+}} 2\pi \lambda\, C_{D,j,p}\;C_{i,p,k}\;\frac{(h_j-h_p-1)!}{(2h_p)_{h_j-h_p}}P^{(-h_i-h_j+h_k,-h_j-h_k+h_i)}_{h_j-h_p}\Big(\frac{z_{21}}{z_{13}}+\frac{z_{23}}{z_{13}}\Big)\ , \nn
\end{align}
\begin{align}
\mathcal C_{i,0}^{\rm hol}&= \sum_{\substack{p, \ti{h}_p=\ti{h}_i \\
h_i-h_p\in {\mathbb Z}^+}}\frac{ 2\pi \lambda \,C_{D,i,p}\;\frac{(-1)^{h_i-h_p}}{(h_i-h_p)\,(2h_p)_{h_i-h_p}}\; \pa_1^{h_i-h_p}\langle\phi_p(z_1,\zb_1)\; \phi_j(z_2,\zb_2)\; \phi_k(z_3,\zb_3)\rangle}
{{(z_{12})^{-(h_i+h_j-h_k)}(z_{13})^{-(h_i+h_k-h_j)}(z_{23})^{-(h_j+h_k-h_i)}}\;\times\;({\rm a.h.})}\nn\\
&= \sum_{\substack{p, \ti{h}_p=\ti{h}_i \\
h_i-h_p\in {\mathbb Z}^+}}2\pi \lambda\, C_{D,i,p}\;C_{p,j,k}\;\frac{(h_i-h_p-1)!}{(2h_p)_{h_i-h_p}}\;P^{(-h_i-h_j+h_k,-h_i-h_k+h_j)}_{h_i-h_p}\Big(\frac{z_{12}}{z_{23}}+\frac{z_{13}}{z_{23}}\Big)\ ,\nn
\end{align}  
along with their anti-holomorphic counterparts (with analogous restrictions on the sums).

\subsection{Conformal blocks and crossing symmetry}\label{app_3pf_hyper}
The ordinary hypergeometric function is defined as:
\be\label{hyper}
_2F_1\big(\genfrac{}{}{0pt}{1}{a,b}{c};z\big)\equiv \sum_{n=0}^\infty \frac{(a)_{n}\,(b)_{n}}{(c)_{n}}\; \frac{z^n}{n!}\ ,
\ee
where $|z|<1$. The 4-point function $G_{3,4}^{2,1}$ defined in eq. \eqref{sl2_4pf} is expressed in terms of hypergeometric function --- see eq. \eqref{3pf_hyper}. Crossing symmetry relations \eqref{cross_sym} yield:
\begin{align}\label{app_3pf_cross}
&G_{3,4}^{2,1}(z,\zb) = \sum_p C_{p,3,4} \;C_{p,2,1}\;z^{-h_p^{3,4}}\;\zb^{-\ti{h}_p^{3,4}}
\,_2F_1\bigg(\genfrac{}{}{0pt}{}{h^{p,2}_1,h^{p,3}_4}{2h_p};z\bigg) \;
_2F_1\bigg(\genfrac{}{}{0pt}{}{\ti{h}^{p,2}_1,\ti{h}^{p,3}_4}{2\ti{h}_p};\zb\bigg)\\[5pt]
&= \sum_p C_{p,3,2}\; C_{p,4,1}\;(1-z)^{-h^{2,3}_p}\;(1-\zb)^{-\ti{h}^{2,3}_p}
\,_2F_1\bigg(\genfrac{}{}{0pt}{}{h^{p+4}_1,h^{p,3}_2}{2h_p};1-z\bigg) \;
_2F_1\bigg(\genfrac{}{}{0pt}{}{\ti{h}^{p,4}_1,\ti{h}^{p,3}_2}{2\ti{h}_p};1-\zb\bigg)\nn\\[5pt]
&= \frac{1}{z^{2h_3}\zb^{2\ti{h}_3}}\sum_p C_{p,3,1} \;C_{p,2,4}\;\Big(\frac1z\Big)^{-h^{1,3}_p}\Big(\frac1{\zb}\Big)^{-\ti{h}^{1,3}_p}
\,_2F_1\bigg(\genfrac{}{}{0pt}{}{h^{p,2}_4,h^{p,3}_1}{2h_p};\frac1z\bigg) \;
_2F_1\bigg(\genfrac{}{}{0pt}{}{\ti{h}^{p,2}_4,\ti{h}^{p,3}_1}{2\ti{h}_p};\frac1{\zb}\bigg)\ ,\nn
\end{align}
where we have used the notation \eqref{hshbars}. For the perturbation integral of our interest \eqref{4ptconvention}, we replace $1\rightarrow i$, $2\rightarrow j$, $3\rightarrow D$, and $4\rightarrow k$. Recalling that the deformation operator has dimensions $h_D=\ti{h}_D=1$, we obtain
\begin{align}\label{app_3pf_cross_d}
& G_{D,k}^{j,i}(z,\zb) = \sum_p C_{p,D,k}\; C_{p,j,i}\;z^{h^p_k-1}\;\zb^{\ti{h}^p_k-1}
\,_2F_1\bigg(\genfrac{}{}{0pt}{}{h^{p,j}_i,h^p_k+1}{2h_p};z\bigg) \;
_2F_1\bigg(\genfrac{}{}{0pt}{}{\ti{h}^{p,j}_i,\ti{h}^p_k+1}{2\ti{h}_p};\zb\bigg) \\[5pt]
&= \sum_p C_{p,D,j}\; C_{p,k,i}\;(1-z)^{h^p_j-1}\;(1-\zb)^{\ti{h}^p_j-1}
\,_2F_1\bigg(\genfrac{}{}{0pt}{}{h^{p,k}_i,h^p_j+1}{2h_p};1-z\bigg) \;
_2F_1\bigg(\genfrac{}{}{0pt}{}{\ti{h}^{p,k}_i,\ti{h}^p_j+1}{2\ti{h}_p};1-\zb\bigg)\nn \\[5pt]
&= \frac{1}{z^{2}\zb^{2}}\sum_p C_{p,D,i}\; C_{p,j,k}\; \Big(\frac1z\Big)^{h^p_i-1}\Big(\frac1{\zb}\Big)^{\ti{h}^p_i-1}
\,_2F_1\bigg(\genfrac{}{}{0pt}{}{h^{p,j}_k,h^p_i+1}{2h_p};\frac1z\bigg) \;
_2F_1\bigg(\genfrac{}{}{0pt}{}{\ti{h}^{p,j}_k,\ti{h}^p_i+1}{2\ti{h}_p};\frac1{\zb}\bigg)\ .\nn
\end{align}

\section{Compact boson warm up}\label{app_bos}
\subsection{\texorpdfstring{$\delta h$}{TEXT}}\label{freebos_2pf}

For the first warm up, we compute the anomalous dimensions, namely the change to the conformal dimensions, of the quasi primary operators $V_{\vec k,a,b}$ \eqref{quasidefine}. We may easily compute exact expressions for the conformal dimensions in the perturbed theory with $R'=\sqrt{1-\lambda}\,R$, and then expand this to all orders in $\lambda$.  To the first order we find:
\begin{align}\label{weightshift}
h=\frac{k^{'2}_L}{2}+a&=\frac{1}{2}\Big(\frac{m^2}{R'^2}+\frac{w^2R'^2}{2}+mw\Big)+a=\frac{k^{2}_L}{2}+a -\frac{\lambda}{2}\Big(\frac{m^2}{R^2}-\frac{w^2R^2}{4}\Big)+O(\lambda^2)\ , \\
\ti{h}=\frac{k^{'2}_R}{2}+b& =\frac{1}{2}\Big(\frac{m^2}{R'^2}+\frac{w^2R'^2}{2}-mw\Big)+b=\frac{k^{2}_R}{2}+b-\frac{\lambda}{2}\Big(\frac{m^2}{R^2}-\frac{w^2R^2}{4}\Big)+O(\lambda^2)\ .\nn
\end{align}
We only expect shifts in the sectors with non-zero momentum $\vec{k}$, given that these are the operators with $R$ dependence in their dimensions.

Let us now compute this shift in conformal perturbation theory. The integrand of the perturbation integral is the 3-point functions $\lambda\,\langle V_{-\vec k,a,b} V_{\vec k,a,b} \mathcal{O}_D\rangle$, where $\mathcal{O}_D =\frac{1}{4\pi} \pa  x \pab  x$ is the deformation operator --- see eqs. \eqref{PIexpand} and \eqref{fbpert}. For simplicity we specialize to $a=b=0$ and consider the vertex operators $V_{\vec k}$ \eqref{vo}. We obtain:
\begin{align}
\langle V_{-\vec k}(z_1,\zb_1)\; V_{\vec k}(z_2,\zb_2)\;\mathcal{O}_D(z,\zb)\rangle
=-\frac{e^{i\pi mw}}{4\pi}\; z_{12}^{-k_L^2+1}\,\zb_{12}^{-k_R^2+1}\,\frac{k_Lk_R}{(z-z_1)(z-z_2)(\zb-\zb_1)(\zb-\zb_2)}\ ,\nn
\end{align}
where the overall phase is resulted from commuting the cocycles.\footnote{In fact, by definition, this is the same phase as appears in the 2-point function (\ref{2ptMomWind}).} The result is symmetric under mapping $\vec{k}\rightarrow -\vec{k}$, and so is diagonalized by the basis (\ref{diagbasis}).  Denoting $W^\pm\!\equiv W^\pm_{\vec k,0,0}$, we obtain
\begin{equation}
C_{D,W^+,W^+}=C_{D,W^-,W^-}=-\frac{1}{4\pi}\, k_L k_R\ .
\end{equation}
The shift to the dimensions (\ref{hshifts})  is then found to be:
\begin{equation}\label{bos_anomdim}
\frac{\pa h}{\pa \lambda}\Big|_{\lambda=0}\!\!\!=\frac{\pa \ti{h}}{\pa \lambda}\Big|_{\lambda=0}\!\!\!=
2\pi\, C_{D,W^\pm,W^\pm}=-\frac{1}{2}\, k_L k_R=-\frac{1}{2}\Big(\frac{m^2}{R^2}-\frac{w^2R^2}{4}\Big)\ ,
\end{equation}
which matches with the exact shifts obtained in eq. (\ref{weightshift}). One may repeat this calculation for the other $W$ operators (\ref{diagbasis}).  The structure constant in all four cases is $C_{D,W^{\pm}_{\vec k,a,b}\!\!\!\!,W^\pm_{\vec k,a,b}}\!\!\!\!\!\!=-\frac{1}{4\pi} k_L k_R$. Therefore, they all experience the same shift \eqref{bos_anomdim}, agreeing with the coefficient of $\lambda$ in (\ref{weightshift}).

\subsection{\texorpdfstring{$\delta C_{i,j,k}$}{TEXT}}\label{app_freebos_3pf}

For the second warm up we compute the shift to the structure constant for an easy example where the simplified domain of integration \eqref{4ptallowed0} is sufficient to evaluate the shift, and there are no corrections to it coming from the exact domain \eqref{4ptallowedfull_first}. Consider the 3-point function:
\begin{align}
\langle V'_{-\vec{k}'}(z_1,\zb_1)\;\pa X\,\pab X(z_2,\zb_2)\;V'_{\vec{k}'}(z_3,\zb_3) \rangle=
\frac{-e^{i\pi mw}\,k'_Lk'_R}{z_{12}\,z_{13}^{k^{'2}_L-1}\,z_{23}\;\;\zb_{12}\,\zb_{13}^{k^{'2}_R-1}\,\zb_{23}}\ ,\nn
\end{align}
where $V'$ is the vertex operator in the perturbed theory \eqref{voprime}. Diagonaizing in the $W$ basis \eqref{diagbasis}, and recalling the notation $W^{'\pm}\!\equiv W^{'\pm}_{\vec k,0,0}$, we may expand the structure constant in $\lambda$
\begin{equation}
C_{\pa X\pab X,W^{'\pm},W^{'\pm}}=-k'_Lk'_R= -k_Lk_R+\lambda\Big(\frac{m^2}{R^2}+\frac{w^2R^2}{4}\Big)
+O(\lambda)\ . \label{CtoLinear}
\end{equation}
Therefore, the exact computation gives $\delta C_{\pa X\pab X,W^{\pm},W^{\pm}}\!\!=\lambda({m^2}/{R^2}+{w^2R^2}/{4})$.

We next compute the shift in conformal perturbation theory. The 3-point function in the unperturbed theory is $\langle V_{-\vec{k}'}(z_1,\zb_1)\;\pa x\,\pab x(z_2,\zb_2)\;V_{\vec{k}'}(z_3,\zb_3) \rangle$. Consider the operator at $z_2$.\footnote{ \label{4pifoot} The overall factor $1/4\pi$ is not included. The inserted operator is not the deformation $\mathcal O_D=\frac{1}{4\pi} \pa  x \pab  x$ defined by the perturbative term \eqref{fbpert}, but rather the canonically normalized operator of dimension $(1,1)$ in the theory.}  The only operator of lower dimension than $\pa x$ in the zero momentum sector is the identity, which has no $sl(2)$ descendants. Therefore, at location $z_2$, there is no quasi primary to take derivatives of to make an operator with $h=1$. (A similar argument holds for $\pab x$). As such, there is no contribution from the crescent region at $z_2$ \eqref{3ptinegralcorrect_1} to the shift $\delta C$. This simplifies matters as the latter contribution is rather involved --- see the discussion below eq. \eqref{3ptinegralcorrect_inf}. 

Similarly, we observe that the operators inserted at $z_1$ and $z_2$ are of the lowest dimension in their momentum class. Therefore, there are no contributions from crescents at $z_1$ and $z_3$ as well --- see eqs. \eqref{3ptinegralcorrectS} and \eqref{3ptinegralcorrect_inf}. Hence, no additional terms arise from integrating over the full domain (\ref{4ptallowedfull_first}) as compared to the simplified domain (\ref{4ptallowed0}). We shall verify this below by an explicit computation.

The perturbation integral \eqref{4ptconvention} is given by:
\begin{equation}
\hat{\mathcal I}=-\lambda\int dz^2 \langle V_{-\vec{k}}(z_1,\zb_1)\;\pa x\,\pab x(z_2,\zb_2)\;
V_{\vec{k}}(z_3,\zb_3)\;\frac{1}{4\pi}\pa x\,\pab x(z,\zb)\rangle\ .\nn
\end{equation}
We would like to evaluate this over the simplified domain (\ref{4ptallowed0}) and extract its constant term, which gives the value of the shift $\delta C$. The 4-point function in the integrand reads:
\begin{align}
&\langle V_{-\vec{k}}(z_1,\zb_1)\;\pa x\,\pab x(z_2,\zb_2)\;V_{\vec{k}}(z_3,\zb_3)\;\pa x\pa x(z,\zb)\rangle = \nn\\
&\qquad\frac{e^{i\pi n w}}{z_{{13}}^{k_L^2-2}z_{23}^2\,(z_1-z)^2\;\;\zb_{{13}}^{k_R^2-2}\zb_{23}^2\,(\zb_1-z)^2}
\bigg(\frac{k_L^2}{\hat z}+\frac{1}{(\hat z-1)^2}\bigg)\bigg(\frac{k_R^2}{\bar{\hat z}}+\frac{1}{(\bar{\hat z}-1)^2}\bigg)\ , \nn
\end{align}
where $\hat z$ is defined in eq. \eqref{zhat}.
The correlator is invariant under $\vec{k}\rightarrow-\vec{k}$, thus we may consider the $W$ basis \eqref{diagbasis}. Integrating over the simplified domain \eqref{4ptallowed0}, $\mathcal I_{\rm s}$ \eqref{perturbstruct} reads:
\begin{equation}\label{IsWOWO}
\mathcal I_{\rm s}=-\frac{\lambda}{4\pi} \int_{\rm simp}\!\!d^2\hat{z}\bigg(\frac{k_L^2}{\hat{z}}+\frac{1}{(\hat{z}-1)^2}\bigg)
\bigg(\frac{k_R^2}{\bar{\hat{z}}}+\frac{1}{(\bar{\hat{z}}-1)^2}\bigg)\ .
\end{equation}
We compute the 4 terms individually and denote them as $\mathcal I_{\rm s}^a$, $a=\{1,\cdots\,4\}$.  We first consider
\begin{equation}
-\frac{4\pi}{\lambda}\,\mathcal I_{\rm s}^1\!:=\!\!\int_{\rm simp}\!\!\!\!\!\!d^2\hat{z}\,\frac{k_L^2k_R^2}{\hat{z}\,\hat{z}}=4\pi k_L^2k_R^2 
\bigg(\!\!\ln\bigg(\frac{|z_{12}||z_{13}|}{|z_{23}|\epsilon}\bigg)\!-\!\ln\bigg(\frac{|z_{{12}}|\epsilon}{|z_{23}||z_{{13}}|}\bigg)\!\bigg)
\!\!-k_L^2k_R^2\!\!\int_{|z-1|\leq0}\!\!\!\!\!\!\!\!\! d^2\hat{z}\; \frac{1}{1-\hat{z}}\;\frac{1}{1-\bar{\hat{z}}}\ ,\nn
\end{equation}
where we have explicitly written out the excised hole at $\hat{z}=1$. The last integral is calculated to be ${2\pi \ln(1-|z_{13}|^2\epsilon^2/(|z_{23}|^2|z_{12}|^2))}$, however, its value is unimportant: the integrand is a function which is finite at $\hat{z}=1$, and is integrated over a vanishingly small region. Hence, it represents a purely perturbative expansion in $\epsilon$. There is therefore no $\epsilon^0$ piece contributing from $\mathcal I_{\rm s}^1$, i.e. $\mathcal I_{\rm s,0}^{1}=0$. In fact, the coefficient of the log terms in the first term on the \textsc{rhs} give the shift to the conformal dimension of $V'$ (see eq. \eqref{Expand3pt} and appendix \ref{freebos_2pf}), and the second integral shows that the shift to the conformal dimension of $\pa x\pab  x$ is 0, as it should. (This is because the exactly marginal operator does not acquire an anomalous dimension). 

We next consider another term in eq. \eqref{IsWOWO}:
\begin{equation}
\mathcal I_{\rm s}^2:=-\frac{\lambda}{4\pi}\int_{\rm simp}\!\!\!\!d^2\hat{z}\;\frac{k_L^2}{\hat{z}}\;\frac{1}{(\bar{\hat{z}}-1)^2} \label{int2}\ ,\nn
\end{equation}
which has singularities only at $\hat{z}=0,1$ and $\infty$. We write the integrand as $-\pa_{\bar{\hat{z}}}(k_L^2/(\hat{z}(\bar{\hat{z}}-1))$. This gives a delta function when $\pa_{\bar{\hat{z}}}$ acts on the simple pole term $1/\hat{z}$.  However, this is in the excised region.  We may therefore use the divergence theorem and write:
\begin{equation}\label{contourC2}
\mathcal I_{\rm s}^2=-\frac{\lambda i}{4\pi}\oint_{\pa{\rm\,simp}}\!\!\!\!\!d\hat{z}\, \bigg(\frac{k_L^2}{\hat{z}}\;\frac{1}{\bar{\hat{z}}-1}\bigg)\ .
\end{equation}
On the boundary ``$\pa{\, \rm simp}$'' we take counterclockwise integration for the excised region of radius $1/\epsilon$ and clockwise for the regions of radius $\epsilon$. The three circles are given by --- {\it c.f.} eq. \eqref{4ptallowed0}:
\begin{align}\label{3circles}
&\hat{z}=\frac{|z_{12}|\epsilon}{|z_{23}||z_{13}|}e^{i \phi}, \qquad  \hat{z}=1+\frac{|z_{13}|\epsilon}{|z_{23}||z_{12}|}e^{i \phi}, \qquad \hat{z}=\frac{|z_{12}||z_{13}|}{|z_{23}|\epsilon}e^{i \phi}.
\end{align}
The contour integral \eqref{contourC2} then reads:
\begin{align}\label{bosC2cons}
\mathcal I_{\rm s}^2=-\frac{\lambda}{4\pi}&\Bigg\{\int_0^{2\pi}\!\!\!d\phi\Bigg(\frac{k_L^2}{\frac{|z_{12}|\epsilon}{|z_{23}||z_{13}|}e^{-i x}-1}\Bigg)
+\int_0^{2\pi}\!\!\!e^{2i\phi}d\phi \Bigg(\frac{k_L^2}{1+\frac{|z_{13}|\epsilon}{|z_{23}||z_{12}|}e^{i\phi}}\Bigg)\\
&\!\!- \int_0^{2\pi}\!\!\!d\phi\; \frac{|z_{23}|\epsilon}{|z_{12}||z_{13}|}\,e^{i\phi}\Bigg( \frac{k_L^2}{1- \frac{|z_{23}|\epsilon}{|z_{12}||z_{13}|}e^{i\phi}}\Bigg)
\Bigg\}=-\frac{\lambda}{4\pi}\Big(-2\pi k_L^2+0+0\Big) =\frac{\lambda}{2}\, k_L^2\ ,\nn
\end{align}
and we find $\mathcal I_{\rm s,0}^{2}=\frac{\lambda}{2}\, k_L^2$. Following the same steps, we obtain:
\begin{equation}\label{bosC3cons}
\mathcal I_{\rm s}^3:=-\frac{\lambda}{4\pi}\int_{\rm simp}\!\!\!\!\! d^2\hat{z}\;\frac{k_R^2}{\bar{\hat{z}}}\;\frac{1}{{(\hat{z}-1)}^2}=
\frac{\lambda}{2}\, k_R^2=\mathcal I_{\rm s,0}^{3}\ .
\end{equation}

Finally, we address the integral
\be
\mathcal I_{\rm s}^4:=-\frac{\lambda}{4\pi}\int_{\rm simp}\!\!\!\!\!d^2 \hat{z}\; \frac{1}{(\hat{z}-1)^2}\;\frac{1}{(\bar{\hat{z}}-1)^2}
=\frac{i\lambda}{4\pi}\oint_{\pa{\rm\; simp}}\!\!\!\!\!d\hat{z} \left( \frac{1}{(\hat{z}-1)^2}\;\frac{1}{(\bar{\hat{z}}-1)}\right) \ ,\nn
\ee
where we used the divergence theorem to get the second equality. Using eq. \eqref{3circles}, we have
\begin{align}
\mathcal I_{\rm s}^4&=-\frac{\lambda}{4\pi}\Bigg\{\int_0^{2\pi}d\phi\;\frac{|z_{12}|\epsilon}{|z_{23}||z_{13}|}\;e^{i\phi}
\left( \frac{1}{(\frac{|z_{12}|\epsilon}{|z_{23}||z_{13}|}e^{i\phi}-1)^2}\frac{1}{(\frac{|z_{12}|\epsilon}{|z_{23}||z_{13}|}e^{-i\phi}-1)}\right)\nn \\
&\qquad\quad +\int_0^{2\pi}d\phi\;\frac{|z_{13}|\epsilon}{|z_{23}||z_{12}|}e^{i\phi}
 \left( \frac{1}{(\frac{|z_{13}|\epsilon}{|z_{23}||z_{12}|}e^{i\phi})^2}\frac{1}{\frac{|z_{13}|\epsilon}{|z_{23}||z_{12}|}e^{-i\phi}}\right) \nn \\
&\qquad\quad - \int_0^{2\pi}d\phi\;\left(\frac{|z_{23}|\epsilon}{|z_{12}||z_{13}|}\right)^2  \left( \frac{1}{(1-\frac{|z_{23}|\epsilon}{|z_{12}||z_{13}|}e^{-i\phi})^2}\frac{1}{(1-\frac{|z_{23}|\epsilon}{|z_{12}||z_{13}|}e^{i\phi})}\right)\Bigg\}\ . \nn
\end{align}
The integrals may be computed exactly, however, it is more instructive to simply extract the constant term. The first integral has one power of $\epsilon e^{i\phi}$ from the measure $d\hat{z}$. We expand the integrand in $\hat{z}$ and $\bar{\hat{z}}$. To cancel the $\epsilon$, only terms of the form $\frac{1}{\hat{z}} \left(\frac{z}{\bar{\hat{z}}}\right)^n$ contribute (for any $n\in\bz$).  However, simultaneously we must cancel the $e^{i\phi}$, and so only the $n=0$ term survives. Thus, only a term proportional to $1/\hat{z}$ in the expansion contributes to a constant. Note that $1/(\hat{z}-1)^2=-\pa (1/(\hat{z}-1))$, and so no such $1/\hat{z}$ term can exist: the function is a total derivative of a well defined function, excluding log terms which would lead to $1/\hat{z}$ terms.  Similar statements hold for the other two integrals.  For completeness, we give the full answer:
\begin{align}
\mathcal I_{\rm s}^4=\frac{\lambda}{2}\Bigg\{\frac{\left(\frac{|z_{12}|\epsilon}{|z_{23}||z_{13}|}\right)^2}
{\Big(1-\big(\frac{|z_{12}|\epsilon}{|z_{23}||z_{13}|}\big)^2\Big)^2}-\left(\frac{|z_{23}||z_{12}|}{|z_{13}|\epsilon}\right)^2+
\frac{\left(\frac{|z_{23}|\epsilon}{|z_{12}||z_{13}|}\right)^2}{\Big(1-\big(\frac{|z_{23}|\epsilon}{|z_{12}||z_{13}|}\big)^2\Big)^2}\Bigg\}\ ,
\end{align}
which clearly has no $O(\epsilon^0)$ terms: $\mathcal I_{\rm s,0}^{4}=0$.

All in all, the only constant terms come from eqs. \eqref{bosC2cons} and \eqref{bosC3cons} and we find:
\begin{equation}
\delta C^{\rm pert}_{\pa X\pab X,W^{\pm},W^{\pm}}\!\!=\sum_{a=1}^4\mathcal I_{\rm s,0}^{a}=
\frac{\lambda}{2}(k_L^2+k_R^2)=\lambda\Big(\frac{m^2}{R^2}+\frac{w^2R^2}{4}\Big)\ ,
\end{equation}
which matches the shift computed exactly in eq. (\ref{CtoLinear}). Thus, in this simple example, the ``minimal subtraction" scheme is sufficient to compute the shift $\delta C$. This, however, is a special case. In general, the results of section \ref{cpt_3pf} show that the crescent regions contribute to $\delta C$ and must be taken into account to compute the correct shift. The example worked out in subsection \ref{freebos_3pf} presents such a case where the integral over the simplified domain is insufficient to calculate the perturbative change to a structure constant.

\small\baselineskip=.87\baselineskip
\bibliographystyle{utphys}
\bibliography{3pf}

\end{document}